%% file: main.tex
\PassOptionsToPackage{expansion=false}{microtype}
\documentclass[acmsmall, screen, nonacm]{acmart}

\setcopyright{acmlicensed}
\copyrightyear{2025}
\acmYear{2025}
\acmDOI{}

\acmJournal{TOSEM}
\acmVolume{1}
\acmNumber{1}
\acmArticle{1}
\acmMonth{1}

\usepackage{tcolorbox}
\usepackage{adjustbox}
\usepackage{needspace}
\usepackage{float}
\usepackage{placeins}
\usepackage{multirow}
\usepackage{tabularx}
\usepackage{makecell}
\usepackage{subcaption}
\usepackage{mathtools,nccmath}
\usepackage{cleveref}
\usepackage{custom}

\makeatletter
\def\@ACM@checkaffil{%
  \if@ACM@instpresent\else
    \ClassWarningNoLine{\@classname}{No institution present for an affiliation}%
  \fi
  \if@ACM@citypresent\else
    \ClassWarningNoLine{\@classname}{No city present for an affiliation}%
  \fi
  \if@ACM@countrypresent\else
    \ClassWarningNoLine{\@classname}{No country present for an affiliation}%
  \fi
}
\makeatother

\begin{document}

\title{Multi-LLM Orchestration for High-Quality Code Generation: Exploiting Complementary Model Strengths}

\author{Huashan Chen}
\authornote{Both authors contributed equally to this research.}
\email{chenhuashan@iie.ac.cn}
\affiliation{%
  \institution{Chinese Academy of Science}
  \city{Beijing}
  \country{China}
}

\author{Zhenyu Qi}
\authornotemark[1]
\authornote{Corresponding author.}
\email{qzydustin@arizona.edu}
\affiliation{%
  \institution{University of Arizona}
  \city{Tucson}
  \country{USA}
}

\author{Haotang Li}
\email{haotangl@arizona.edu}
\affiliation{%
  \institution{University of Arizona}
  \city{Tucson}
  \country{USA}
}

\author{Hong Chen}
\email{chenhong2024@iie.ac.cn}
\affiliation{%
  \institution{Chinese Academy of Science}
  \city{Beijing}
  \country{China}
}

\author{Jinfu Chen}
\email{jinfuchen@whu.edu.cn}
\affiliation{%
  \institution{Wuhan University}
  \city{Wuhan}
  \country{China}
}

\author{Kebin Peng}
\email{pengk24@ecu.edu}
\affiliation{%
  \institution{East Carolina University}
  \city{Greenville}
  \country{USA}
}

\author{In Kee Kim}
\email{inkee.kim@uga.edu}
\affiliation{%
  \institution{University of Georgia}
  \city{Athens}
  \country{USA}
}

\author{Kyu Hyung Lee}
\email{kyuhlee@uga.edu}
\affiliation{%
  \institution{University of Georgia}
  \city{Athens}
  \country{USA}
}

\author{Sen He}
\email{senhe@arizona.edu}
\affiliation{%
  \institution{University of Arizona}
  \city{Tucson}
  \country{USA}
}

\author{Weiyi Shang}
\email{wshang@uwaterloo.ca}
\affiliation{%
  \institution{University of Waterloo}
  \city{Waterloo}
  \country{Canada}
}

\renewcommand{\shortauthors}{Chen et al.}

\begin{abstract}

Large Language Models (LLMs) are central to automated code generation, yet most approaches use one model at every stage, from writing code to repairing and optimizing it.
LLMs have complementary strengths: their failures do not fully overlap across programming languages, algorithmic categories, and development stages, allowing one model to succeed where another fails.
For example, in Python, the model with the highest code-generation accuracy ranks last at code repair.
These complementary strengths suggest that using one model to generate code, another to repair it, and a third to refine it can achieve higher correctness and better execution performance than using a single model throughout.
To utilize these complementary strengths, we propose PerfOrch, a multi-LLM-based multi-agent system consisting of four agents for categorization, generation, debugging, and refinement.
The generation, debugging, and refinement agents select models using a Memory module that ranks candidate LLMs by their measured performance for each stage, programming language, and algorithmic category.
We evaluate PerfOrch with five candidate LLMs on 4{,}640 problems from four benchmarks covering five programming languages: HumanEval-X, EffiBench-X, BigCodeBench, and APPS.
We compare PerfOrch with three types of baselines: single-model direct generation, single-model generate--debug--refine pipelines (PerfCodeGen), and the state-of-the-art multi-agent systems (MapCoder and CodeSIM).
PerfOrch achieves the highest pass@1 in all twelve benchmark--language settings, exceeding the strongest baseline in each setting by an average of 8.19 percentage points.
For execution performance, PerfOrch reduces execution time for an average of 75.03\% of solved problems across the 12 settings, with an average speedup of 1.18$\times$.
Further evaluation shows that the Memory rankings remain effective on three unseen benchmarks without re-profiling.
Sequential refinement matches exhaustive execution-time refinement coverage with 66\% fewer refinement-stage tokens; the corresponding reduction for average-memory optimization is 57\%.
Finally, complementary model strengths persist after all five LLMs are replaced with their current-generation successors.

\end{abstract}



\maketitle

\input{section/01.introduction}

\input{section/02.background}

\input{section/03.methodology}

\input{section/04.experiment}

\input{section/05.evaluation}

\input{section/06.modelupdatecasestudy}

\input{section/07.discussion}

\input{section/08.relatedwork}

\input{section/09.conclusion}

\bibliographystyle{ACM-Reference-Format}
\bibliography{main}

\input{section/Appendix}

\end{document}

%% file: section/01.introduction.tex
\section{Introduction}

Large Language Models (LLMs) have become integral to modern software engineering workflows, powering a wide range of tasks from code completion and program repair to test generation and documentation~\cite{li2025model,xu2025licoeval}.
Coding assistants such as GitHub Copilot~\cite{donato2025studying}, Amazon Q Developer~\cite{amazonQ}, and Cursor~\cite{cursor} embed these capabilities directly into development environments; GitHub Copilot alone exceeds 22 million installations~\cite{donato2025studying}, and 92\% of U.S. developers report using AI coding tools daily~\cite{xu2025licoeval}.
Of the many tasks these tools support, function-level code generation, producing a complete function body from a natural-language specification or function signature, has emerged as one of the most widely adopted applications~\cite{li2025model,xu2025licoeval}.
Capabilities in this area have improved steadily, from early systems such as Codex~\cite{chen2021evaluating} to recent models including GPT-4o~\cite{gpt4o}, Claude 3.7 Sonnet~\cite{claude37sonnet}, and Gemini 2.0 Flash~\cite{gemini20flash}, which achieve high pass rates on standard benchmarks~\cite{evalplusleaderboard}.

Despite this progress, \textit{most approaches for code generation rely on a single LLM}.
These include direct generation techniques (few-shot prompting~\cite{li2024assessing,kouemo2024chain}, Chain-of-Thought reasoning~\cite{wei2022chain,niu2024evaluating}, and retrieval-augmented generation~\cite{acharya2025optimizing,gao2023makes}), iterative pipeline approaches~\cite{huang2024effilearner,peng2025perfcodegen,xia2024agentless}, and agentic systems~\cite{bouzenia2025repairagent,zhang2024paircoder,hosain2025xolver,zhou2024lats,zhang2025thinkcoder}.
For example, MapCoder~\cite{islam2024mapcoder} and CodeSIM~\cite{islam2025codesim} assign planning, coding, and debugging to separate agents, but all agents query the same backbone model.
A second line of work draws on a pool of models and selects among them at inference time.
Query-level routers such as OmniRouter~\cite{mei2025omnirouter} and Pick-and-Spin~\cite{vangala2025efficient} select one model for the whole request.
DAAO~\cite{su2025difficulty} allocates workflow operators and models using estimated query difficulty and operator context, balancing task success against inference cost.
Its routing policy does not use separately measured generation, debugging, and refinement capabilities or execution-based acceptance of individual code candidates to determine which model to try next.
Section~\ref{sec:empirical_study} demonstrates that the strongest model varies across programming languages, problem categories, and development stages.
We refer to this property as \textit{complementary model strengths}: LLMs have partly non-overlapping blind spots, allowing one model to address errors or inefficiencies that another leaves unresolved.
Combining these strengths across generation, debugging, and refinement can produce higher-quality code than using a single model throughout.
For example, \textit{a strong code generator may be a weak debugger}.
A model that excels at generating Python code may underperform on Rust.
A model that excels at array-manipulation problems may struggle with number-theoretic tasks in the same programming language.
These \textit{complementary strengths} therefore raise the question: \textbf{how can multiple LLMs effectively work together to improve the quality of generated code?}

Our preliminary study (Section~\ref{sec:empirical_study}) investigates this question by comparing two strategies for combining three LLMs on Python and Rust.
The first is an ensemble~\cite{jiang2023llm,su2025difficulty}, in which every model answers the same prompt and the most common output is kept.
The second is per-stage cross-model composition, which assigns code generation and debugging to the models with the highest measured performance at the respective stages.
Per-stage cross-model composition outperforms all single-model pipelines and the ensemble on both languages, achieving pass@1 of 98.17\% on Python and 85.98\% on Rust, compared with 97.56\% and 84.76\% for the ensemble, respectively.
These results show that \textit{assigning models to their strongest stages is more effective than aggregating their answers to the same prompt}.
We further observe that model rankings shift across algorithmic problem categories within the same language and stage (Table~\ref{tab:per_stage_best_category_python_rust}).
Adding category-level model selection to per-stage cross-model composition increases pass@1 on Rust from 85.98\% to 88.41\% (Table~\ref{tab:paper_python_rust_final_ranked}), motivating \textit{category-aware routing}.

Based on these observations, we propose PerfOrch, a multi-agent orchestration system that leverages complementary LLM strengths for high-quality code generation.
In this paper, we characterize code quality along two dimensions: \textit{functional correctness} (whether a function passes its test suite) and \textit{execution performance} (the resource consumption of a correct implementation, measured by execution time or average memory utilization).
PerfOrch consists of four collaborative agents: 1) a \textbf{Categorizing Agent} that identifies each problem's algorithmic categories; 2) a \textbf{Generation Agent} that produces initial code from the problem specification; 3) a \textbf{Debugging Agent} that repairs failing solutions; and 4) a \textbf{Refinement Agent} that optimizes correct solutions for runtime performance.
The Generation, Debugging, and Refinement agents each maintain a distinct \textbf{Memory module}, a ranking matrix indexed by programming language and problem categories, constructed offline from profiling data and updatable as new models/versions become available.
The Categorizing Agent maintains a fixed category vocabulary used for multi-LLM consensus labeling.
At runtime, the Generation, Debugging, and Refinement agents consult their respective Memory modules and try LLMs in ranked order.
The Generation and Debugging agents accept the first candidate that passes all tests.
The Refinement Agent accepts the first candidate that passes all tests and whose relative performance improvement exceeds the threshold for the target metric.
If a refinement candidate fails either requirement, the agent retains the input solution and tries the next-ranked model.

\begin{figure}[!ht]
    \vspace{-0.1cm}
    \centering
    \begin{subfigure}{0.32\linewidth}
        \centering
        \includegraphics[width=\textwidth]{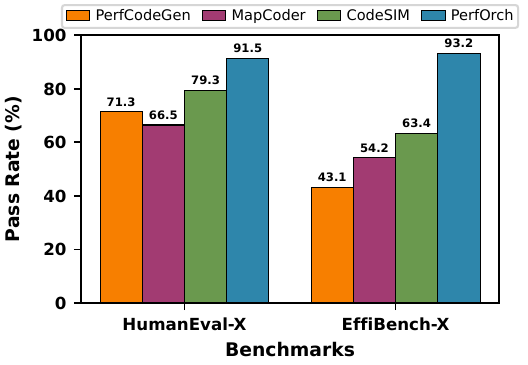}
        \caption{Pass@1 rate (\%)}
        \label{fig:intro_pass_rate}
    \end{subfigure}
    \hfill
    \begin{subfigure}{0.32\linewidth}
        \centering
        \includegraphics[width=\textwidth]{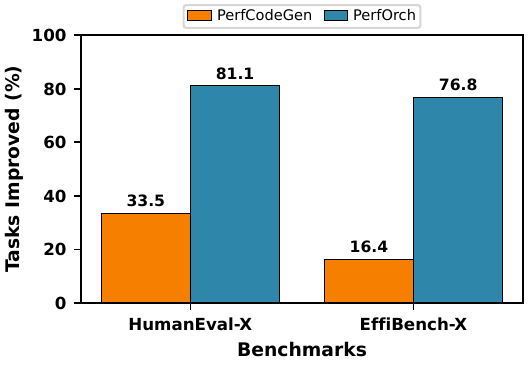}
        \caption{Performance improved (\%)}
        \label{fig:intro_improved_share}
    \end{subfigure}
    \hfill
    \begin{subfigure}{0.32\linewidth}
        \centering
        \includegraphics[width=\textwidth]{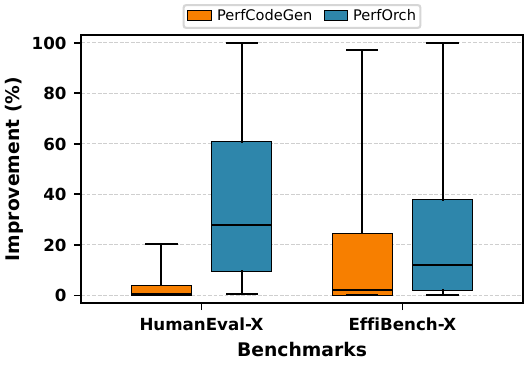}
        \caption{Improvement distribution (\%)}
        \label{fig:intro_improvement_boxplot}
    \end{subfigure}
    \vspace{-0.3cm}
    \caption{
        PerfOrch evaluation on Rust.
        (a)~Pipeline pass@1.
        (b)~Fraction of correct solutions whose execution time is reduced after refinement.
        (c)~Per-solution exec-time improvement (\%) among refined solutions.
        All baselines use GPT-4o.
        MapCoder and CodeSIM have no refinement stage and appear only in (a).
    }
    \Description{
        Three panels on Rust for HumanEval-X and EffiBench-X.
        Panel (a) compares the pass@1 of PerfCodeGen, MapCoder, and CodeSIM, all on GPT-4o, with PerfOrch.
        Panels (b) and (c) compare PerfCodeGen on GPT-4o with PerfOrch on the share of correct solutions whose execution time is reduced and on the per-solution improvement.
    }
    \label{fig:introduction}
    \vspace{-0.2cm}
\end{figure}

Figure~\ref{fig:introduction} visualizes PerfOrch's effectiveness on Rust, a language where function-level code generation remains particularly challenging for individual LLMs.
We compare it with the single-model generate--debug--refine pipeline PerfCodeGen~\cite{peng2025perfcodegen} and with the multi-agent systems MapCoder and CodeSIM, all three running on GPT-4o~\cite{gpt4o}.
On HumanEval-X~\cite{codegeex} and EffiBench-X~\cite{qing2025effibench}, PerfOrch reaches 91.46\% and 93.15\% pass@1, while the strongest of the three baselines reaches 79.27\% and 63.39\% (Figure~\ref{fig:intro_pass_rate}).
PerfOrch lowers the execution time of 81.1\% and 76.8\% of its correct Rust solutions on the two benchmarks, against 33.5\% and 16.4\% for PerfCodeGen (Figure~\ref{fig:intro_improved_share}).
Among the refined solutions, PerfOrch's median execution-time reduction is about 28\% on HumanEval-X and 12\% on EffiBench-X (Figure~\ref{fig:intro_improvement_boxplot}).

We evaluate PerfOrch with five LLMs across five programming languages (Python, Java, C++, Go, Rust) on four benchmarks: the multilingual HumanEval-X and EffiBench-X, and the Python-only BigCodeBench and APPS.
The Memory rankings profiled on HumanEval-X are reused without re-profiling on three benchmarks unseen during profiling: EffiBench-X, BigCodeBench, and APPS.
PerfOrch achieves average pass@1 rates of 97.19\% on HumanEval-X and 95.83\% on EffiBench-X.
On BigCodeBench and APPS, it achieves pass@1 rates of 80.44\% and 93.40\%, respectively.
Beyond correctness, PerfOrch reduces execution time for 21--90\% of correctly solved problems across benchmarks and languages, with mean speedups of 1.05--1.43$\times$.
PerfOrch matches the execution-time refinement coverage of exhaustive multi-model evaluation, which tries every candidate and keeps the best, with 66\% fewer refinement-stage tokens; the corresponding reduction for average-memory optimization is 57\%.
The main contributions are as follows:

\begin{itemize}
\item We present empirical evidence that LLMs exhibit complementary strengths across languages, algorithmic categories, and development stages, and that multi-LLM per-stage collaboration surpasses both individual models and ensembling in code generation quality.

\item We propose PerfOrch, a multi-LLM-based multi-agent orchestration system with four agents for categorization, generation, debugging, and refinement, where the latter three select LLMs using Memory rankings based on measured performance for each stage, programming language, and algorithmic category.

\item We evaluate PerfOrch on 4{,}640 problems across five languages and four benchmarks, and across two generations of the candidate model pool, demonstrating consistent gains in both correctness and execution performance over single-model and multi-agent baselines at a fraction of the pipeline token cost of brute-force search across all stages.
\end{itemize}

The remainder of this paper is organized as follows.
Section~\ref{sec:background_overall} provides background on LLM-based code generation.
Section~\ref{sec:methodology_overall} presents our preliminary study and describes the design and implementation of PerfOrch.
Section~\ref{sec:experiment_overall} shows experiment settings.
Section~\ref{sec:evaluation_overall} reports experimental results.
Section~\ref{sec:model_update_case_study} conducts a case study of PerfOrch on model update.
Section~\ref{sec:discussion_overall} discusses findings and threats to validity.
Section~\ref{sec:related_overall} reviews related work.
And Section~\ref{sec:conclusion_overall} concludes the paper.

%% file: section/02.background.tex
\section{Background}
\label{sec:background_overall}

This section provides the technical foundations for our study.
Section~\ref{sec:background_stages} introduces the three-stage code generation pipeline that forms the basis of our orchestration framework.
Section~\ref{sec:background_agent} introduces multi-agent code generation systems.
Section~\ref{sec:background_category} discusses the role of algorithmic categories in LLM-based code generation.
Section~\ref{sec:background_bench} describes the four benchmarks used for evaluation, and Section~\ref{sec:background_tools} presents the external testing tools for code correctness and performance.

\subsection{Iterative Code Generation Pipeline}
\label{sec:background_stages}

Studies of how programmers write and debug code show that development proceeds through repeated edit–run–diagnose cycles rather than one-shot authoring~\cite{Alaboudi2021EditRun, Hirsch2021Debug}.
Mirroring this reality, recent approaches to LLM-based code generation have demonstrated that decomposing the task into distinct stages (generation, debugging, and refinement) improves both correctness and efficiency over one-shot generation~\cite{Hirsch2021Debug, peng2025perfcodegen}.
In the \textit{generation} stage, an LLM produces an initial solution from a natural-language specification~\cite{mathews2024test}.
Because these initial outputs frequently contain syntax errors, logical flaws, or semantic mismatches~\cite{trentini2025advancing,wang2025towards}, the subsequent \textit{debugging} stage supplies the model with failed test execution traces to elicit a corrected solution~\cite{bouzenia2025repairagent, peng2025perfcodegen}.
Finally, a \textit{refinement} stage optimizes functionally correct code for a target metric such as execution time or memory utilization~\cite{huang2024effilearner, li2024assessing}.

PerfCodeGen~\cite{peng2025perfcodegen} instantiates this three-stage pipeline with one LLM without any additional training.
To repair an incorrect solution, it prompts the model with the code, a failing test case, and execution feedback.
Once a correct solution is obtained, the model first optimizes it without execution feedback.
PerfCodeGen then measures the optimized solution's execution time on each test and asks the model to optimize it again using the most time-consuming test as feedback.
If an optimization attempt fails the correctness tests, PerfCodeGen returns the fastest correct solution obtained so far.
It limits repair and performance optimization with execution feedback to one iteration each to reduce inference cost.

\subsection{Multi-Agent Code Generation}
\label{sec:background_agent}

Multi-agent code generation systems assign tasks such as recalling similar problems, planning, coding, and debugging to different LLM-based agents.
Each agent uses a task-specific prompt and receives intermediate results from other agents.
MapCoder~\cite{islam2024mapcoder} uses four agents for retrieval, planning, coding, and debugging.
The retrieval agent prompts the LLM to recall similar problems, together with their solutions and algorithms.
For each recalled example, the planning agent develops a plan for the current problem and assigns it a confidence score.
The coding agent implements the highest-scoring plan and tests the code on the sample cases provided in the problem statement.
If the code fails these tests, the debugging agent uses the test feedback to revise it for a fixed number of rounds.
If debugging remains unsuccessful, MapCoder tries the next plan in confidence order.

CodeSIM~\cite{islam2025codesim} uses simulation to guide planning and debugging.
Its planning agent first generates a similar problem with a plan and a solution, then develops a plan for the current problem.
The agent simulates this plan step by step on a sample input and revises it if the simulated output differs from the expected output.
The coding agent implements the verified plan and tests the code on the sample cases.
If a test fails, the debugging agent simulates the code on the failing input to locate the error and then revises it.
If debugging remains unsuccessful after a fixed number of rounds, CodeSIM returns to the planning agent.
The process stops when the code passes the sample tests or the maximum number of planning cycles is reached.
In both systems, the agents perform different roles using the same backbone LLM.

\subsection{Algorithmic Categories}
\label{sec:background_category}
LLM performance on code generation varies not only across programming languages but also across algorithmic problem categories~\cite{hossain2025llm, huang2024knowledge, xia2025leetcodedataset}.
Hossain et al.~\cite{hossain2025llm} show that model rankings on competitive programming problems shift substantially across category types (e.g., dynamic programming vs.\ graph traversal).
Xia et al.~\cite{xia2025leetcodedataset} and Huang et al.~\cite{huang2024effibench} further demonstrate that temporal shifts in problem categories affect LLM evaluation reliability, confirming that category is a meaningful axis of performance variation.
However, existing work treats category-level variation as an evaluation concern rather than as a signal for model selection; no prior approach uses per-category profiling to route problems to the most capable LLM.

\subsection{Benchmarks}
\label{sec:background_bench}

\begin{figure}[ht]
    \vspace{0.1cm}
    \centering
    \includegraphics[width=\textwidth]{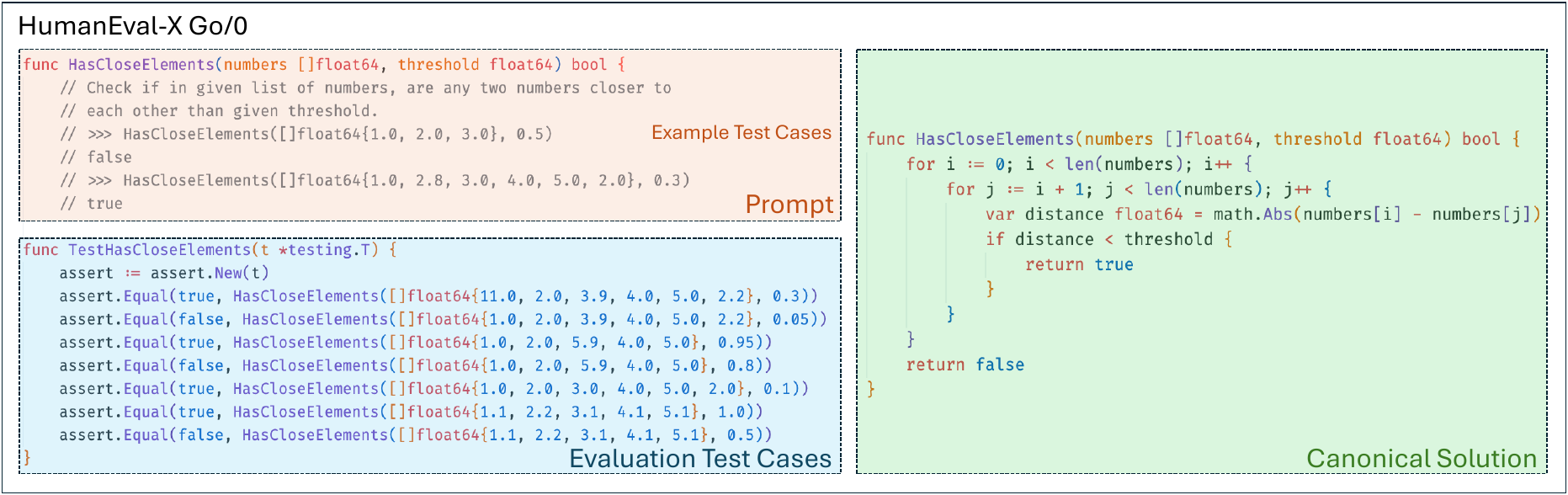}
    \caption{Structure of a HumanEval-X problem (Go/0). Each problem comprises three components: the \textit{prompt} containing the function signature and a natural-language docstring with example inputs/outputs; the \textit{evaluation test cases} used for correctness verification (hidden from the LLM); and the \textit{canonical solution}.}
    \Description{This figure presents an illustrative example from the Humaneval-X benchmark in the Go programming language (problem go/0). Each task is defined by three components: (i) the prompt, which specifies the problem statement and function signature; (ii) the test cases, which serve as the evaluation harness for verifying correctness; and (iii) the canonical solution, a reference implementation that ensures solvability and provides a ground truth for validation.}
    \label{fig:humaneval-x}
\end{figure}

HumanEval-X~\cite{codegeex} extends OpenAI's HumanEval benchmark~\cite{chen2021evaluating} to five languages: Python, C++, Java, Go, and Rust.
Each of the 164 problems includes a prompt (function signature and docstring), a canonical solution, and a comprehensive test suite for correctness verification (Figure~\ref{fig:humaneval-x}).
For debugging, we use the corresponding buggy solutions provided by HumanEval-Pack~\cite{muennighoff2023octopack}.
We distinguish between \textit{example test cases} embedded in the prompt and \textit{evaluation test cases} used for verification.
The prompt's docstring contains one or two illustrative input-output examples that serve as specification hints.
The evaluation test suite covers edge cases and corner conditions and is not part of the generation prompt.

EffiBench-X~\cite{qing2025effibench} extends EffiBench~\cite{huang2024effibench} to the same five languages, comprising 336 problems per language sourced from competitive programming contests that demand efficient algorithmic solutions under strict time and memory limits.
Like HumanEval-X, each problem includes a prompt, evaluation test suite, and canonical solution; however, EffiBench-X problems tend to require more complex algorithmic reasoning (e.g., dynamic programming, graph traversal) and are, on average, more difficult, as reflected in the substantially lower single-model pass rates (Table~\ref{tab:agent_correctness}).

BigCodeBench~\cite{zhuo2024bigcodebench} is a Python benchmark of 1{,}140 tasks that use 139 libraries across seven application domains.
Each task is released in two prompt variants.
The Complete variant provides a function signature and docstring and asks the model to complete the function body.
The Instruct variant expresses the task as a short natural-language instruction.
A solution typically composes calls to several third-party APIs rather than implementing an algorithm from scratch.

APPS~\cite{hendrycks2021measuring} is a Python benchmark of 10{,}000 problems collected from open-access programming-contest and interview-practice sites, divided evenly into training and test splits.
The problems span three difficulty levels: introductory, interview, and competition.
Most problems ask for a complete program that reads from standard input, with input and output formats and sample cases provided in the problem statement.

\subsection{Correctness-Verification and Performance-Profiling Tools}
\label{sec:background_tools}

We test the correctness of LLM-produced code via \textit{correctness verification}, which follows a compile-then-execute protocol. For each candidate solution, the source file is written into an isolated temporary directory, compiled using the language's standard toolchain (\texttt{g++} for C++, \texttt{javac} for Java, and \texttt{cargo} for Rust), and executed against the full evaluation test suite (Section~\ref{sec:background_bench}) under a 180-second timeout enforced by the Linux \texttt{timeout} utility. Python is interpreted directly, and Go's \texttt{go test} runner compiles and executes in a single invocation. A solution is accepted only when the process exits with code~0 (all assertions pass); any compilation failure, runtime error, or timeout is treated as a failure, and the resulting error log is available to downstream agents.

Beyond correctness, we measure two runtime metrics to evaluate the execution performance of correct solutions.~\cite{chen2017exploratory,alshoaibi2019price,li2024assessing,he2019statistics,he2021performance}. \textit{Execution time} is measured using Linux \texttt{perf stat}~\cite{linuxperfstat}, and it is reported as wall-clock time in milliseconds. \textit{Memory utilization} (average memory utilization) is sampled at 1\,ms intervals using cmdbench~\cite{cmdbench}, which reads \texttt{VmData} and \texttt{VmStk} from Linux \texttt{/proc/<pid>/status} to approximate the process's heap and stack footprint~\cite{linuxprocstatus}.

%% file: section/03.methodology.tex
\section{PerfOrch: Motivation and Framework}
\label{sec:methodology_overall}

This section presents PerfOrch, a multi-agent orchestration system that dynamically routes tasks to profiled candidate LLMs to produce high-quality code (in terms of functional correctness and execution performance).
Section~\ref{sec:empirical_study} reports a motivating study showing that different LLMs exhibit complementary strengths across programming languages and coding task categories.
Section~\ref{sec:overview} gives an architectural overview, and Sections~\ref{sec:category}--\ref{sec:refinement} describe the four specialized agents.

\subsection{Motivating Observations}
\label{sec:empirical_study}

We conduct a preliminary study to test whether multi-LLM collaboration improves the correctness of code generation compared with single-model approaches.
We evaluate using three LLMs, Claude 3.7 Sonnet~\cite{claude37sonnet}, GPT-4o~\cite{gpt4o}, and Grok 3~\cite{grok3}, on two languages: Python and Rust.
Each model is profiled on two standalone tasks: code generation (pass@1~\cite{chen2021evaluating} on HumanEval-X~\cite{codegeex}) and debugging (fix@1 on HumanEval-Pack~\cite{muennighoff2023octopack} buggy solutions).
We then compare four combination strategies: (1)~\emph{ensemble}~\cite{shi2022natural,jiang2023llm,su2025difficulty}, which selects the most common output from all three models; (2)~\emph{single-model pipelines}, which apply one fixed model to both generation and debugging, following PerfCodeGen~\cite{peng2025perfcodegen}; (3)~\emph{per-stage cross-model composition}, which assigns the best-profiled model to each stage independently; and (4)~\emph{per-stage, per-category cross-model composition}, which further conditions model selection on algorithmic problem category.

\begin{table}[ht!]
    \vspace{0.2cm}
    \caption{Standalone generation pass@1 and debugging fix@1 (\%) of three LLMs on Python and Rust, using HumanEval-X and HumanEval-Pack buggy solutions, respectively. Bold marks the best model per column. No single model dominates across both languages and stages.}
    \centering
    \small
    \setlength{\tabcolsep}{6pt}
    \renewcommand{\arraystretch}{1.08}
    \begin{tabular}{lcccc}
    \toprule
    & \multicolumn{2}{c}{\textbf{Python}} & \multicolumn{2}{c}{\textbf{Rust}} \\
    \cmidrule(lr){2-3}\cmidrule(lr){4-5}
    \textbf{Model} & \textbf{Generate} & \textbf{Debug} & \textbf{Generate} & \textbf{Debug} \\
    \midrule
    Claude~3.7        & 94.51          & \textbf{97.56} & 79.27          & \textbf{80.49} \\
    GPT-4o            & \textbf{95.12} & 88.41          & 69.51          & 65.85          \\
    Grok~3            & 93.29          & 90.24          & \textbf{83.54} & 70.73          \\
    \bottomrule
    \end{tabular}
    \vspace{0.2cm}
    \label{tab:paper_python_rust_ranked}
\end{table}

\begin{table}[!ht]
    \caption{Category-level generation pass@1 and debugging fix@1 (\%) of three LLMs on Python and Rust, using the four-category taxonomy~\cite{intisar2019classification}. Generation is measured on HumanEval-X; debugging on HumanEval-Pack. The Graph category contains a single problem.}
    \vspace{-0.3cm}
    \centering
    \renewcommand{\arraystretch}{0.85}
    \setlength{\aboverulesep}{0.2ex}
    \setlength{\belowrulesep}{0.2ex}
    \adjustbox{max width=\textwidth}{
    \begin{tabular}{l|l|l|cccc}
    \toprule
    \textbf{Lang.} & \textbf{Stage} & \textbf{Model} & \textbf{Implementation} & \textbf{Basic Info.} & \textbf{Math} & \textbf{Graph} \\
    \midrule
    \multirow{6}{*}{\rotatebox{90}{Python}} & \multirow{3}{*}{\textbf{Generate}} & Claude & 94.38 & 96.51 & 92.00 & 100.00 \\
     &  & GPT-4o & 95.62 & 96.51 & 94.00 & 0.00 \\
     &  & Grok 3 & 93.75 & 96.51 & 88.00 & 100.00 \\
    \cmidrule{2-7}
     & \multirow{3}{*}{\textbf{Debug}} & Claude & 97.50 & 97.67 & 98.00 & 100.00 \\
     &  & GPT-4o & 88.75 & 89.53 & 86.00 & 100.00 \\
     &  & Grok 3 & 90.00 & 89.53 & 94.00 & 0.00 \\
    \midrule
    \multirow{6}{*}{\rotatebox{90}{Rust}} & \multirow{3}{*}{\textbf{Generate}} & Claude & 79.38 & 81.40 & 76.00 & 0.00 \\
     &  & GPT-4o & 69.38 & 70.93 & 74.00 & 0.00 \\
     &  & Grok 3 & 83.12 & 83.72 & 84.00 & 100.00 \\
    \cmidrule{2-7}
     & \multirow{3}{*}{\textbf{Debug}} & Claude & 80.00 & 81.40 & 82.00 & 0.00 \\
     &  & GPT-4o & 66.25 & 70.93 & 60.00 & 0.00 \\
     &  & Grok 3 & 70.62 & 72.09 & 68.00 & 100.00 \\
    \bottomrule
    \end{tabular}
    }
    \label{tab:per_stage_best_category_python_rust}
\end{table}

\begin{table}[!ht]
    \caption{Pass@1 (\%) under four model-combination strategies on Python and Rust (generate $\rightarrow$ debug). Cross-model composition assigns the best-profiled model to each stage, and its per-category form further conditions that choice on algorithmic category.}
    \centering
    \small
    \setlength{\tabcolsep}{6pt}
    \renewcommand{\arraystretch}{1.08}
    \begin{tabular}{llcc}
    \toprule
    \textbf{Strategy} & \textbf{Model} & \textbf{Python} & \textbf{Rust} \\
    \midrule
    Ensemble & GPT+Claude+Grok & 97.56 & 84.76 \\
    \midrule
    \multirow{3}{*}{Single-model pipeline} & Claude+Claude & 97.56 & 79.88 \\
                                           & GPT+GPT & 96.95 & 71.34 \\
                                           & Grok+Grok & 95.12 & 84.15 \\
    \midrule
    \multirow{2}{*}{Cross-model composition} & Python: GPT+Claude & \multirow{2}{*}{98.17} & \multirow{2}{*}{85.98} \\
                                       & Rust: Grok+Claude &                         &                         \\
    \midrule
    Cross-model composition, per category & Category-aware selection & \textbf{98.17} & \textbf{88.41} \\
    \bottomrule
    \end{tabular}
    \label{tab:paper_python_rust_final_ranked}
\end{table}

\begingroup
\emergencystretch=0pt
\tolerance=1000
\textbf{Observation 1: Multi-LLM collaboration via per-stage routing outperforms single-model approaches and the ensemble.}
Table~\ref{tab:paper_python_rust_ranked} reports standalone capability profiles, showing that no single model dominates across both languages and stages.
GPT-4o leads Python generation (95.12\%) but drops to last place in Rust (69.51\%), where Grok 3 ranks first (83.54\%).
In debugging, Claude achieves the highest fix@1 for both languages (Python: 97.56\%; Rust: 80.49\%), while GPT-4o, the top Python generator, ranks last among debuggers.
Table~\ref{tab:paper_python_rust_final_ranked} tests whether exploiting these complementary strengths improves end-to-end correctness.
Under a generation-debugging pipeline, per-stage cross-model composition, which assigns the best-profiled generator and best-profiled debugger to their respective stages, outperforms the ensemble on both languages (Python: 98.17\% vs.\ 97.56\%; Rust: 85.98\% vs.\ 84.76\%) and all single-model pipelines.
Aligning each stage with the most capable model, therefore, yields gains beyond those provided by redundancy-based aggregation.
\emph{This observation motivates selecting models independently for generation and debugging.}
\par\endgroup

\textbf{Observation 2: Category-aware selection provides additional improvement.} We classify HumanEval-X problems into four algorithmic categories (Implementation, Basic Information Processing, Math, Graph) following~\cite{intisar2019classification}.
Table~\ref{tab:per_stage_best_category_python_rust} reports category-level generation pass@1 and debugging fix@1.
Rankings shift across categories within the same language and stage: for Python generation, GPT-4o leads Implementation (95.62\%) and Math (94.00\%), while Claude and Grok lead Graph (100.00\%).
For Rust debugging, Grok repairs the sole Graph problem (100.00\%), while Claude leads the remaining three categories.
The last row of Table~\ref{tab:paper_python_rust_final_ranked} quantifies the effect of category-aware model selection.
Conditioning on problem category improves Rust pass@1 over language-level composition by 2.43 percentage points (88.41\% vs.\ 85.98\%).
While Python pass@1 remains unchanged at 98.17\%, the Rust result indicates that category-level routing captures complementary strengths that language-level selection misses.
\emph{This observation motivates category-aware routing: the Generation, Debugging, and Refinement agents maintain Memory modules indexed by both programming language and algorithmic category, enabling fine-grained model selection (Sections~\ref{sec:generation}--\ref{sec:refinement}) for better code quality}.

\subsection{PerfOrch Overview}
\label{sec:overview}

\begin{figure}[!ht]
    \centering
    \includegraphics[width=\textwidth]{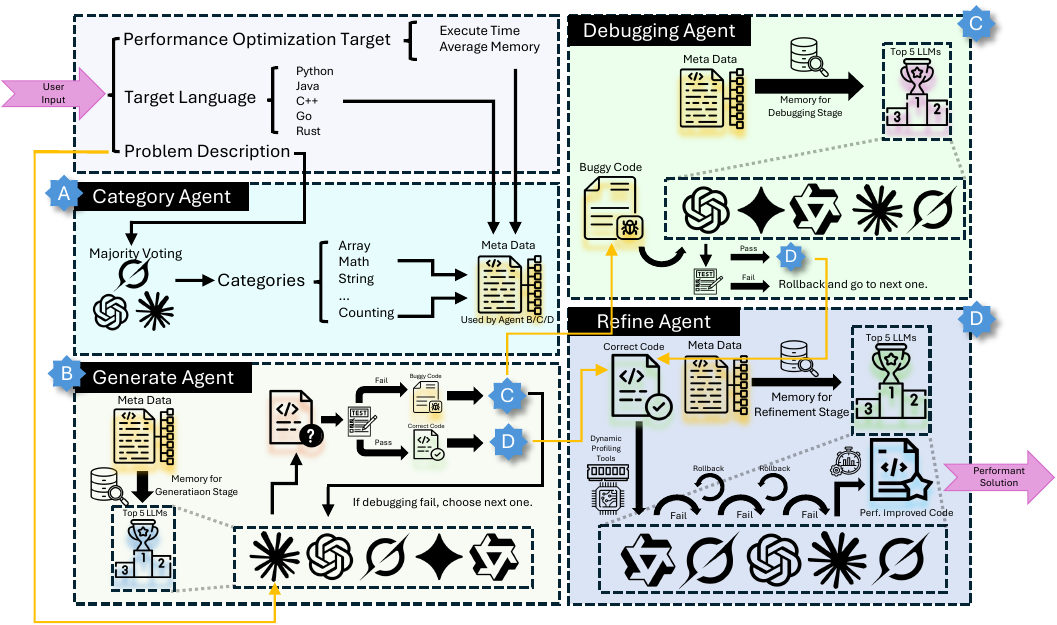}
    \caption{PerfOrch architecture. The pipeline comprises four agents: the Categorizing Agent assigns algorithmic categories; the Generation Agent produces initial solutions; the Debugging Agent repairs faulty code; and the Refinement Agent optimizes execution performance. The Generation, Debugging, and Refinement agents each maintain a Memory module of per-language, per-category LLM rankings from offline profiling.}
    \Description{}
    \label{fig:agent}
\end{figure}

Based on these observations, we design PerfOrch as a multi-agent orchestration system with four specialized agents (Figure~\ref{fig:agent} shows the system architecture), including:
a \textbf{Categorizing Agent} that identifies each problem's algorithmic categories.
We start with this agent because model performance varies across algorithmic categories and user-specified programming languages.
The category labels are then used by downstream agents for task routing (Section~\ref{sec:category}).
A \textbf{Generation Agent} that produces initial code solutions.
It queries its own memory of ranked candidate LLMs for each (language, category) pair, where the rankings are derived from offline profiling (Section~\ref{sec:generation}).
A \textbf{Debugging Agent} repairs faulty code.
It selects ranked LLMs from its own memory and applies sequential retry logic until it finds a correct solution (Section~\ref{sec:debugging}).
And finally, a \textbf{Refinement Agent} optimizes correctly generated/debugged solutions for the target performance metric chosen by the user, such as execution time or average memory utilization.
It uses metric-specific memory rankings and applies acceptance rules that preserve correctness (Section~\ref{sec:refinement}).

PerfOrch follows the generate--debug--refine pipeline of PerfCodeGen~\cite{peng2025perfcodegen}, but selects models independently for the three stages using rankings measured through offline code execution.
Two principles govern this selection.
First, model competence is measured separately for generation, debugging, and refinement in each programming language and algorithmic category.
Second, execution results determine whether to continue searching: a debugging candidate must pass the tests, and a refinement candidate must preserve correctness and exceed the improvement threshold for the target performance metric.
In practice, the orchestration unfolds iteratively: if an initial generation attempt fails, the flawed candidate is forwarded to the Debugging Agent.
If all five targeted debugging attempts fail, control loops back to the Generation Agent to invoke the next-ranked generator.
Once a functionally correct solution is established, it advances to the Refinement Agent.
Here, up to five optimization candidates are sequentially evaluated and accepted strictly if they preserve correctness and exceed the improvement threshold for the target performance metric.
Crucially, because \emph{this multi-agent routing logic is decoupled from the underlying candidate pool, integrating a newly released LLM requires only updating the offline memory rankings, leaving the orchestration framework entirely unmodified}.

\subsection{Categorizing Agent}
\label{sec:category}

The Categorizing Agent assigns algorithmic categories to coding problems.
These categories support category-aware model selection in downstream agents.
Prior work shows that LLM performance varies across problem categories~\cite{hossain2025llm,xia2025leetcodedataset,huang2024knowledge}.
This variation is linked to differences in training data and model specialization~\cite{huang2024effibench,hossain2025llm}.
Our preliminary study (Section~\ref{sec:empirical_study}) shows the same pattern.

However, the coarse-grained four-category taxonomy used in Table~\ref{tab:per_stage_best_category_python_rust} has two limitations.
First, each problem is assigned to exactly one category, yet coding problems typically involve multiple algorithmic concepts simultaneously.
For example, a problem requiring sorting an array while counting inversions spans at least \emph{Array}, \emph{Sorting}, and \emph{Counting}.
Competitive programming platforms such as LeetCode routinely assign multiple tags to such problems~\cite{leetcode,huang2024effibench}.
A single-label scheme discards this multi-faceted structure and loses information that downstream model selection could exploit.
Second, with only four categories, each covers a broad and heterogeneous problem space, amplifying the impact of any miscategorization on routing accuracy.
A finer-grained, multi-label vocabulary mitigates both issues: each category is more specific, and routing combines scores across the categories assigned to each problem.

\textbf{Category Vocabulary Construction.} We therefore use the LeetCode taxonomy as a \emph{multi-label, fine-grained} category vocabulary~\cite{leetcode,huang2024effibench}.
To build the Categorizing Agent's \textit{category memory}, two human experts independently assign categories from the full LeetCode taxonomy to each HumanEval-X problem; we retain categories selected by both, with a third expert adjudicating disagreements.
We quantify the reliability of the annotation using Cohen's Kappa statistic~\cite{mchugh2012interrater}, which yielded a considerable agreement score of 0.9024.
However, the full LeetCode taxonomy produces many low-frequency categories: several carry fewer than three problems in the HumanEval-X corpus.
Such sparse categories undermine routing reliability, because a model's profiled success rate on one or two problems is too noisy to support meaningful ranking (e.g., 100\% pass@1 on a single problem conveys no generalizable signal).
We therefore select the ten most frequent categories from the annotation results (Table~\ref{Tab:category}: Category Labels) as the \textit{category memory}.
These ten categories provide full coverage of all 164 HumanEval-X problems per language with at least ten problems in each category for profiling.

\begin{table}[!ht]
\caption{Ten algorithmic categories selected as the Categorizing Agent's vocabulary, ranked by frequency in the human-annotated HumanEval-X corpus (164 problems per language). Each problem may carry multiple labels; counts therefore sum to more than 164. Each category contains at least ten problems, ensuring stable per-category profiling.}
\adjustbox{max width=0.7\textwidth}{
\centering
\small
\begin{tabular}{ccccc}
\toprule
\textbf{1. Array} & \textbf{2. Math} & \textbf{3. String} & \textbf{4. Counting} & \textbf{5. Num Theory} \\
77 & 76 & 74 & 50 & 33 \\
\midrule
\textbf{6. Simulation} & \textbf{7. Sorting} & \textbf{8. Enumeration} & \textbf{9. Greedy} & \textbf{10. Hash Table} \\
32 & 29 & 12 & 11 & 11 \\
\bottomrule
\end{tabular}
}
\label{Tab:category}
\end{table}

\textbf{Automated Categorization.} At runtime, the agent performs automated categorization via multi-LLM consensus over this category memory.
Three LLMs (GPT-4o, Claude 3.7 Sonnet, Grok 3) independently assign categories from Table~\ref{Tab:category} to each problem.
A category is retained only when at least two of three models select it.
This conservative majority-voting filter trades recall for routing safety by suppressing uncertain categories.
Since downstream agents rely on these labels for model selection, we verify the reliability of the automated categorization procedure against the human baseline (Top-10 categories) using two complementary metrics.
Let $Y_p$ denote the set of human-annotated categories for problem $p$, let $\hat{Y}_p$ denote the set assigned by the Categorizing Agent, and let $N$ denote the total number of problems.
\emph{Subset accuracy} (Equation~\ref{eq:subset_accuracy_category}) counts a problem as correctly labeled only when all agent-annotated categories fall within the human-annotated set, emphasizing routing safety by penalizing any false positive.
\emph{Micro-recall} (Equation~\ref{eq:micro_recall_category}) measures coverage of the agent-annotated categories against the human baseline across all problems.

\begin{equation}
    \text{Subset-Accuracy} = \frac{1}{N}\sum_p \mathbf{1}[\hat{Y}_p \subseteq Y_p] \times 100\%
    \label{eq:subset_accuracy_category}
\end{equation}

\begin{equation}
    \text{Micro-Recall} = \frac{\sum_p |\hat{Y}_p \cap Y_p|}{\sum_p |Y_p|} \times 100\%
    \label{eq:micro_recall_category}
\end{equation}

The subset accuracy is 96.95\% (159/164 problems), confirming that the agent rarely introduces incorrect categories into downstream routing.
The micro-recall is 94.81\% (384/405 category labels), indicating that the majority-voting scheme retains most valid categories despite its conservative filtering.

\textbf{Single-Model Categorization.} Running one model reduces categorization from three calls per problem to one and lowers subset accuracy by 1.22 percentage points.
On the human-annotated corpus, the three-model consensus reaches 96.95\% (159/164), while the best single model, GPT-4o, reaches 95.73\% (157/164), ahead of Claude 3.7 Sonnet at 94.51\% (155/164) and Grok 3 at 92.07\% (151/164).
GPT-4o is therefore the default when the Categorizing Agent runs on one model.
In consensus mode, when no category receives at least two votes, the agent falls back to the default model's labels so that downstream routing can still use the category dimension (Sections~\ref{sec:generation}--\ref{sec:refinement}).

\subsection{Generation Agent}
\label{sec:generation}

The Generation Agent produces initial code solutions from the problem prompt, the user-specified programming language, and the problem categories determined by the Categorizing Agent (Section~\ref{sec:category}).
Specifically, at runtime, the Generation Agent queries its memory for the top-ranked generation model for that language-category combination, and issues the problem prompt together with a generation-specific system instruction.
After the selected model produces an initial solution, the agent validates the solution with the \emph{full evaluation test cases} (Section~\ref{sec:background_bench}) via the correctness-verification tools (Section~\ref{sec:background_tools}).
Crucially, the selected LLM never receives the full evaluation test cases or the testing feedback; this separation prevents data leakage and ensures that the LLM's output reflects generalization from the specification rather than memorization of test cases.
The validation result (pass/fail) is then used by the agent for future routing: If all tests pass, the solution advances to the Refinement Agent.
Otherwise, the Generation Agent forwards the faulty code to the Debugging Agent.
The agent invokes one generation model per attempt and moves to the next-ranked generator only after all debugging candidates for the current generator's output are exhausted.
This \emph{fix-before-regenerate} ordering is motivated on two grounds.
First, our motivating study (Section~\ref{sec:empirical_study}) shows that generation and debugging strengths differ across models, so routing faulty code to a stronger debugger can increase the chance of a successful repair by exploiting these complementary strengths.
Second, a fix call reuses the existing faulty code as a structured starting point, giving the model a concrete artifact to localize and correct rather than synthesizing from a natural-language specification alone; this contextual grounding could reduce the average inference cost per final correct solution compared to repeated cold regeneration~\cite{xia2023automated, hidvegi2024cigar}.
Section~\ref{sec:discussion_overall} discusses the budget trade-offs in this retry policy.

\textbf{Memory Construction.} The Generation Agent maintains its own memory that stores ranked lists of LLMs by programming language and problem categories.
We build the rankings from profiling on HumanEval-X~\cite{codegeex}.
We profile all five candidate LLMs for each combination of programming language and problem category, and store them by the per-partition pass@1 rate (Equation~\ref{eq:correctness}).
Section~\ref{sec:experiment_memory} reports the full memory for the Generation Agent (Table~\ref{tab:generate_language_tag}).

\begin{equation}
    \text{pass@1}(llm, language, category) = \frac{N_{\text{correct}}(llm, language, category)}{N_{\text{total}}(language, category)} \times 100\%
    \label{eq:correctness}
\end{equation}

\textbf{Memory-Guided Routing.} Since each problem usually carries multiple categories (e.g., array and sorting), for each LLM and language, we aggregate $\text{pass@1}(llm, language, category)$ across all categories $\mathcal{C}_p$ of problem $p$ using a product (Equation~\ref{eq:multi_cat_genfix}).
The product penalizes models with low performance on any constituent category, favoring consistently capable models over those with high variance across categories.
This risk-averse aggregation is appropriate because a multi-tagged problem may require competence in all associated algorithmic skills; a model that excels on array problems but fails on sorting could be unreliable for a problem tagged with both.
At runtime, the agent ranks all candidate LLMs by $S_{\text{gen}}$.

\begin{equation}
    S_{\text{gen}}(llm, language, \mathcal{C}_p) = \prod_{category \in \mathcal{C}_p} \text{pass@1}(llm, language, category)
    \label{eq:multi_cat_genfix}
\end{equation}

\subsection{Debugging Agent}
\label{sec:debugging}
The Debugging Agent repairs faulty code produced by the Generation Agent.
At runtime, this agent receives \textit{faulty code}, retrieves a ranked list of $k$ debugging candidates for the current language and categories from its memory (default $k{=}5$), and tries each candidate in rank order.
Each model receives the faulty code and a debugging-specific system instruction.
The instruction asks the model to localize and fix the defect rather than rewrite from scratch.
After each repair attempt, the agent invokes the same correctness-verification tools (Section~\ref{sec:background_tools}).
The agent then uses the verification result (pass/fail) for routing decisions (accepting if it passes the full evaluation test cases, rejecting otherwise), without exposing any test cases to LLMs.
An accepted patched code is forwarded to the Refinement Agent, thereby ending the debugging loop.
A rejected patch is discarded; on rejection, the agent restores the original faulty code rather than chaining from the failed patch, since iterative patching inflates input token cost and risks corrupting correct logic unrelated to the original defect~\cite{hidvegi2024cigar, olausson2023self}.
The agent then tries the next candidate LLM.
If all $k$ candidates fail, control returns to the Generation Agent.

\textbf{Memory Construction.} The Debugging Agent maintains its own memory of $k$ LLMs indexed by programming language and problem category.
We construct this memory by profiling \textit{buggy solutions} on HumanEval-Pack~\cite{muennighoff2023octopack}, which provides faulty code for each problem in the HumanEval-X benchmark.
Each model receives buggy solutions and the debugging-specific prompt during profiling.
For each combination of language and category, we store the per-partition fix@1 rate (Equation~\ref{eq:fix}) for each model.
Analogous to pass@1~\cite{chen2021evaluating}, fix@1 measures the fraction of buggy programs that a model successfully repairs on its first attempt.
Section~\ref{sec:experiment_memory} reports the Debugging Agent's memory (Table~\ref{tab:fix_language_tag}).

\begin{equation}
    \text{fix@1}(llm, language, category) = \frac{N_{\text{fixed}}(llm, language, category)}{N_{\text{total\_buggy}}(language, category)} \times 100\%
    \label{eq:fix}
\end{equation}

\textbf{Memory-Guided Routing.} As with the Generation Agent, each problem carries multiple categories, we aggregate $\text{fix@1}(llm, language, category)$ across all categories $\mathcal{C}_p$ of problem $p$ using a product (Equation~\ref{eq:multi_cat_fix}).
The product applies the same risk-averse logic: a debugging model must demonstrate consistent ability to fix bugs across all relevant algorithmic categories.
At runtime, the agent ranks all candidate LLMs by $S_{\text{fix}}$ and tries them in rank order.
Because fix@1 is measured on a fixed set of buggy programs, debugging rankings are constructed separately from generation rankings, allowing the repairing model to differ from the model that produced the code.

\begin{equation}
    S_{\text{fix}}(llm, language, \mathcal{C}_p) = \prod_{category \in \mathcal{C}_p} \text{fix@1}(llm, language, category)
    \label{eq:multi_cat_fix}
\end{equation}

\subsection{Refinement Agent}
\label{sec:refinement}

The Refinement Agent optimizes functionally correct solutions for a user-specified performance-optimization target (e.g., execution time or average memory utilization).
At runtime, this agent receives a functionally correct solution.
It queries its memory for $k$ refinement models ranked by the user-specified target metric (default $k{=}5$), and tries them in rank order.
Each model receives the functionally correct code and a refinement-specific system instruction.
The instruction requests optimization while preserving the correct functional behavior.
The agent then evaluates each optimized output on two axes by invoking verification and profiling tools (Section~\ref{sec:background_tools}).
First, using the correctness-verification tools, if the correctness check passes, then performance is measured by the performance-profiling tools.
The agent uses these results to make accept/reject decisions.
An optimized version is accepted only if it preserves correctness and its relative reduction in the target metric exceeds a threshold $\tau$, thereby terminating the refinement loop and outputting the final version.
We use $\tau=0.004$ for average memory utilization and $\tau=0.012$ for execution time, based on the profiler validation in Section~\ref{sec:threats}.
Otherwise, the agent reverts to the original input solution and tries the next-ranked model.
This early-termination strategy trades potentially larger gains from later candidates for reduced token cost; Section~\ref{sec:refinement_strategy} quantifies this trade-off by comparing sequential acceptance against the exhaustive strategy, justifying our design choice.
Finally, if no candidate is accepted, the pipeline outputs the input solution unchanged.

\textbf{Memory Construction.} The Refinement Agent maintains its own memory of LLM rankings indexed by programming language, problem category, and user-specified performance-optimization target.
We construct the memory by requesting each model to optimize the execution performance of canonical solutions (Section~\ref{sec:background_bench}) from the HumanEval-X benchmark.
For each (LLM, language, category) partition, we quantify improvement depth via the mean improvement ratio $\overline{IMP}$ (Equation~\ref{eq:imp}), the average relative reduction from baseline performance $P_{\text{base}}$ to refined performance $P_{\text{refined}}$ across all $N_{\text{total}}$ problems in the partition.
When a refinement attempt fails correctness verification or its relative improvement does not exceed the target metric's threshold, its contribution is 0.
The same formula applies to both execution time and average memory utilization.

\begin{equation}
    \overline{IMP}(llm, language, category) = \frac{1}{N_{\text{total}}} \sum_{k=1}^{N_{\text{total}}} \frac{P_{\text{base},k} - P_{\text{refined},k}}{P_{\text{base},k}}
    \label{eq:imp}
\end{equation}

We complement $\overline{IMP}$ with the percent-optimized metric: $OPT$~\cite{shypula2024learning} (Equation~\ref{eq:opt}), which captures improvement breadth: the fraction of problems in a partition with an accepted improvement on the target metric.
Like $\overline{IMP}$, $OPT$ applies to both execution time and average memory utilization.

\begin{equation}
    OPT = \frac{N_{\text{improved}}}{N_{\text{total}}} \times 100\%
    \label{eq:opt}
\end{equation}

For execution time specifically, we convert $\overline{IMP}$ into a speedup multiplier $SP$ (Equation~\ref{eq:sp}), which expresses the same relative improvement as a speedup factor more naturally interpretable in performance analysis.
We do not apply this conversion to average memory utilization, where $\overline{IMP}$ already serves as a directly meaningful normalized ratio.

\begin{equation}
    SP = \frac{1}{1 - \overline{IMP}}
    \label{eq:sp}
\end{equation}

We therefore populate two distinct sets of memory tables.
For execution-time refinement, we store $OPT$ and $SP$ per (language, category) partition for each LLM.
For average-memory refinement, we store $OPT$ and $\overline{IMP}$.
Section~\ref{sec:experiment_memory} reports the category-level execution-time memory (Tables~\ref{tab:memory_refine_opt_execution_time_by_language_category}--\ref{tab:memory_refine_sp_execution_time_by_language_category}); the corresponding average-memory tables appear in Appendix~\ref{sec:appendix_profiling}.

\textbf{Memory-Guided Routing.} Unlike generation and debugging, where model selection is based on binary pass/fail outcomes, refinement involves both the number of improved problems and their improvement magnitudes.
Moreover, since each problem carries multiple categories, a routing score must aggregate these two dimensions across all categories $\mathcal{C}_p$ of problem $p$.
We multiply improvement breadth ($OPT$) by improvement depth to obtain a per-category score, then sum these scores with equal weights across categories~\cite{yoon1995multiple}.
The form of the depth term varies by target metric: for execution time, the agent computes $SP \times OPT$ (Equation~\ref{eq:score_ref_time}); for average memory, it computes $\overline{IMP} \times OPT$ (Equation~\ref{eq:score_ref_mem}).
Because $OPT$ and $SP$/$\overline{IMP}$ are computed within each partition rather than pooled across problems, a category with more problems carries no more weight in $S_{\text{ref}}$ than a smaller one.
At runtime, the agent ranks all candidate LLMs by $S_{\text{ref}}^{\text{exec}}$ or $S_{\text{ref}}^{\text{mem}}$, and selects the top-$k$ models as its ordered candidate list.

\begin{equation}
    \begin{multlined}
    S_{\text{ref}}^{\text{exec}}(llm, language, \mathcal{C}_p) = \\
    \sum_{category \in \mathcal{C}_p} SP(llm, language, category) \times OPT(llm, language, category)
    \end{multlined}
    \label{eq:score_ref_time}
\end{equation}

\begin{equation}
    \begin{multlined}
    S_{\text{ref}}^{\text{mem}}(llm, language, \mathcal{C}_p) = \\
    \sum_{category \in \mathcal{C}_p} \overline{IMP}(llm, language, category) \times OPT(llm, language, category)
    \end{multlined}
    \label{eq:score_ref_mem}
\end{equation}

%% file: section/04.experiment.tex
\section{Experiment Setup}
\label{sec:experiment_overall}

We evaluate PerfOrch on five programming languages (Python, Java, C++, Go, and Rust) across four benchmarks.
Section~\ref{sec:experiment_settings} describes the benchmark configurations and evaluation settings.
Section~\ref{sec:experiment_memory} reports the agent memory tables derived from offline profiling.

\subsection{Benchmarks and Settings}
\label{sec:experiment_settings}

We use the four benchmarks described in Section~\ref{sec:background_bench}.
The full HumanEval-X problem set (164 problems per language) is used to construct the agent-specific memory rankings (Section~\ref{sec:methodology_overall}) and to evaluate in-distribution performance.
We evaluate PerfOrch on EffiBench-X (336 problems per language), BigCodeBench (1{,}140 tasks), and APPS (1{,}000 sampled problems) using the Memory rankings obtained from HumanEval-X without re-profiling.
All three benchmarks are unseen during profiling, allowing us to assess whether the rankings generalize to efficiency-oriented problems, library-based tasks, and contest problems.
For BigCodeBench, we use the Complete prompts, which provide a function signature and docstring consistent with the other function-level tasks in this study.
To limit evaluation cost, we use a fixed stratified random sample of 1{,}000 problems from APPS's 5{,}000-problem test split: 200 introductory, 600 interview, and 200 competition problems.
We manually convert APPS standard-input programs (Section~\ref{sec:background_bench}) to the function-level interface using wrappers that supply test inputs and capture outputs.
This adaptation allows all configurations to share the same evaluation harness for a fair comparison.
The four benchmarks contain a total of 4{,}640 problems for evaluation.

We evaluate using five LLMs: Claude 3.7 Sonnet (Anthropic), Gemini 2.0 Flash (Google), GPT-4o-2024-08-06 (OpenAI), Grok~3 (xAI), and Qwen 2.5 72B (Alibaba).
The first four are proprietary frontier models accessed via their respective APIs; Qwen 2.5 72B is an open-source model deployed locally.
The pool spans five distinct providers and includes both API-based and self-hosted models, providing architectural and training-data diversity while remaining small enough for exhaustive cross-stage profiling across five languages and ten categories.

We compare PerfOrch with three types of baselines: single-model one-shot generation, single-model pipelines, and multi-agent systems.
For one-shot generation, we evaluate each of the five candidate models directly on the problem prompts.
For the single-model pipelines, we run PerfCodeGen~\cite{peng2025perfcodegen} with each candidate model performing generation, debugging, and refinement.
We also compare with the state-of-the-art multi-agent systems MapCoder~\cite{islam2024mapcoder} and CodeSIM~\cite{islam2025codesim}, both using GPT-4o as the backbone model.
MapCoder and CodeSIM are evaluated for correctness and token cost, while PerfCodeGen also serves as the baseline for performance refinement.

We use greedy decoding ($temperature=0$, $top-p=0.9$) for all five candidate LLMs across all stages and benchmarks, following prior code-generation evaluations~\cite{liu2023your, zhuo2024bigcodebench, xu2025licoeval, wang2025towards, peng2025perfcodegen, li2026performance}.
This setting limits sampling variation in offline profiling and the resulting Memory rankings.
We apply the same decoding configuration to MapCoder and CodeSIM and retain the other hyperparameters reported in their original papers.
Section~\ref{sec:sampling_strategies} examines the effect of using provider-default sampling.

Functional correctness is measured by pass@1 rate~\cite{chen2021evaluating} (Equation~\ref{eq:correctness}).
Performance improvement is quantified across execution time and average memory utilization.
For execution time, we report $OPT$ and $SP$ (Equations~\ref{eq:opt} and~\ref{eq:sp}); for average memory utilization, we report $OPT$ and $\overline{IMP}$ (Equations~\ref{eq:opt} and~\ref{eq:imp}).

Open-source LLM Qwen 2.5 72B is hosted on a server equipped with two Intel Xeon Silver 4509Y CPUs (2.6\,GHz, 8C/16T each), 512\,GB RDIMM (8$\times$64\,GB, 5600\,MT/s), and two NVIDIA H100 GPUs.
Code execution and performance measurements are conducted on a Lenovo ThinkPad T480 (Intel i5-8250U, 16 GB RAM) with strict configurations: Intel Turbo Boost, dynamic frequency scaling, C-states, and SMT disabled; CPU pinned to maximum frequency in performance mode.

\subsection{Agent Memory}
\label{sec:experiment_memory}

This subsection reports the category-level profiling results that populate the agent-specific memory rankings described in Sections~\ref{sec:generation}--\ref{sec:refinement}.
Table~\ref{tab:generate_language_tag} presents category-level pass@1 rates for each model across all five languages.
Table~\ref{tab:fix_language_tag} reports the analogous fix@1 breakdowns.

Extending the preliminary observations in Section~\ref{sec:empirical_study} to five LLMs, five languages, and ten categories, three patterns emerge from the profiling data that motivate the category-aware, stage-wise design of PerfOrch.
\textit{First}, no single model dominates all language--category cells across stages: for generation (Table~\ref{tab:generate_language_tag}), the top-ranked model shifts across categories even within the same language (e.g., GPT-4o leads Python--Array but falls behind in Python--Simulation).
\textit{Second}, a model's ranking can change substantially between stages: Grok 3 leads Java generation in most categories (Table~\ref{tab:generate_language_tag}) but ranks fourth or fifth in Java fixing for the same categories (Table~\ref{tab:fix_language_tag}).

Tables~\ref{tab:memory_refine_opt_execution_time_by_language_category} and~\ref{tab:memory_refine_sp_execution_time_by_language_category} report category-level execution-time OPT and SP, which provide directly interpretable refinement outcomes and are used to populate the refinement-stage memory rankings (Section~\ref{sec:refinement}).
The corresponding average-memory OPT/IMP results are provided in Appendix~\ref{sec:appendix_profiling}.
\textit{Third}, refinement OPT and SP (Tables~\ref{tab:memory_refine_opt_execution_time_by_language_category}--\ref{tab:memory_refine_sp_execution_time_by_language_category}) exhibit high variance across categories and languages.

\begin{table}[!htbp]
    \caption{Category-level generation pass@1 (\%) of the five candidate LLMs across five languages and ten algorithmic categories on HumanEval-X. Shading marks \cellfirst{first}, \cellsecond{second}, and \cellthird{third} per column. These values populate the Generation Agent memory rankings.}
    \vspace{-0.3cm}
    \centering
    \renewcommand{\arraystretch}{0.85}
    \setlength{\aboverulesep}{0.2ex}
    \setlength{\belowrulesep}{0.2ex}
    \adjustbox{max width=0.98\textwidth}{
    \begin{tabular}{l|l|cccccccccc}
    \toprule
    \textbf{Lang.} & \textbf{Model} & \textbf{Array} & \textbf{Math} & \textbf{String} & \textbf{Count.} & \makecell{\textbf{Num.} \\\textbf{Theory}} & \textbf{Sim.} & \textbf{Sort.} & \textbf{Enum.} & \makecell{\textbf{Greedy} \\\textbf{Algo.}} & \makecell{\textbf{Hash} \\\textbf{Table}} \\
    \midrule
    \multirow{5}{*}{\rotatebox{90}{Python}} & Claude 3.7 Sonnet & \cellsecond{96.10} & \cellsecond{93.42} & \cellsecond{93.24} & \cellfirst{96.00} & \cellfirst{90.91} & \cellsecond{93.75} & \cellfirst{96.55} & \cellfirst{100.00} & \cellfirst{100.00} & \cellfirst{100.00} \\
     & Gemini 2.0 Flash & \cellthird{92.21} & \cellthird{89.47} & \cellthird{90.54} & 88.00 & \cellfirst{90.91} & \cellthird{87.50} & \cellthird{89.66} & \cellsecond{91.67} & \cellsecond{90.91} & \cellsecond{90.91} \\
     & GPT-4o & \cellfirst{97.40} & \cellfirst{94.74} & \cellfirst{95.95} & \cellfirst{96.00} & \cellfirst{90.91} & \cellsecond{93.75} & \cellsecond{93.10} & \cellfirst{100.00} & \cellfirst{100.00} & \cellsecond{90.91} \\
     & Grok 3 & \cellsecond{96.10} & \cellthird{89.47} & \cellsecond{93.24} & \cellsecond{94.00} & \cellsecond{87.88} & \cellsecond{93.75} & \cellfirst{96.55} & \cellsecond{91.67} & \cellsecond{90.91} & \cellfirst{100.00} \\
     & Qwen 2.5 72B & \cellsecond{96.10} & \cellsecond{93.42} & \cellthird{90.54} & \cellthird{90.00} & \cellfirst{90.91} & \cellfirst{96.88} & \cellsecond{93.10} & \cellsecond{91.67} & \cellfirst{100.00} & \cellsecond{90.91} \\
    \midrule
    \multirow{5}{*}{\rotatebox{90}{Java}} & Claude 3.7 Sonnet & \cellsecond{92.21} & 82.89 & 82.43 & 88.00 & \cellthird{69.70} & \cellfirst{90.62} & \cellfirst{89.66} & \cellsecond{83.33} & \cellfirst{100.00} & \cellfirst{100.00} \\
     & Gemini 2.0 Flash & \cellfirst{93.51} & \cellthird{85.53} & \cellthird{83.78} & \cellthird{90.00} & \cellsecond{84.85} & \cellthird{84.38} & \cellfirst{89.66} & \cellfirst{91.67} & \cellsecond{90.91} & \cellthird{81.82} \\
     & GPT-4o & 72.73 & 51.32 & 56.76 & 56.00 & 39.39 & 71.88 & 75.86 & \cellthird{66.67} & \cellthird{63.64} & 72.73 \\
     & Grok 3 & \cellsecond{92.21} & \cellfirst{93.42} & \cellsecond{89.19} & \cellfirst{96.00} & \cellfirst{96.97} & \cellsecond{87.50} & \cellsecond{86.21} & \cellfirst{91.67} & \cellsecond{90.91} & \cellfirst{100.00} \\
     & Qwen 2.5 72B & \cellthird{89.61} & \cellsecond{86.84} & \cellfirst{90.54} & \cellsecond{94.00} & \cellsecond{84.85} & \cellfirst{90.62} & \cellthird{82.76} & \cellsecond{83.33} & \cellfirst{100.00} & \cellsecond{90.91} \\
    \midrule
    \multirow{5}{*}{\rotatebox{90}{C++}} & Claude 3.7 Sonnet & \cellsecond{92.21} & \cellfirst{90.79} & \cellfirst{93.24} & \cellfirst{94.00} & \cellfirst{90.91} & \cellsecond{90.62} & \cellthird{86.21} & \cellsecond{83.33} & \cellsecond{90.91} & \cellsecond{90.91} \\
     & Gemini 2.0 Flash & \cellsecond{92.21} & \cellthird{88.16} & \cellsecond{89.19} & \cellthird{88.00} & \cellsecond{87.88} & \cellfirst{93.75} & \cellfirst{93.10} & \cellfirst{91.67} & \cellsecond{90.91} & \cellfirst{100.00} \\
     & GPT-4o & \cellthird{89.61} & 85.53 & \cellthird{85.14} & 82.00 & \cellthird{84.85} & 81.25 & \cellsecond{89.66} & \cellfirst{91.67} & \cellfirst{100.00} & \cellthird{81.82} \\
     & Grok 3 & \cellfirst{93.51} & \cellfirst{90.79} & \cellsecond{89.19} & \cellsecond{92.00} & \cellfirst{90.91} & \cellfirst{93.75} & \cellsecond{89.66} & \cellsecond{83.33} & \cellsecond{90.91} & \cellsecond{90.91} \\
     & Qwen 2.5 72B & 88.31 & \cellsecond{89.47} & \cellsecond{89.19} & \cellthird{88.00} & \cellfirst{90.91} & \cellthird{87.50} & \cellthird{86.21} & \cellsecond{83.33} & \cellfirst{100.00} & \cellfirst{100.00} \\
    \midrule
    \multirow{5}{*}{\rotatebox{90}{Go}} & Claude 3.7 Sonnet & \cellsecond{85.71} & \cellthird{84.21} & \cellsecond{82.43} & \cellsecond{88.00} & 75.76 & \cellthird{81.25} & \cellfirst{82.76} & \cellsecond{75.00} & \cellfirst{100.00} & \cellsecond{72.73} \\
     & Gemini 2.0 Flash & \cellfirst{87.01} & 78.95 & \cellfirst{87.84} & \cellthird{84.00} & \cellthird{81.82} & \cellfirst{87.50} & \cellfirst{82.76} & \cellsecond{75.00} & \cellfirst{100.00} & \cellfirst{81.82} \\
     & GPT-4o & 77.92 & \cellfirst{90.79} & 78.38 & 76.00 & \cellfirst{93.94} & \cellsecond{84.38} & \cellsecond{75.86} & \cellfirst{83.33} & \cellsecond{90.91} & \cellsecond{72.73} \\
     & Grok 3 & \cellthird{83.12} & \cellsecond{86.84} & \cellthird{81.08} & \cellfirst{90.00} & \cellsecond{90.91} & \cellfirst{87.50} & \cellthird{72.41} & \cellsecond{75.00} & \cellthird{81.82} & \cellfirst{81.82} \\
     & Qwen 2.5 72B & 79.22 & \cellsecond{86.84} & \cellthird{81.08} & 76.00 & \cellfirst{93.94} & \cellfirst{87.50} & \cellsecond{75.86} & \cellsecond{75.00} & \cellsecond{90.91} & \cellthird{63.64} \\
    \midrule
    \multirow{5}{*}{\rotatebox{90}{Rust}} & Claude 3.7 Sonnet & \cellsecond{77.92} & \cellsecond{80.26} & \cellsecond{78.38} & \cellthird{80.00} & \cellthird{81.82} & \cellthird{81.25} & \cellthird{58.62} & \cellsecond{75.00} & \cellfirst{90.91} & \cellsecond{72.73} \\
     & Gemini 2.0 Flash & \cellsecond{77.92} & 75.00 & \cellsecond{78.38} & 76.00 & 78.79 & 78.12 & \cellsecond{62.07} & \cellsecond{75.00} & \cellsecond{81.82} & \cellfirst{81.82} \\
     & GPT-4o & 67.53 & 72.37 & 72.97 & 68.00 & \cellthird{81.82} & 68.75 & \cellthird{58.62} & \cellsecond{75.00} & \cellthird{54.55} & \cellfirst{81.82} \\
     & Grok 3 & \cellfirst{80.52} & \cellfirst{81.58} & \cellfirst{87.84} & \cellfirst{84.00} & \cellsecond{90.91} & \cellsecond{87.50} & \cellfirst{68.97} & \cellfirst{83.33} & \cellfirst{90.91} & \cellfirst{81.82} \\
     & Qwen 2.5 72B & \cellthird{74.03} & \cellthird{78.95} & \cellthird{77.03} & \cellsecond{82.00} & \cellfirst{93.94} & \cellfirst{90.62} & \cellthird{58.62} & \cellthird{58.33} & \cellfirst{90.91} & \cellthird{45.45} \\
    \bottomrule
    \end{tabular}
    }
    \label{tab:generate_language_tag}
\end{table}

\begin{table}[!htbp]
    \caption{Category-level fix@1 (\%) of the five candidate LLMs across five languages and ten algorithmic categories on HumanEval-Pack. Shading marks \cellfirst{first}, \cellsecond{second}, and \cellthird{third} per column. These values populate the Debugging Agent memory rankings.}
    \vspace{-0.3cm}
    \centering
    \renewcommand{\arraystretch}{0.85}
    \setlength{\aboverulesep}{0.2ex}
    \setlength{\belowrulesep}{0.2ex}
    \adjustbox{max width=0.98\textwidth}{
    \begin{tabular}{l|l|cccccccccc}
    \toprule
    \textbf{Lang.} & \textbf{Model} & \textbf{Array} & \textbf{Math} & \textbf{String} & \textbf{Count.} & \makecell{\textbf{Num.} \\\textbf{Theory}} & \textbf{Sim.} & \textbf{Sort.} & \textbf{Enum.} & \makecell{\textbf{Greedy} \\\textbf{Algo.}} & \makecell{\textbf{Hash} \\\textbf{Table}} \\
    \midrule
    \multirow{5}{*}{\rotatebox{90}{Python}} & Claude 3.7 Sonnet & \cellfirst{98.70} & \cellfirst{97.37} & \cellfirst{97.30} & \cellfirst{98.00} & \cellfirst{96.97} & \cellfirst{100.00} & \cellfirst{96.55} & \cellfirst{91.67} & \cellfirst{100.00} & \cellsecond{90.91} \\
     & Gemini 2.0 Flash & \cellthird{94.81} & 90.79 & \cellsecond{95.95} & \cellsecond{94.00} & \cellthird{87.88} & \cellsecond{96.88} & \cellsecond{93.10} & \cellfirst{91.67} & \cellfirst{100.00} & \cellfirst{100.00} \\
     & GPT-4o & 88.31 & 88.16 & 83.78 & 90.00 & \cellthird{87.88} & 90.62 & 82.76 & \cellfirst{91.67} & \cellfirst{100.00} & \cellthird{81.82} \\
     & Grok 3 & 92.21 & \cellsecond{93.42} & 87.84 & \cellthird{92.00} & \cellsecond{93.94} & 84.38 & \cellthird{86.21} & \cellfirst{91.67} & \cellsecond{81.82} & \cellfirst{100.00} \\
     & Qwen 2.5 72B & \cellsecond{96.10} & \cellthird{92.11} & \cellthird{89.19} & \cellsecond{94.00} & \cellsecond{93.94} & \cellthird{93.75} & \cellthird{86.21} & \cellfirst{91.67} & \cellfirst{100.00} & \cellthird{81.82} \\
    \midrule
    \multirow{5}{*}{\rotatebox{90}{Java}} & Claude 3.7 Sonnet & \cellfirst{93.51} & \cellfirst{90.79} & \cellfirst{89.19} & \cellfirst{94.00} & \cellfirst{81.82} & \cellsecond{84.38} & \cellfirst{89.66} & \cellfirst{91.67} & \cellfirst{90.91} & \cellfirst{100.00} \\
     & Gemini 2.0 Flash & \cellsecond{89.61} & \cellsecond{85.53} & \cellsecond{86.49} & \cellsecond{86.00} & \cellsecond{78.79} & \cellfirst{90.62} & \cellfirst{89.66} & \cellsecond{83.33} & \cellsecond{81.82} & \cellfirst{100.00} \\
     & GPT-4o & \cellthird{85.71} & \cellthird{84.21} & 81.08 & \cellthird{84.00} & \cellfirst{81.82} & \cellsecond{84.38} & \cellthird{75.86} & \cellfirst{91.67} & \cellfirst{90.91} & \cellsecond{90.91} \\
     & Grok 3 & \cellthird{85.71} & 82.89 & 78.38 & 82.00 & \cellthird{75.76} & \cellsecond{84.38} & \cellsecond{79.31} & 50.00 & \cellfirst{90.91} & \cellthird{81.82} \\
     & Qwen 2.5 72B & 79.22 & 82.89 & \cellthird{83.78} & 78.00 & \cellsecond{78.79} & \cellthird{81.25} & 72.41 & \cellthird{66.67} & \cellsecond{81.82} & \cellthird{81.82} \\
    \midrule
    \multirow{5}{*}{\rotatebox{90}{C++}} & Claude 3.7 Sonnet & \cellfirst{85.71} & \cellfirst{84.21} & \cellfirst{94.59} & \cellfirst{88.00} & \cellfirst{87.88} & \cellfirst{93.75} & \cellthird{72.41} & \cellfirst{83.33} & \cellfirst{90.91} & \cellfirst{90.91} \\
     & Gemini 2.0 Flash & 79.22 & \cellsecond{82.89} & \cellthird{83.78} & \cellthird{76.00} & \cellsecond{81.82} & \cellfirst{93.75} & \cellsecond{79.31} & \cellsecond{75.00} & \cellfirst{90.91} & \cellfirst{90.91} \\
     & GPT-4o & \cellsecond{83.12} & 78.95 & 82.43 & \cellsecond{86.00} & \cellthird{78.79} & \cellthird{78.12} & \cellfirst{82.76} & \cellfirst{83.33} & \cellfirst{90.91} & \cellfirst{90.91} \\
     & Grok 3 & \cellsecond{83.12} & \cellthird{80.26} & \cellsecond{86.49} & \cellfirst{88.00} & \cellthird{78.79} & \cellsecond{87.50} & \cellfirst{82.76} & \cellsecond{75.00} & \cellsecond{81.82} & \cellsecond{81.82} \\
     & Qwen 2.5 72B & \cellthird{80.52} & \cellsecond{82.89} & 79.73 & \cellthird{76.00} & 72.73 & \cellthird{78.12} & \cellthird{72.41} & \cellsecond{75.00} & \cellsecond{81.82} & \cellfirst{90.91} \\
    \midrule
    \multirow{5}{*}{\rotatebox{90}{Go}} & Claude 3.7 Sonnet & \cellsecond{93.51} & \cellfirst{90.79} & \cellfirst{89.19} & \cellfirst{98.00} & \cellfirst{90.91} & \cellfirst{90.62} & \cellsecond{82.76} & \cellfirst{91.67} & \cellfirst{100.00} & \cellthird{81.82} \\
     & Gemini 2.0 Flash & \cellfirst{94.81} & \cellthird{84.21} & \cellsecond{87.84} & \cellsecond{96.00} & \cellsecond{81.82} & \cellfirst{90.62} & \cellfirst{86.21} & \cellfirst{91.67} & \cellfirst{100.00} & \cellthird{81.82} \\
     & GPT-4o & 85.71 & \cellthird{84.21} & 78.38 & \cellthird{84.00} & \cellsecond{81.82} & 68.75 & \cellthird{75.86} & \cellfirst{91.67} & \cellfirst{100.00} & \cellsecond{90.91} \\
     & Grok 3 & \cellthird{87.01} & 77.63 & 77.03 & 80.00 & \cellsecond{81.82} & \cellthird{81.25} & \cellsecond{82.76} & \cellthird{75.00} & \cellsecond{90.91} & \cellfirst{100.00} \\
     & Qwen 2.5 72B & 83.12 & \cellsecond{85.53} & \cellthird{83.78} & \cellthird{84.00} & \cellsecond{81.82} & \cellsecond{87.50} & 68.97 & \cellsecond{83.33} & \cellfirst{100.00} & \cellsecond{90.91} \\
    \midrule
    \multirow{5}{*}{\rotatebox{90}{Rust}} & Claude 3.7 Sonnet & \cellfirst{77.92} & \cellfirst{80.26} & \cellfirst{83.78} & \cellfirst{84.00} & \cellfirst{87.88} & \cellfirst{87.50} & \cellsecond{58.62} & \cellfirst{91.67} & \cellfirst{90.91} & \cellfirst{81.82} \\
     & Gemini 2.0 Flash & \cellsecond{76.62} & \cellsecond{76.32} & \cellsecond{79.73} & \cellfirst{84.00} & \cellsecond{78.79} & \cellsecond{84.38} & \cellfirst{65.52} & \cellsecond{75.00} & \cellsecond{81.82} & \cellfirst{81.82} \\
     & GPT-4o & 66.23 & 63.16 & 68.92 & \cellthird{64.00} & 66.67 & \cellthird{75.00} & \cellsecond{58.62} & \cellthird{66.67} & \cellthird{72.73} & \cellfirst{81.82} \\
     & Grok 3 & 68.83 & \cellthird{69.74} & \cellthird{72.97} & \cellsecond{74.00} & \cellthird{69.70} & 71.88 & \cellsecond{58.62} & \cellthird{66.67} & \cellsecond{81.82} & \cellfirst{81.82} \\
     & Qwen 2.5 72B & \cellthird{75.32} & 68.42 & \cellthird{72.97} & \cellthird{64.00} & 66.67 & \cellsecond{84.38} & \cellsecond{58.62} & 50.00 & \cellfirst{90.91} & \cellsecond{72.73} \\
    \bottomrule
    \end{tabular}
    }
    \label{tab:fix_language_tag}
\end{table}

\FloatBarrier

\begin{table}[!htbp]
    \caption{Category-level execution-time OPT (\%) of the five candidate LLMs across five languages and ten algorithmic categories on HumanEval-X. OPT measures the fraction of problems with an accepted execution-time improvement (Equation~\ref{eq:opt}). Shading marks \cellfirst{first}, \cellsecond{second}, and \cellthird{third} per column. Together with Table~\ref{tab:memory_refine_sp_execution_time_by_language_category}, these values populate the Refinement Agent memory rankings for exec. time.}
    \vspace{-0.3cm}
    \centering
    \renewcommand{\arraystretch}{0.82}
    \setlength{\aboverulesep}{0.2ex}
    \setlength{\belowrulesep}{0.2ex}
    \adjustbox{max width=0.98\textwidth}{
    \begin{tabular}{l|l|cccccccccc}
    \toprule
    \textbf{Lang.} & \textbf{Model} & \textbf{Array} & \textbf{Math} & \textbf{String} & \textbf{Count.} & \makecell{\textbf{Num.} \\ \textbf{Theory}} & \textbf{Sim.} & \textbf{Sort.} & \textbf{Enum.} & \makecell{\textbf{Greedy} \\ \textbf{Algo.}} & \makecell{\textbf{Hash} \\ \textbf{Table}} \\
    \midrule
    \multirow{5}{*}{\rotatebox{90}{Python}} & GPT-4o & \cellthird{54.55} & \cellthird{55.26} & 47.30 & \cellsecond{58.00} & \cellfirst{60.61} & 40.62 & \cellsecond{65.52} & \cellfirst{91.67} & \cellsecond{63.64} & 36.36 \\
     & Claude 3.7 Sonnet & \cellsecond{59.74} & 53.95 & \cellsecond{52.70} & \cellsecond{58.00} & \cellsecond{54.55} & \cellsecond{56.25} & \cellfirst{68.97} & \cellthird{66.67} & \cellfirst{72.73} & \cellsecond{72.73} \\
     & Gemini 2.0 Flash & \cellfirst{62.34} & \cellsecond{56.58} & \cellfirst{58.11} & \cellfirst{60.00} & \cellfirst{60.61} & \cellfirst{59.38} & \cellsecond{65.52} & \cellsecond{83.33} & \cellthird{54.55} & \cellfirst{81.82} \\
     & Grok 3 & \cellthird{54.55} & 51.32 & 48.65 & 48.00 & \cellthird{45.45} & 46.88 & \cellthird{55.17} & 58.33 & \cellsecond{63.64} & \cellthird{45.45} \\
     & Qwen 2.5 72B & 46.75 & \cellfirst{59.21} & \cellthird{50.00} & \cellthird{54.00} & \cellfirst{60.61} & \cellthird{50.00} & 51.72 & 58.33 & \cellsecond{63.64} & 36.36 \\
    \midrule
    \multirow{5}{*}{\rotatebox{90}{Java}} & GPT-4o & \cellthird{68.83} & \cellfirst{67.11} & \cellthird{63.51} & 66.00 & \cellfirst{66.67} & \cellsecond{81.25} & \cellsecond{65.52} & \cellfirst{66.67} & \cellsecond{81.82} & \cellthird{36.36} \\
     & Claude 3.7 Sonnet & \cellsecond{70.13} & \cellthird{60.53} & \cellsecond{70.27} & \cellsecond{76.00} & \cellthird{51.52} & \cellthird{71.88} & \cellfirst{68.97} & \cellfirst{66.67} & \cellfirst{90.91} & \cellsecond{63.64} \\
     & Gemini 2.0 Flash & \cellfirst{79.22} & \cellsecond{63.16} & \cellfirst{72.97} & \cellfirst{82.00} & \cellfirst{66.67} & \cellfirst{84.38} & \cellsecond{65.52} & \cellfirst{66.67} & \cellsecond{81.82} & \cellfirst{81.82} \\
     & Grok 3 & 58.44 & \cellthird{60.53} & 60.81 & \cellthird{68.00} & \cellsecond{60.61} & 62.50 & 44.83 & \cellthird{33.33} & 54.55 & \cellthird{36.36} \\
     & Qwen 2.5 72B & 57.14 & 55.26 & 45.95 & 58.00 & \cellthird{51.52} & 46.88 & \cellthird{51.72} & \cellsecond{41.67} & \cellthird{72.73} & 27.27 \\
    \midrule
    \multirow{5}{*}{\rotatebox{90}{C++}} & GPT-4o & \cellsecond{54.55} & \cellsecond{55.26} & 66.22 & 56.00 & \cellsecond{60.61} & \cellthird{53.12} & \cellsecond{58.62} & \cellsecond{33.33} & 45.45 & \cellthird{45.45} \\
     & Claude 3.7 Sonnet & \cellfirst{67.53} & \cellfirst{63.16} & \cellfirst{82.43} & \cellfirst{78.00} & \cellfirst{72.73} & \cellfirst{65.62} & \cellfirst{65.52} & \cellfirst{75.00} & \cellfirst{72.73} & \cellsecond{54.55} \\
     & Gemini 2.0 Flash & 50.65 & \cellthird{52.63} & \cellsecond{70.27} & \cellthird{58.00} & \cellthird{54.55} & \cellfirst{65.62} & 41.38 & \cellthird{25.00} & \cellsecond{63.64} & \cellthird{45.45} \\
     & Grok 3 & \cellthird{51.95} & 44.74 & \cellthird{68.92} & \cellsecond{62.00} & \cellthird{54.55} & \cellsecond{56.25} & \cellthird{55.17} & \cellthird{25.00} & \cellthird{54.55} & \cellfirst{72.73} \\
     & Qwen 2.5 72B & 46.75 & 44.74 & 63.51 & 48.00 & \cellthird{54.55} & 43.75 & 44.83 & \cellsecond{33.33} & 36.36 & \cellthird{45.45} \\
    \midrule
    \multirow{5}{*}{\rotatebox{90}{Go}} & GPT-4o & \cellfirst{63.64} & \cellsecond{60.53} & \cellsecond{58.11} & \cellthird{58.00} & \cellsecond{66.67} & \cellthird{53.12} & \cellthird{65.52} & \cellfirst{75.00} & \cellfirst{63.64} & \cellfirst{72.73} \\
     & Claude 3.7 Sonnet & \cellthird{59.74} & 55.26 & \cellsecond{58.11} & \cellfirst{66.00} & 57.58 & \cellfirst{71.88} & \cellfirst{79.31} & \cellthird{33.33} & \cellsecond{54.55} & \cellfirst{72.73} \\
     & Gemini 2.0 Flash & 51.95 & 47.37 & \cellthird{55.41} & \cellsecond{62.00} & 54.55 & \cellsecond{65.62} & \cellthird{65.52} & \cellthird{33.33} & \cellsecond{54.55} & \cellfirst{72.73} \\
     & Grok 3 & \cellsecond{62.34} & \cellfirst{64.47} & \cellfirst{62.16} & \cellfirst{66.00} & \cellfirst{69.70} & \cellfirst{71.88} & \cellsecond{68.97} & \cellsecond{50.00} & \cellsecond{54.55} & \cellsecond{54.55} \\
     & Qwen 2.5 72B & 54.55 & \cellthird{56.58} & 50.00 & 50.00 & \cellthird{63.64} & 50.00 & 48.28 & 25.00 & \cellthird{45.45} & \cellsecond{54.55} \\
    \midrule
    \multirow{5}{*}{\rotatebox{90}{Rust}} & GPT-4o & \cellsecond{58.44} & \cellthird{57.89} & \cellthird{60.81} & \cellthird{60.00} & \cellfirst{75.76} & \cellsecond{62.50} & \cellsecond{48.28} & \cellsecond{33.33} & \cellsecond{63.64} & \cellthird{54.55} \\
     & Claude 3.7 Sonnet & \cellfirst{61.04} & \cellsecond{61.84} & \cellfirst{74.32} & \cellsecond{64.00} & \cellsecond{66.67} & \cellfirst{71.88} & \cellfirst{51.72} & \cellfirst{50.00} & \cellsecond{63.64} & \cellfirst{72.73} \\
     & Gemini 2.0 Flash & \cellfirst{61.04} & \cellfirst{67.11} & \cellfirst{74.32} & \cellfirst{72.00} & \cellfirst{75.76} & 46.88 & \cellthird{41.38} & \cellfirst{50.00} & \cellfirst{72.73} & \cellsecond{63.64} \\
     & Grok 3 & \cellthird{50.65} & 55.26 & \cellsecond{63.51} & 56.00 & \cellthird{63.64} & \cellthird{59.38} & 37.93 & \cellsecond{33.33} & \cellfirst{72.73} & 45.45 \\
     & Qwen 2.5 72B & 41.56 & 46.05 & 47.30 & 48.00 & 57.58 & 43.75 & 31.03 & \cellsecond{33.33} & \cellthird{54.55} & 36.36 \\
    \bottomrule
    \end{tabular}
    }
    \label{tab:memory_refine_opt_execution_time_by_language_category}
\end{table}

\begin{table}[!htbp]
    \caption{Category-level execution-time SP (speedup factor, Equation~\ref{eq:sp}) of the five candidate LLMs across five languages and ten algorithmic categories on HumanEval-X. Shading marks \cellfirst{first}, \cellsecond{second}, and \cellthird{third} per column. Tied values share a shade.}
    \vspace{-0.3cm}
    \centering
    \renewcommand{\arraystretch}{0.82}
    \setlength{\aboverulesep}{0.2ex}
    \setlength{\belowrulesep}{0.2ex}
    \adjustbox{max width=0.98\textwidth}{
    \begin{tabular}{l|l|cccccccccc}
    \toprule
    \textbf{Lang.} & \textbf{Model} & \textbf{Array} & \textbf{Math} & \textbf{String} & \textbf{Count.} & \makecell{\textbf{Num.} \\ \textbf{Theory}} & \textbf{Sim.} & \textbf{Sort.} & \textbf{Enum.} & \makecell{\textbf{Greedy} \\ \textbf{Algo.}} & \makecell{\textbf{Hash} \\ \textbf{Table}} \\
    \midrule
    \multirow{5}{*}{\rotatebox{90}{Python}} & GPT-4o & \cellfirst{1.15} & \cellfirst{1.24} & \cellthird{1.08} & \cellfirst{1.15} & \cellfirst{1.33} & \cellthird{1.11} & \cellfirst{1.18} & \cellfirst{1.28} & \cellthird{1.18} & 1.07 \\
     & Claude 3.7 Sonnet & \cellfirst{1.15} & \cellfirst{1.24} & \cellsecond{1.10} & \cellfirst{1.15} & \cellsecond{1.30} & \cellfirst{1.13} & \cellsecond{1.17} & 1.17 & \cellfirst{1.24} & \cellsecond{1.15} \\
     & Gemini 2.0 Flash & \cellthird{1.13} & \cellsecond{1.23} & \cellfirst{1.11} & \cellsecond{1.14} & \cellfirst{1.33} & \cellfirst{1.13} & 1.15 & \cellsecond{1.22} & 1.17 & \cellfirst{1.16} \\
     & Grok 3 & \cellsecond{1.14} & \cellthird{1.21} & \cellthird{1.08} & \cellthird{1.13} & \cellthird{1.23} & \cellthird{1.11} & \cellthird{1.16} & \cellthird{1.20} & 1.15 & 1.07 \\
     & Qwen 2.5 72B & 1.12 & \cellthird{1.21} & \cellthird{1.08} & 1.12 & \cellsecond{1.30} & \cellsecond{1.12} & 1.14 & 1.14 & \cellsecond{1.20} & \cellthird{1.08} \\
    \midrule
    \multirow{5}{*}{\rotatebox{90}{Java}} & GPT-4o & \cellsecond{1.11} & \cellfirst{1.21} & 1.11 & \cellsecond{1.15} & \cellfirst{1.25} & 1.14 & \cellthird{1.08} & \cellfirst{1.17} & \cellthird{1.08} & \cellthird{1.07} \\
     & Claude 3.7 Sonnet & \cellfirst{1.13} & \cellfirst{1.21} & \cellfirst{1.15} & \cellfirst{1.18} & \cellthird{1.22} & \cellfirst{1.19} & \cellfirst{1.10} & \cellsecond{1.12} & \cellsecond{1.09} & \cellfirst{1.15} \\
     & Gemini 2.0 Flash & \cellsecond{1.11} & \cellthird{1.17} & \cellthird{1.12} & 1.13 & \cellsecond{1.23} & \cellthird{1.15} & \cellsecond{1.09} & \cellthird{1.11} & \cellthird{1.08} & \cellsecond{1.11} \\
     & Grok 3 & \cellthird{1.10} & \cellsecond{1.19} & \cellsecond{1.13} & \cellthird{1.14} & \cellthird{1.22} & \cellsecond{1.18} & \cellsecond{1.09} & 1.06 & \cellthird{1.08} & 1.06 \\
     & Qwen 2.5 72B & 1.08 & 1.16 & 1.09 & 1.12 & 1.21 & 1.09 & \cellthird{1.08} & \cellthird{1.11} & \cellfirst{1.14} & 1.06 \\
    \midrule
    \multirow{5}{*}{\rotatebox{90}{C++}} & GPT-4o & \cellthird{1.17} & \cellsecond{1.23} & \cellthird{1.34} & \cellthird{1.25} & \cellsecond{1.36} & \cellthird{1.25} & \cellsecond{1.19} & \cellsecond{1.11} & 1.05 & \cellsecond{1.30} \\
     & Claude 3.7 Sonnet & \cellfirst{1.24} & \cellfirst{1.34} & \cellfirst{1.59} & \cellfirst{1.40} & \cellfirst{1.52} & \cellfirst{1.33} & \cellfirst{1.20} & \cellfirst{1.36} & \cellfirst{1.34} & \cellthird{1.23} \\
     & Gemini 2.0 Flash & 1.15 & 1.18 & 1.32 & 1.23 & 1.26 & \cellsecond{1.26} & 1.08 & \cellthird{1.04} & \cellsecond{1.17} & 1.21 \\
     & Grok 3 & \cellsecond{1.19} & \cellthird{1.20} & \cellsecond{1.43} & \cellsecond{1.32} & \cellsecond{1.36} & 1.21 & \cellsecond{1.19} & \cellsecond{1.11} & \cellthird{1.14} & \cellfirst{1.58} \\
     & Qwen 2.5 72B & 1.14 & \cellthird{1.20} & 1.32 & 1.23 & \cellthird{1.35} & 1.20 & \cellthird{1.10} & \cellsecond{1.11} & 1.02 & 1.15 \\
    \midrule
    \multirow{5}{*}{\rotatebox{90}{Go}} & GPT-4o & \cellfirst{1.05} & \cellfirst{1.08} & \cellfirst{1.05} & \cellfirst{1.07} & \cellfirst{1.15} & \cellfirst{1.06} & \cellsecond{1.03} & \cellfirst{1.12} & \cellfirst{1.03} & \cellsecond{1.04} \\
     & Claude 3.7 Sonnet & \cellsecond{1.04} & \cellthird{1.06} & \cellfirst{1.05} & \cellsecond{1.06} & \cellsecond{1.12} & \cellfirst{1.06} & \cellfirst{1.05} & 1.00 & \cellsecond{1.02} & \cellfirst{1.08} \\
     & Gemini 2.0 Flash & \cellthird{1.03} & \cellthird{1.06} & \cellthird{1.03} & 1.04 & \cellsecond{1.12} & \cellsecond{1.04} & \cellsecond{1.03} & \cellthird{1.09} & \cellthird{1.01} & \cellsecond{1.04} \\
     & Grok 3 & \cellthird{1.03} & \cellsecond{1.07} & \cellsecond{1.04} & \cellthird{1.05} & \cellfirst{1.15} & \cellsecond{1.04} & \cellsecond{1.03} & \cellsecond{1.10} & \cellthird{1.01} & \cellthird{1.03} \\
     & Qwen 2.5 72B & \cellsecond{1.04} & \cellthird{1.06} & \cellthird{1.03} & 1.04 & \cellthird{1.11} & \cellsecond{1.04} & \cellthird{1.02} & 1.01 & \cellsecond{1.02} & \cellsecond{1.04} \\
    \midrule
    \multirow{5}{*}{\rotatebox{90}{Rust}} & GPT-4o & \cellthird{1.22} & \cellsecond{1.38} & \cellthird{1.47} & 1.39 & \cellthird{1.64} & \cellthird{1.38} & \cellthird{1.16} & 1.10 & \cellsecond{1.28} & \cellthird{1.26} \\
     & Claude 3.7 Sonnet & \cellfirst{1.27} & \cellfirst{1.45} & \cellfirst{1.68} & \cellfirst{1.61} & \cellfirst{1.72} & \cellfirst{1.40} & \cellfirst{1.25} & \cellfirst{1.43} & 1.23 & \cellfirst{1.62} \\
     & Gemini 2.0 Flash & \cellsecond{1.23} & \cellsecond{1.38} & \cellsecond{1.59} & \cellsecond{1.53} & \cellsecond{1.65} & 1.31 & \cellsecond{1.17} & \cellsecond{1.20} & \cellthird{1.25} & \cellsecond{1.28} \\
     & Grok 3 & 1.19 & \cellthird{1.34} & 1.46 & \cellthird{1.43} & 1.56 & \cellsecond{1.39} & 1.12 & \cellthird{1.17} & \cellfirst{1.36} & 1.24 \\
     & Qwen 2.5 72B & 1.12 & 1.23 & 1.26 & 1.28 & 1.41 & 1.23 & 1.08 & 1.10 & 1.18 & 1.10 \\
    \bottomrule
    \end{tabular}
    }
    \label{tab:memory_refine_sp_execution_time_by_language_category}
\end{table}

\FloatBarrier
\Needspace{5\baselineskip}

%% file: section/05.evaluation.tex
\section{Evaluation}
\label{sec:evaluation_overall}
We evaluate PerfOrch on functional correctness and performance refinement using the candidate models and baselines defined in Section~\ref{sec:experiment_settings}.
Section~\ref{sec:quantitative_results} compares these outcomes on HumanEval-X and three unseen benchmarks that extend the evaluation to an efficiency-oriented multilingual workload and to more demanding library-based and contest workloads.
Section~\ref{sec:qualitative_analysis} examines successful repairs, remaining failures, and algorithmic optimizations, while Section~\ref{sec:token_analysis} measures the token cost of the evaluated configurations.

\subsection{Quantitative Analysis}
\label{sec:quantitative_results}

\begin{table}[!ht]
    \caption{Pass@1 (\%) on the four benchmarks. Shading marks \cellfirst{first}, \cellsecond{second}, and \cellthird{third} per column.}
    \centering
    \adjustbox{max width=\textwidth}{
    \begin{tabular}{l|ccccc|ccccc|c|c}
    \toprule
    \multirow{2}{*}{\textbf{Model}} & \multicolumn{5}{c|}{\textbf{HumanEval-X}} & \multicolumn{5}{c|}{\textbf{EffiBench-X}} & \multirow{2}{*}{\textbf{BigCodeBench}} & \multirow{2}{*}{\textbf{APPS}} \\
     & \textbf{Python} & \textbf{Java} & \textbf{C++} & \textbf{Go} & \textbf{Rust} & \textbf{Python} & \textbf{Java} & \textbf{C++} & \textbf{Go} & \textbf{Rust} & & \\
    \midrule
    Claude 3.7 Sonnet & 94.51 & 85.98 & 90.85 & 84.76 & 79.27 & 63.99 & 79.17 & 68.45 & 64.88 & 68.75 & 55.88 & 67.10 \\
    GPT-4o & 95.12 & 59.76 & 86.59 & 82.32 & 69.51 & 41.67 & 74.70 & 47.32 & 40.48 & 41.37 & 56.58 & 48.90 \\
    Gemini 2.0 Flash & 90.24 & 87.20 & 90.24 & 85.37 & 77.44 & 61.61 & 76.49 & 63.39 & 48.81 & 54.17 & 63.33 & 61.60 \\
    Grok 3 & 93.29 & 91.46 & 90.24 & 83.54 & \cellthird{83.54} & 73.81 & 85.12 & 69.94 & 58.04 & 59.23 & 57.63 & \cellthird{73.80} \\
    Qwen 2.5 72B & 92.68 & 89.63 & 89.63 & 82.93 & 77.44 & \cellthird{86.01} & 80.06 & \cellthird{77.68} & 61.61 & 69.94 & 56.58 & 67.70 \\
    \midrule
    PCG + Claude 3.7 Sonnet & \cellthird{97.56} & 85.98 & 92.07 & 85.98 & 79.88 & 72.92 & 83.33 & 70.83 & \cellthird{72.62} & \cellthird{75.89} & 57.37 & 68.40 \\
    PCG + GPT-4o & 96.95 & 84.76 & 88.41 & 84.15 & 71.34 & 50.60 & 76.19 & 47.92 & 44.05 & 43.15 & 58.95 & 49.10 \\
    PCG + Gemini 2.0 Flash & 90.85 & 89.02 & \cellthird{92.68} & \cellthird{88.41} & 77.44 & 64.58 & 80.36 & 65.18 & 53.27 & 57.44 & \cellthird{65.35} & 62.40 \\
    PCG + Grok 3 & 95.12 & \cellsecond{93.29} & 92.07 & 85.37 & \cellsecond{84.15} & 76.19 & 86.31 & 72.62 & 65.18 & 61.90 & 59.56 & \cellsecond{74.90} \\
    PCG + Qwen 2.5 72B & \cellsecond{98.17} & \cellthird{92.68} & \cellsecond{93.29} & 86.59 & 80.49 & \cellsecond{92.86} & \cellsecond{92.86} & \cellsecond{91.37} & \cellsecond{84.23} & \cellsecond{78.57} & 60.26 & 68.10 \\
    \midrule
    MapCoder + GPT-4o & 93.29 & 83.54 & 89.63 & \cellsecond{90.85} & 66.46 & 60.42 & \cellthird{90.48} & 59.52 & 54.76 & 54.17 & \cellsecond{66.14} & 55.10 \\
    CodeSIM + GPT-4o & 95.12 & 91.46 & 92.07 & \cellsecond{90.85} & 79.27 & 66.37 & 88.10 & 60.42 & 63.39 & 63.39 & 64.65 & 52.40 \\
    \midrule
    PerfOrch & \cellfirst{99.39} & \cellfirst{98.17} & \cellfirst{98.78} & \cellfirst{98.17} & \cellfirst{91.46} & \cellfirst{96.13} & \cellfirst{99.40} & \cellfirst{96.13} & \cellfirst{94.35} & \cellfirst{93.15} & \cellfirst{80.44} & \cellfirst{93.40} \\
    \bottomrule
    \end{tabular}
    }
    \label{tab:agent_correctness}
\end{table}

\textbf{Overall Correctness.}
Table~\ref{tab:agent_correctness} compares PerfOrch with one-shot generation, single-model PerfCodeGen (PCG) pipelines, and the multi-agent systems MapCoder and CodeSIM.
The PCG comparison tests whether selecting models across stages improves over using one candidate throughout the pipeline, while the multi-agent comparison tests PerfOrch against alternative agent designs.
PerfOrch achieves the highest pass@1 in all twelve benchmark--language settings, exceeding the strongest baseline in each setting by 1.22--18.50 percentage points (pp), an average of 8.19~pp.

Even with a single model, PCG consistently lifts correctness over raw one-shot generation: on HumanEval-X, for example, Claude~3.7~Sonnet on Python rises from 94.51\% to 97.56\%, and GPT-4o on Java from 59.76\% to 84.76\%.
On EffiBench-X, PCG~+~Qwen~2.5~72B, the strongest single-model configuration on this benchmark, improves over raw Qwen~2.5~72B by 7--23~percentage points (pp) across languages.
On the more challenging BigCodeBench and APPS workloads, PCG improves over one-shot generation by 1.49--3.68~pp and 0.20--1.30~pp, respectively, while PerfOrch exceeds the strongest PCG configurations by 15.09~pp and 18.50~pp.
Single-model iteration adds less than 2~pp on HumanEval-X Rust (0.61~pp), APPS (1.10~pp), and HumanEval-X Java (1.83~pp).
Per-stage assignment adds 7.31, 18.50, and 4.88~pp in these settings, respectively.
PerfOrch's pass@1 ranges from 91.46\% to 99.39\% on HumanEval-X and from 93.15\% to 99.40\% on EffiBench-X.
On Rust, PerfOrch's margin over the strongest PCG baseline grows from 7.31~pp on HumanEval-X to 14.58~pp on EffiBench-X, a more demanding efficiency-oriented benchmark that retains five-language coverage.
The strongest PCG model also varies: Qwen leads on EffiBench-X, Gemini leads on BigCodeBench, and Grok leads on APPS.
Comparing PCG with one-shot generation shows that part of PerfOrch's correctness gain comes from the \textit{iterative generation pipeline}. Additionally, comparing PerfOrch with PCG shows that \textit{exploiting complementary model strengths} across stages adds a further gain, which is the larger of the two on harder tasks.

The multi-agent baselines test whether these gains also hold against systems that coordinate several agents using one backbone.
MapCoder uses retrieved exemplars for planning, and CodeSIM uses simulation to guide planning and debugging.
MapCoder improves over one-shot GPT-4o in every setting except HumanEval-X Python and Rust, while CodeSIM improves in every setting except HumanEval-X Python, where it ties one-shot GPT-4o.
Both MapCoder and CodeSIM surpass every PCG configuration on HumanEval-X Go, but PerfOrch improves further to 98.17\%, compared with 90.85\% for both baselines.
Across the twelve settings, PerfOrch exceeds the stronger multi-agent baseline by 18.74~pp on average.
On BigCodeBench, MapCoder is the strongest baseline overall, and PerfOrch exceeds it by 14.30~pp (80.44\% vs.\ 66.14\%).
The 1{,}140 BigCodeBench tasks and 1{,}000 APPS tasks extend the comparison to larger evaluations of challenging library-based and contest problems.
By selecting LLMs separately for each stage, PerfOrch's agents exploit complementary generation and debugging strengths, supporting our design of multi-model orchestration (Section~\ref{sec:qualitative_analysis}).

EffiBench-X, BigCodeBench, and APPS reuse the Memory rankings profiled on HumanEval-X without re-profiling.
PerfOrch leads every setting on these three benchmarks, showing that the measured model capabilities transfer to new workloads and remain useful for multi-agent routing.
The model-pool replacement in Section~\ref{sec:model_update_case_study} separately examines whether the gains persist when the candidate models change.

\textbf{Performance Refinement.} Tables~\ref{tab:agent_opt_sp_combined} and~\ref{tab:agent_opt_imp_average_memory} report refinement results along two performance metrics, execution time and average memory utilization, each measured on two dimensions: refinement breadth (OPT, the fraction of tasks with an accepted improvement) and refinement depth (SP for execution-time speedup; $\overline{IMP}$ for average-memory reduction).

\begin{table}[!ht]
    \caption{Execution-time refinement results on the four benchmarks. OPT and SP follow Equations~\ref{eq:opt} and~\ref{eq:sp}. Shading marks \cellfirst{first}, \cellsecond{second}, and \cellthird{third} per column.}
    \centering
    \adjustbox{max width=\textwidth}{
    \begin{tabular}{l|ccccc|ccccc|c|c}
    \toprule
    \multirow{2}{*}{\textbf{Model}} & \multicolumn{5}{c|}{\textbf{HumanEval-X}} & \multicolumn{5}{c|}{\textbf{EffiBench-X}} & \multirow{2}{*}{\textbf{BigCodeBench}} & \multirow{2}{*}{\textbf{APPS}} \\
     & \textbf{Python} & \textbf{Java} & \textbf{C++} & \textbf{Go} & \textbf{Rust} & \textbf{Python} & \textbf{Java} & \textbf{C++} & \textbf{Go} & \textbf{Rust} & & \\
    \midrule
    \multicolumn{13}{c}{\textbf{OPT (Execution Time)}} \\
    \midrule
    PCG + Claude 3.7 Sonnet & 34.15 & 10.98 & \cellsecond{64.02} & \cellthird{37.80} & \cellsecond{46.34} & 24.11 & \cellthird{42.86} & \cellsecond{38.10} & \cellsecond{29.46} & \cellsecond{37.20} & 14.30 & 31.50 \\
    PCG + GPT-4o & \cellthird{51.22} & 39.02 & 46.95 & 35.98 & 33.54 & 21.13 & 32.14 & 15.48 & 12.80 & 16.37 & 10.26 & \cellsecond{58.60} \\
    PCG + Gemini 2.0 Flash & \cellsecond{52.44} & \cellthird{51.22} & \cellthird{52.44} & \cellsecond{45.12} & \cellthird{41.46} & \cellsecond{28.27} & 35.12 & \cellthird{36.31} & \cellthird{26.19} & 33.04 & \cellsecond{17.37} & 48.10 \\
    PCG + Grok 3 & 50.61 & \cellsecond{54.27} & 38.41 & 37.20 & \cellsecond{46.34} & 26.19 & \cellsecond{47.32} & \cellsecond{38.10} & 24.70 & \cellthird{36.01} & 13.51 & 47.50 \\
    PCG + Qwen 2.5 72B & \cellsecond{52.44} & 46.95 & 36.59 & 31.10 & 37.20 & \cellthird{27.98} & 38.10 & \cellthird{36.31} & 20.83 & 29.17 & \cellthird{15.53} & \cellthird{53.80} \\
    \midrule
    PerfOrch & \cellfirst{78.05} & \cellfirst{78.05} & \cellfirst{86.59} & \cellfirst{85.37} & \cellfirst{81.10} & \cellfirst{77.08} & \cellfirst{90.18} & \cellfirst{80.95} & \cellfirst{60.71} & \cellfirst{76.79} & \cellfirst{21.32} & \cellfirst{84.20} \\
    \midrule
    \midrule
    \multicolumn{13}{c}{\textbf{SP (Execution Time)}} \\
    \midrule
    PCG + Claude 3.7 Sonnet & 1.019 & 1.007 & \cellsecond{1.032} & \cellthird{1.021} & \cellsecond{1.037} & \cellsecond{1.037} & \cellthird{1.073} & \cellsecond{1.111} & \cellthird{1.022} & \cellsecond{1.126} & \cellthird{1.014} & \cellthird{1.067} \\
    PCG + GPT-4o & 1.023 & 1.022 & 1.018 & 1.017 & 1.014 & 1.025 & 1.058 & 1.037 & 1.012 & 1.052 & 1.011 & 1.037 \\
    PCG + Gemini 2.0 Flash & \cellsecond{1.028} & \cellthird{1.024} & \cellthird{1.021} & 1.020 & 1.018 & \cellthird{1.035} & \cellsecond{1.079} & \cellthird{1.108} & \cellsecond{1.029} & \cellthird{1.112} & \cellsecond{1.019} & 1.045 \\
    PCG + Grok 3 & 1.024 & \cellsecond{1.025} & 1.013 & \cellsecond{1.024} & \cellthird{1.031} & 1.032 & 1.065 & 1.071 & 1.017 & 1.093 & \cellthird{1.014} & 1.048 \\
    PCG + Qwen 2.5 72B & \cellthird{1.025} & 1.009 & 1.000 & 1.016 & 1.000 & 1.032 & 1.048 & 1.040 & 1.010 & 1.001 & 1.012 & \cellsecond{1.094} \\
    \midrule
    PerfOrch & \cellfirst{1.141} & \cellfirst{1.146} & \cellfirst{1.335} & \cellfirst{1.050} & \cellfirst{1.427} & \cellfirst{1.075} & \cellfirst{1.139} & \cellfirst{1.240} & \cellfirst{1.046} & \cellfirst{1.250} & \cellfirst{1.124} & \cellfirst{1.179} \\
    \bottomrule
    \end{tabular}
    }
    \label{tab:agent_opt_sp_combined}
\end{table}

\emph{Execution time.} Single-model PCG refinement is unevenly effective: on HumanEval-X, the best PCG baseline for execution-time OPT varies by language and reaches at most about 64\% (Claude on C++); and on several languages, the top single-model configuration improves fewer than half of all tasks.
Refinement depth is similarly limited, with the strongest single-model speedup reaching only 1.037$\times$ (Claude on Rust).
The top-performing model shifts across languages (Claude leads on C++ and Rust, Gemini on Python, Grok on Java), confirming that refinement capability, like generation and debugging, is distributed across models rather than concentrated in any one.
PerfOrch exploits this distribution: on HumanEval-X, it achieves execution-time OPT of 78--87\% and speedups of 1.050--1.427$\times$, ranking first in every column.
The largest gains appear on C++ (1.335$\times$) and Rust (1.427$\times$).
On EffiBench-X, PerfOrch maintains execution-time OPT of 61--90\% with speedups of 1.046--1.250$\times$, exceeding the strongest PCG configuration by 31--49~pp in OPT across all five languages.
The execution-time gains also transfer to BigCodeBench and APPS without re-profiling the refinement rankings.
PerfOrch achieves OPT of 21.32\% and 84.20\%, compared with the best PCG values of 17.37\% and 58.60\%, respectively.
Its SP values are 1.124$\times$ and 1.179$\times$, compared with the best PCG values of 1.019$\times$ and 1.094$\times$.
PerfOrch leads both refinement breadth and depth in all twelve benchmark--language settings.

\begin{table}[!ht]
    \caption{Average-memory refinement results on the four benchmarks. OPT and $\overline{IMP}$ follow Equations~\ref{eq:opt} and~\ref{eq:imp}. Shading marks \cellfirst{first}, \cellsecond{second}, and \cellthird{third} per column.}
    \centering
    \adjustbox{max width=\textwidth}{
    \begin{tabular}{l|ccccc|ccccc|c|c}
    \toprule
    \multirow{2}{*}{\textbf{Model}} & \multicolumn{5}{c|}{\textbf{HumanEval-X}} & \multicolumn{5}{c|}{\textbf{EffiBench-X}} & \multirow{2}{*}{\textbf{BigCodeBench}} & \multirow{2}{*}{\textbf{APPS}} \\
     & \textbf{Python} & \textbf{Java} & \textbf{C++} & \textbf{Go} & \textbf{Rust} & \textbf{Python} & \textbf{Java} & \textbf{C++} & \textbf{Go} & \textbf{Rust} & & \\
    \midrule
    \multicolumn{13}{c}{\textbf{OPT (Average Memory)}} \\
    \midrule
    PCG + Claude 3.7 Sonnet & 44.51 & 28.05 & \cellsecond{3.05} & \cellthird{42.07} & \cellsecond{2.44} & \cellsecond{37.50} & \cellthird{48.81} & \cellsecond{20.54} & \cellsecond{29.76} & \cellsecond{18.45} & \cellthird{26.14} & 37.40 \\
    PCG + GPT-4o & 45.12 & 39.02 & \cellthird{1.22} & 40.85 & 1.22 & 27.08 & 36.61 & 8.33 & 14.88 & 8.33 & \cellsecond{27.28} & 42.30 \\
    PCG + Gemini 2.0 Flash & 31.71 & \cellsecond{49.39} & \cellsecond{3.05} & \cellthird{42.07} & 1.22 & \cellthird{36.31} & 41.37 & \cellthird{18.15} & \cellthird{25.60} & \cellthird{16.37} & 24.65 & 36.50 \\
    PCG + Grok 3 & \cellthird{48.17} & \cellthird{41.46} & 0.61 & \cellsecond{49.39} & \cellthird{1.83} & 35.42 & \cellsecond{52.08} & 14.88 & \cellthird{25.60} & 15.18 & 21.40 & \cellsecond{47.30} \\
    PCG + Qwen 2.5 72B & \cellsecond{49.39} & 35.37 & 0.61 & 31.71 & 1.22 & \cellthird{36.31} & 43.15 & 13.69 & 21.73 & 11.61 & 25.79 & \cellthird{45.00} \\
    \midrule
    PerfOrch & \cellfirst{81.71} & \cellfirst{80.49} & \cellfirst{51.22} & \cellfirst{76.83} & \cellfirst{54.27} & \cellfirst{65.77} & \cellfirst{81.25} & \cellfirst{47.92} & \cellfirst{64.29} & \cellfirst{41.96} & \cellfirst{31.67} & \cellfirst{76.80} \\
    \midrule
    \midrule
    \multicolumn{13}{c}{\textbf{IMP (Average Memory)}} \\
    \midrule
    PCG + Claude 3.7 Sonnet & 0.38 & 0.30 & \cellsecond{1.36} & \cellsecond{0.41} & \cellsecond{0.98} & \cellsecond{1.30} & \cellsecond{1.84} & \cellsecond{4.30} & \cellsecond{0.38} & \cellsecond{3.54} & 1.06 & \cellsecond{1.41} \\
    PCG + GPT-4o & 0.38 & \cellthird{0.71} & 0.38 & 0.34 & 0.03 & 0.84 & 1.44 & 1.75 & 0.18 & 1.75 & \cellsecond{1.29} & 0.62 \\
    PCG + Gemini 2.0 Flash & 0.23 & \cellsecond{0.82} & \cellthird{0.68} & 0.32 & 0.06 & 1.04 & \cellthird{1.70} & 3.67 & \cellthird{0.32} & \cellthird{2.94} & \cellthird{1.19} & \cellthird{1.24} \\
    PCG + Grok 3 & \cellthird{0.39} & 0.65 & 0.26 & \cellthird{0.37} & \cellthird{0.45} & 1.06 & 1.46 & \cellthird{4.10} & 0.24 & 2.55 & 0.97 & 1.13 \\
    PCG + Qwen 2.5 72B & \cellsecond{0.42} & 0.51 & 0.24 & 0.34 & 0.32 & \cellthird{1.14} & 1.13 & 3.82 & 0.23 & 1.79 & 1.12 & 1.05 \\
    \midrule
    PerfOrch & \cellfirst{1.37} & \cellfirst{1.97} & \cellfirst{4.81} & \cellfirst{1.08} & \cellfirst{8.13} & \cellfirst{1.80} & \cellfirst{2.21} & \cellfirst{6.39} & \cellfirst{0.48} & \cellfirst{7.16} & \cellfirst{1.54} & \cellfirst{2.83} \\
    \bottomrule
    \end{tabular}
    }
    \label{tab:agent_opt_imp_average_memory}
\end{table}

\emph{Average memory utilization.}
Table~\ref{tab:agent_opt_imp_average_memory} tests whether the refinement gains extend to a second optimization target.
The largest coverage differences on HumanEval-X occur on C++ and Rust: PerfOrch achieves OPT of 51.22\% and 54.27\%, compared with the best PCG values of 3.05\% and 2.44\%.
On EffiBench-X, PerfOrch achieves OPT of 41.96--81.25\% and leads every language.
On BigCodeBench and APPS, its OPT is 31.67\% and 76.80\%, compared with the best PCG values of 27.28\% and 47.30\%.
Its mean memory-improvement ratios are 1.54\% and 2.83\%, compared with the best PCG values of 1.29\% and 1.41\%, respectively.
PerfOrch also achieves the highest mean memory-improvement ratio in all twelve settings, so the advantage extends to both the number of improved tasks and the magnitude of improvement.

For C++ and Rust, where short execution times can increase variability in memory measurements, we use Wilcoxon signed-rank tests to compare PerfOrch's memory-improvement ratios with those of the PCG configuration with the highest pass@1 in each setting.
Each comparison includes only tasks that both configurations solve correctly and for which both have valid profiling data.
The comparators are PCG~+~Qwen 2.5 72B for both C++ settings and EffiBench-X Rust, and PCG~+~Grok 3 for HumanEval-X Rust.
We use a rank-based test because the ratios are heavy-tailed.
All four comparisons yield $p<0.01$: HumanEval-X C++ ($p=1.1\times10^{-15}$, $n=122$), HumanEval-X Rust ($p=3.2\times10^{-12}$, $n=136$), EffiBench-X C++ ($p=4.9\times10^{-3}$, $n=225$), and EffiBench-X Rust ($p=7.8\times10^{-7}$, $n=204$).
Section~\ref{sec:threats} reports separate checks of profiler accuracy.

\subsection{Qualitative Analysis}
\label{sec:qualitative_analysis}

Section~\ref{sec:quantitative_results} shows that PerfOrch achieves near-saturating correctness on two multi-lingual benchmarks, HumanEval-X and EffiBench-X, and substantially broader performance refinement than single-model pipelines across all four benchmarks.
This section uses representative examples to illustrate \emph{how} these gains arise and \emph{where} PerfOrch reaches its limits.
We examine three aspects: successful cross-model debugging, failure cases, and performance refinement.
Each draws one example from HumanEval-X and one from EffiBench-X.

\begin{figure}[!htbp]
\centering
\includegraphics[width=\textwidth]{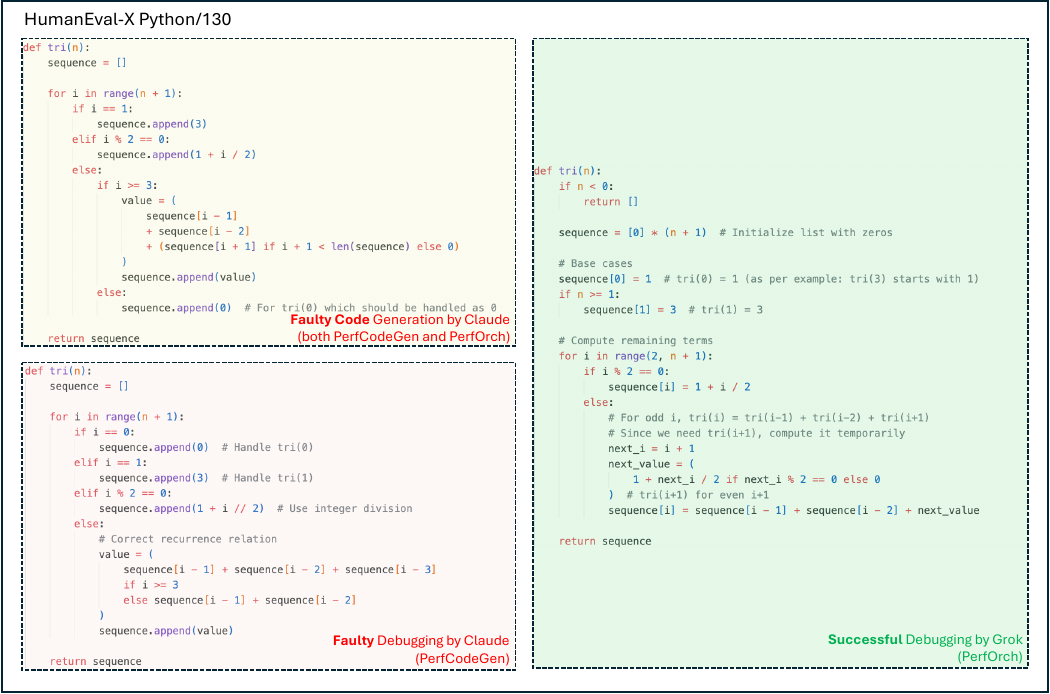}
\caption{Cross-model debugging success on HumanEval-X Python/130 (\texttt{tri}). Top-left: Claude generates code that mishandles base cases and forward references (used by both PerfCodeGen and PerfOrch). Bottom-left: under PerfCodeGen, Claude's own fix corrects the base case but introduces a wrong recurrence relation. Right: PerfOrch routes to Grok, which pre-allocates the sequence array and resolves the forward dependency.}
\Description{Pipeline flow diagram for HumanEval-X Python/130 showing Claude generate fail, Claude fix fail, and Grok fix success with code excerpts at each stage.}
\label{fig:case1-h}
\end{figure}

\begin{figure}[!htbp]
\centering
\includegraphics[width=\textwidth]{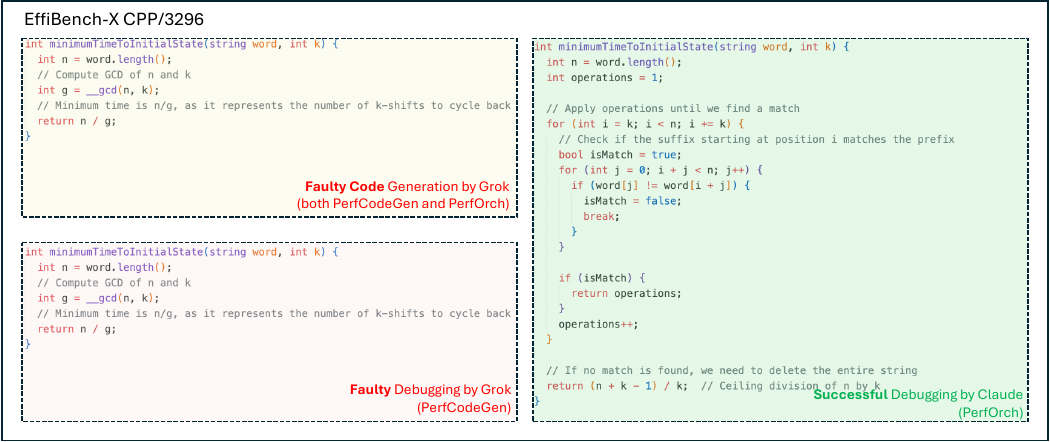}
\caption{Cross-model debugging success on EffiBench-X C++/3296 (\texttt{minimumTimeToInitialState}). Top-left: Grok generates an incorrect GCD-based formula (used by both PerfCodeGen and PerfOrch). Bottom-left: under PerfCodeGen, Grok's fix reproduces the same formula. Right: PerfOrch routes to Claude, which replaces it with a suffix--prefix matching loop that correctly handles all cases.}
\Description{Pipeline flow diagram for EffiBench-X C++/3296 showing Grok generate fail, Grok fix fail, and Claude fix success with code excerpts at each stage.}
\label{fig:case1-e}
\end{figure}

\textbf{Successful Cross-Model Debugging.} Figures~\ref{fig:case1-h} and~\ref{fig:case1-e} show two examples where cross-model debugging succeeds while single-model debugging fails.
In both cases, PerfOrch and PerfCodeGen start from the same faulty code produced by the same generator.
PerfCodeGen then uses that same model to attempt a fix and fails.
PerfOrch instead routes the faulty code to the top-ranked debugging model for the problem's language and category, which produces a correct solution.

\Needspace{8\baselineskip}

HumanEval-X Python/130 (\texttt{tri}) asks for the first $n+1$ terms of a Tribonacci-like sequence where even-indexed terms follow a closed-form rule and odd-indexed terms depend on their neighbors, including the \emph{next} even term.
As shown in Figure~\ref{fig:case1-h}, Claude generates an initial attempt that mishandles the base case for index~0 and tries to reference a value that has not been computed yet.
Under PerfCodeGen, Claude attempts to fix its own code: it corrects the base case but changes the recurrence relation, producing wrong values for odd indices.
The two Claude attempts stay within the same family of partial fixes.
PerfOrch instead routes the faulty code to Grok, which takes a different approach: it pre-allocates the full sequence array, computes even terms first, then fills in odd terms by calculating the next even value on the fly.
This passes all test cases.

\FloatBarrier

\Needspace{6\baselineskip}

EffiBench-X C++/3296 (\texttt{minimumTimeToInitialState}) asks for the minimum number of left-shift operations of length~$k$ needed to restore a string to its original form.
As shown in Figure~\ref{fig:case1-e}, Grok generates a solution based on $n / \gcd(n, k)$, a formula that works for cyclic block shifts but ignores character-level matching.
Under PerfCodeGen, Grok tries to fix its own code but reproduces the same GCD-based formula verbatim.
PerfOrch routes the faulty code to Claude, which takes an entirely different approach: an explicit loop that checks, at each multiple of~$k$, whether the remaining suffix matches the corresponding prefix.
This correctly handles strings whose period does not divide~$k$.

\FloatBarrier

In both debugging examples, the original model retains an incorrect solution strategy, while switching models produces a successful repair.
This illustrates complementary model strengths: a different model corrects a solution strategy that the original model fails to repair.

\Needspace{8\baselineskip}

\textbf{Failure Cases for PerfOrch.}
PerfOrch solves the vast majority of problems, but some remain unsolved after multi-model orchestration.
On HumanEval-X Python, PerfOrch achieves 99.39\% pass@1 (Table~\ref{tab:agent_correctness}), meaning exactly one problem out of 164 fails.
We use this single failure as our first example, and draw a second from EffiBench-X.
Together, they reveal two types of problems that PerfOrch fails to solve in the evaluated setting.

\begin{figure}[!htbp]
\centering
\includegraphics[width=\textwidth]{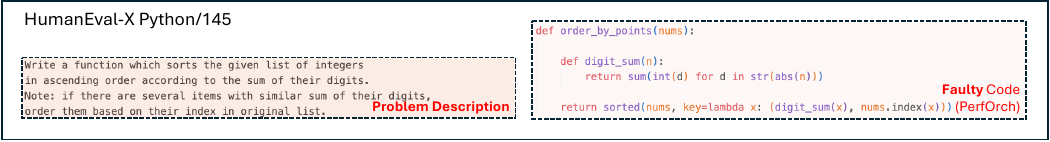}
\captionsetup{justification=raggedright,singlelinecheck=false}
\caption{Failure case: shared specification misunderstanding on HumanEval-X Python/145 (\texttt{order\_by\_points}). The specification defines digit sum with sign-sensitive first-digit handling for negative numbers. All five candidate models compute digit sums using \texttt{abs(...)}, producing wrong sort keys. Because every model makes the same misreading, cross-model routing cannot resolve the failure.}
\Description{Problem description for HumanEval-X Python/145 alongside PerfOrch's final wrong-answer code that uses abs-based digit sum computation.}
\label{fig:case2-h}
\end{figure}

The first type is when all candidate models misread the specification in the same way.
HumanEval-X Python/145 (\texttt{order\_by\_points}) asks for integers to be sorted by digit sum, with ties broken by original index (Figure~\ref{fig:case2-h}).
The specification defines digit sum for negative numbers in an unusual way: the leading negative sign negates only the first digit, rather than being stripped entirely.
All five candidate models compute digit sums using \texttt{abs(...)}, which gives the standard interpretation but the wrong answer.
Because all five candidate models make the same plausible misreading, switching among them does not help, as they all produce the same family of incorrect solutions.

\FloatBarrier

\Needspace{4\baselineskip}

\begin{figure}[!htbp]
\centering
\includegraphics[width=\textwidth]{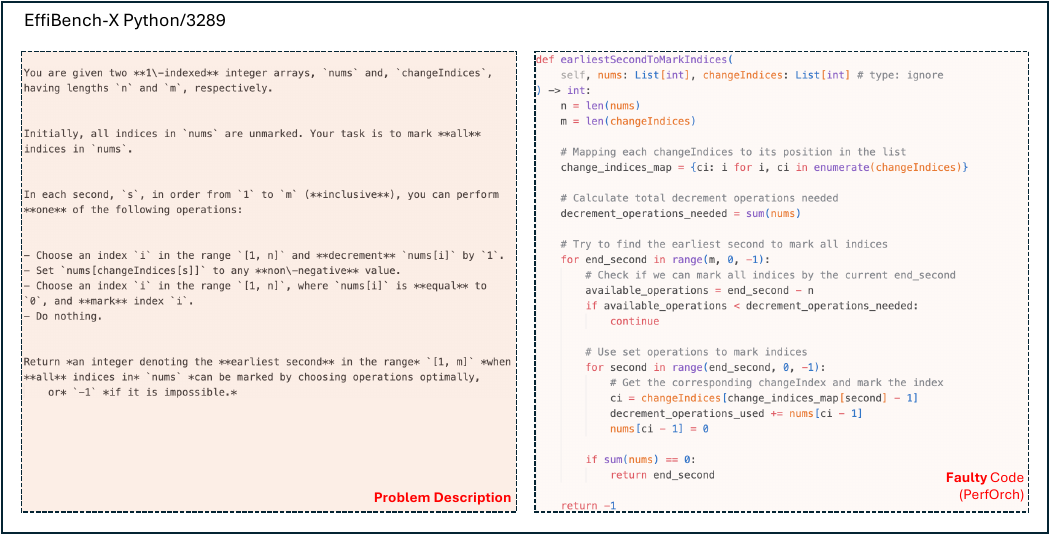}
\captionsetup{justification=raggedright,singlelinecheck=false}
\caption{Failure case: algorithmic complexity barrier on EffiBench-X Python/3289 (\texttt{earliestSecondToMarkIndices}). The correct solution requires binary search combined with greedy feasibility checking. PerfOrch's final attempt reduces the problem to a capacity check with destructive array mutation, failing on non-trivial inputs. No candidate model discovers the required two-level structure.}
\Description{Problem description for EffiBench-X Python/3289 alongside PerfOrch's final wrong-answer code that uses an incorrect greedy reduction.}
\label{fig:case2-e}
\end{figure}

In the second example, EffiBench-X Python/3289 (\texttt{earliestSecondToMarkIndices}) requires binary search combined with a greedy feasibility check.
The candidate models try different approaches, but none constructs this structure within the retry budget.
As shown in Figure~\ref{fig:case2-e}, PerfOrch's final attempt reduces the problem to a simple capacity check with destructive array mutation, which fails on non-trivial inputs.

\FloatBarrier

\Needspace{8\baselineskip}

\begin{figure}[!htbp]
\centering
\includegraphics[width=\textwidth]{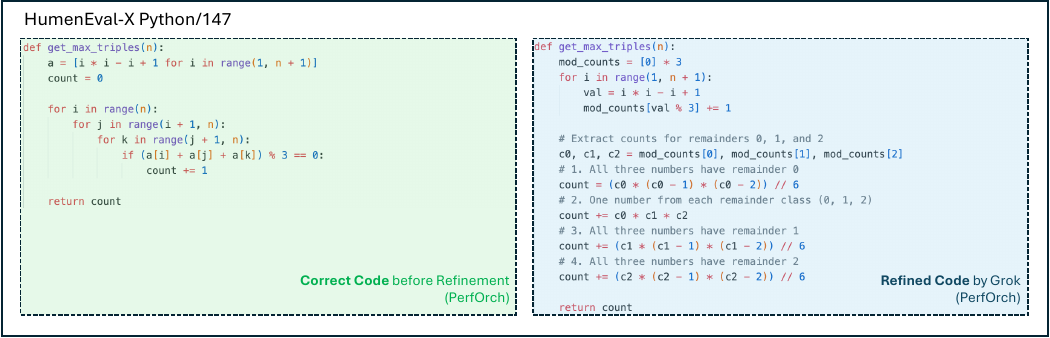}
\caption{Performance refinement on HumanEval-X Python/147 (\texttt{get\_max\_triples}). Left: the original $O(n^3)$ solution enumerates all index triples (22{,}160\,ms). Right: PerfOrch routes to Grok, which replaces the enumeration with $O(n)$ residue-class counting (136\,ms), achieving a 163$\times$ speedup.}
\Description{Side-by-side Python code comparison for HumanEval-X Python/147. The original uses three nested for-loops. The Grok-refined version counts residue classes modulo 3 and computes valid triples combinatorially.}
\label{fig:case3-h}
\end{figure}

\begin{figure}[!htbp]
\centering
\includegraphics[width=\textwidth]{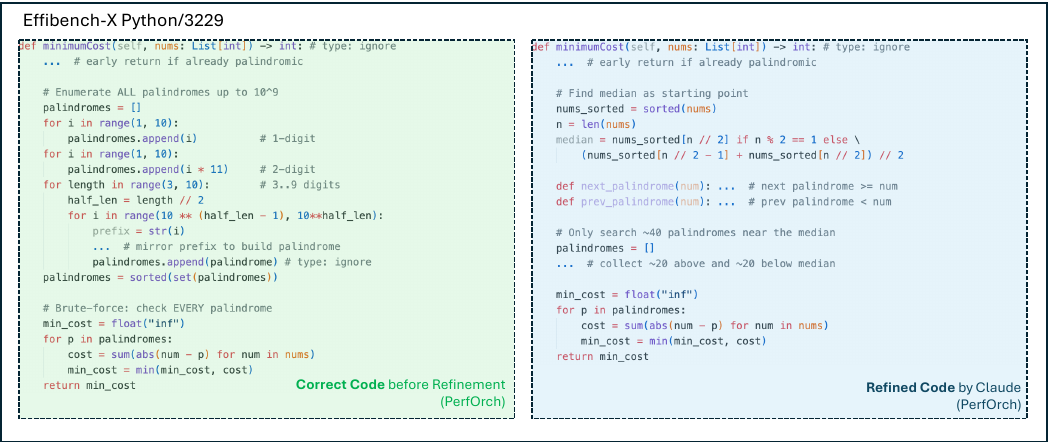}
\caption{Performance refinement on EffiBench-X Python/3229 (\texttt{minimumCost}). Left: the original solution enumerates all palindromes up to $10^9$ (111{,}372\,ms). Right: PerfOrch routes to Claude, which exploits cost-function convexity to search only ${\sim}40$ palindromes near the median (789\,ms), achieving a 141$\times$ speedup.}
\Description{Side-by-side Python code comparison for EffiBench-X Python/3229. The original enumerates all palindromes up to 10 to the 9. The Claude-refined version finds the median and searches only nearby palindromes.}
\label{fig:case3-e}
\end{figure}

\textbf{Performance Refinement.}
The aggregate SP values in Table~\ref{tab:agent_opt_sp_combined} are derived from mean relative runtime reductions, including zero contributions from unsuccessful refinements.
On individual problems with algorithmic headroom, PerfOrch can find much larger improvements by leveraging multiple models to explore diverse optimization strategies.
Figures~\ref{fig:case3-h} and~\ref{fig:case3-e} show two such examples, each refined by a different model.

HumanEval-X Python/147 (\texttt{get\_max\_triples}) asks for the number of index triples $(i,j,k)$ with $i < j < k$ such that $a_i + a_j + a_k$ is divisible by~3, where $a_i = i^2 - i + 1$.
As shown in Figure~\ref{fig:case3-h}, the original solution uses three nested loops to check all triples, running in $O(n^3)$ time (22{,}160\,ms).
PerfOrch routes the refinement to Grok, which recognizes that divisibility by~3 depends only on the remainder of each value modulo~3.
The refined version counts how many values fall into each remainder class and computes the answer combinatorially in $O(n)$ time (136\,ms), a 163$\times$ speedup.

\FloatBarrier

\Needspace{6\baselineskip}

EffiBench-X Python/3229 (\texttt{minimumCost}) asks for the minimum cost to make all array elements equal to some palindromic number.
As shown in Figure~\ref{fig:case3-e}, the original solution enumerates \emph{all} palindromes up to $10^9$ and evaluates the cost for each (111{,}372\,ms).
PerfOrch routes the refinement to Claude, which takes advantage of the cost function's convexity: it finds the median of the input array and searches only a small window of nearby palindromes (roughly 40 candidates), reducing the runtime to 789\,ms, representing a 141$\times$ speedup.

\FloatBarrier

In both examples, the refinement is not a surface-level tweak like loop unrolling; it is a fundamental change in the algorithm: brute-force enumeration replaced by combinatorial counting, and exhaustive search replaced by local search near the median.
In each case, the selected model was the top-ranked refiner for the problem's language and category, linking the observed algorithmic changes to the measured refinement rankings.
Together, the debugging and refinement examples show how PerfOrch exploits complementary model strengths: one model can repair errors or identify algorithmic improvements that another leaves in its code.

\FloatBarrier

\Needspace{6\baselineskip}
\subsection{Cost Analysis}
\label{sec:token_analysis}

\begin{table}[!ht]
    \caption{Mean pipeline output code-token count per task on the four benchmarks. The brute-force row queries all five models at every stage and retains the best result.}
    \centering
    \adjustbox{max width=\textwidth}{
    \begin{tabular}{l|ccccc|ccccc|c|c}
    \toprule
    \multirow{2}{*}{\textbf{Configuration}} & \multicolumn{5}{c|}{\textbf{HumanEval-X}} & \multicolumn{5}{c|}{\textbf{EffiBench-X}} & \multirow{2}{*}{\textbf{BigCodeBench}} & \multirow{2}{*}{\textbf{APPS}} \\
     & \textbf{Python} & \textbf{Java} & \textbf{C++} & \textbf{Go} & \textbf{Rust} & \textbf{Python} & \textbf{Java} & \textbf{C++} & \textbf{Go} & \textbf{Rust} & & \\
    \midrule
    PCG + Claude 3.7 Sonnet & 395.2 & 301.6 & 320.9 & 291.9 & 294.0 & 442.3 & 568.5 & 540.2 & 529.7 & 546.6 & 778.1 & 421.6 \\
    PCG + GPT-4o & 185.1 & 235.4 & 219.2 & 202.1 & 181.1 & 259.6 & 387.2 & 318.6 & 294.7 & 298.0 & 451.0 & 405.5 \\
    PCG + Gemini 2.0 Flash & 317.4 & 216.7 & 227.0 & 186.8 & 194.2 & 343.1 & 395.2 & 378.8 & 324.1 & 343.2 & 792.8 & 397.3 \\
    PCG + Grok 3 & 380.9 & 295.7 & 300.1 & 270.9 & 265.3 & 405.9 & 596.1 & 520.9 & 473.7 & 446.7 & 851.7 & 338.2 \\
    PCG + Qwen 2.5 72B & 403.6 & 347.4 & 403.2 & 386.6 & 420.7 & 2366.9 & 2521.2 & 2602.5 & 3140.7 & 1937.4 & 876.7 & 3194.7 \\
    \midrule
    MapCoder + GPT-4o & 2713.8 & 3667.0 & 3916.6 & 3212.2 & 1579.0 & 3222.5 & 1992.3 & 3079.4 & 2303.2 & 1625.2 & 3122.6 & 2738.0 \\
    CodeSIM + GPT-4o & 1583.0 & 2720.7 & 2435.8 & 2056.3 & 1321.0 & 5605.5 & 4837.6 & 6319.4 & 7098.1 & 5019.4 & 4170.4 & 4852.1 \\
    \midrule
    PerfOrch (exec time) & 443.0 & 650.7 & 524.5 & 597.1 & 855.2 & 1857.0 & 1186.7 & 2073.3 & 2306.2 & 2288.2 & 3524.0 & 3368.9 \\
    PerfOrch (avg memory) & 420.7 & 663.5 & 660.5 & 663.7 & 928.1 & 1935.7 & 1305.2 & 2293.6 & 2266.7 & 2538.8 & 3491.4 & 3452.0 \\
    \midrule
    Brute-force search & 6357.7 & 5495.8 & 5982.1 & 5168.8 & 5060.8 & 8661.7 & 9124.9 & 9414.7 & 9679.5 & 11165.8 & 8862.6 & 15831.9 \\
    \bottomrule
    \end{tabular}
    }
    \label{tab:token_comparison}
\end{table}

\textbf{Cost measurement.} Table~\ref{tab:token_comparison} reports pipeline \emph{output} code-token usage per task for PerfCodeGen (PCG) variants, MapCoder, CodeSIM, PerfOrch, and brute-force search across all four benchmarks, including unsuccessful attempts.
The comparisons below use PerfOrch's execution-time configuration, whose token count includes performance refinement.

\textbf{Single-model pipelines and brute-force search.} Brute-force search uses all five models for generation, applies all five debugging models to each faulty output, selects one correct candidate, and then tries all five refinement models.
This comparison tests how much token usage is reduced by trying ranked candidates sequentially and stopping when a candidate meets the stage's requirements.
On HumanEval-X, PerfOrch uses 443--855 tokens per task, 1.10--2.03$\times$ the cost of PCG~+~Qwen, the strongest PCG configuration averaged across languages, and only 7--17\% of the brute-force budget.
This additional cost over PCG accompanies the highest correctness and refinement results across all five languages (Section~\ref{sec:quantitative_results}).
On EffiBench-X, PerfOrch's cost advantage is even clearer.
The strongest single-model baseline, PCG~+~Qwen~2.5~72B, consumes 1{,}937--3{,}141 tokens per task, which is more than PerfOrch on four of five languages (e.g., Java: 1{,}187 vs.\ 2{,}521; Go: 2{,}306 vs.\ 3{,}141).
Despite being cheaper than PCG~+~Qwen on most languages, PerfOrch surpasses it by 3--15~pp in correctness (Table~\ref{tab:agent_correctness}) and achieves substantially higher execution-time OPT and SP (Table~\ref{tab:agent_opt_sp_combined}).
PerfOrch's token cost is higher than PCG~+~Qwen only on Rust (2{,}288 vs.\ 1{,}937), where the added cost produces 93.15\% correctness against PCG~+~Qwen's 78.57\%, and a 1.25$\times$ speedup against 1.00$\times$.
Relative to brute-force search (8{,}662--11{,}166 tokens), PerfOrch uses only 13--24\% of the budget.
On BigCodeBench, PerfOrch uses 3{,}524 tokens per task, compared with 451--877 for the PCG configurations and 8{,}863 for brute-force search.
The added cost over PCG yields a 15.09~pp correctness gain over its strongest configuration, while PerfOrch uses 40\% of the brute-force budget.
The 223 unsolved tasks exhaust the generate--debug retry budget.
These tasks average 13{,}442 tokens each, compared with 1{,}112 for the 917 solved tasks.
These unsuccessful attempts account for approximately 75\% of the output tokens, explaining much of the higher average cost on this benchmark.
On APPS, PerfOrch uses 3{,}369 tokens per task, compared with 338--422 for four PCG configurations, 3{,}195 for PCG~+~Qwen, and 15{,}832 for brute-force search.
It exceeds the strongest PCG configuration by 18.50~pp in correctness while using 21\% of the brute-force budget.
The 66 unsolved tasks average 12{,}080 tokens each, compared with 2{,}753 for the 934 solved tasks.
Unsolved tasks are more expensive individually, but solved tasks account for approximately 76\% of the total output tokens.

Across all four benchmarks, PerfOrch's execution-time configuration uses 7--40\% of the brute-force budget.
A separate refinement experiment in Section~\ref{sec:refinement_strategy} shows that sequential acceptance matches exhaustive execution-time refinement coverage with 66\% fewer refinement-stage tokens; the corresponding reduction for average-memory refinement is 57\%.

\textbf{Multi-agent systems.} On HumanEval-X, MapCoder and CodeSIM use 1.5--7.5$\times$ as many output tokens as PerfOrch, which achieves higher correctness in every language.
On EffiBench-X, they use approximately 1.0--4.1$\times$ as many tokens, except for MapCoder on Rust (1{,}625 vs.\ 2{,}288), where PerfOrch achieves 93.15\% correctness compared with MapCoder's 54.17\%.
On BigCodeBench, PerfOrch uses 3{,}524 tokens, compared with 3{,}123 for MapCoder and 4{,}170 for CodeSIM, while achieving 80.44\% correctness compared with 66.14\% and 64.65\%, respectively.
On APPS, PerfOrch uses 3{,}369 tokens, compared with 2{,}738 for MapCoder and 4{,}852 for CodeSIM, while achieving 93.40\% correctness compared with 55.10\% and 52.40\%, respectively.
PerfOrch thus combines higher correctness with lower token usage than CodeSIM on all four benchmarks and than MapCoder in most multilingual settings, while also refining execution performance.
Together, these comparisons show how ranked model selection makes multi-LLM-based multi-agent collaboration more economical than exhaustive search and improves correctness over single-backbone agent systems.

%% file: section/06.modelupdatecasestudy.tex
\section{A Case Study on Model Update}
\label{sec:model_update_case_study}

In this section, we replace all five models of Section~\ref{sec:evaluation_overall} with newer models from the same providers.
Section~\ref{sec:incremental_memory} examines the cost and accuracy of incremental Memory updates, and Section~\ref{sec:model_update_correctness} evaluates whether PerfOrch's advantage over the baselines persists across model generations.

\subsection{Incremental Memory Maintenance for Model Updates}
\label{sec:incremental_memory}

LLM providers regularly release new models or update existing ones, affecting the rankings in Memory.
We propose annual full profiling, with incremental updates between full runs to reduce maintenance cost.
Users need only profile the candidate models and programming languages they use.
We handle model changes in two scenarios.

\textit{Adding a previously unprofiled model.} When a user adds a model with no prior profiling data (e.g., from a new provider or a fine-tuned variant), full profiling on the stages and the user's selected languages is required as a one-time cost.

\textit{Updating an existing model's version.} A single re-profiling per version update is manageable, but with five model families releasing updates on different schedules, repeated full re-profiling becomes expensive.
For version updates, we use a cheaper incremental approach: run the new version only on problems the old version failed on, and estimate updated rankings from the combined results.
We test this approach on the generation stage of HumanEval-X (820 problems across five languages).
Table~\ref{tab:incremental_model_update_pool} compares full and incremental profiling for the five model transitions used in Section~\ref{sec:model_update_correctness}.
The incremental estimates exceed the fully measured pass@1 values by 1.70--4.51~pp because they carry forward predecessor successes without testing for regressions.
Qwen3.5-27B has the largest gap and the most regressions on previously solved tasks.
The aggregate model order is preserved except that Gemini 3.6 Flash and Grok 4.1 tie at 97.68\%.
For this generation-stage update, incremental profiling reduces output-token usage by 77--89\%, an average reduction of 83\%, leaving roughly one sixth of the full-profiling cost.
Annual full re-profiling checks for regressions on previously successful tasks.
If HumanEval-X becomes too easy to distinguish candidate models, a harder benchmark can replace it for profiling.
The purpose of profiling is to measure coding capabilities for model selection, and Section~\ref{sec:quantitative_results} shows that rankings measured on one benchmark remain useful on three unseen benchmarks.

\begin{table}[t!]
    \caption{Incremental profiling for the upgraded pool on HumanEval-X (generation stage, 820 problems).
The estimate re-tests only predecessor-failed tasks and carries successes forward.}
    \centering
    \adjustbox{max width=\textwidth}{
    \begin{tabular}{lcccccc}
    \toprule
    & \textbf{Old} & \multicolumn{2}{c}{\textbf{New}} & \multicolumn{2}{c}{\textbf{Estimated New}} \\
    \cmidrule(lr){3-4} \cmidrule(lr){5-6}
    \textbf{Transition} & \textbf{Pass@1 (\%)} & \textbf{Pass@1 (\%)} & \textbf{Tokens} & \textbf{Pass@1 (\%)} & \textbf{Tokens} \\
    \midrule
    GPT-4o $\rightarrow$ GPT-5.6-Luna & 78.66 & 93.78 & 198,670 & 96.46 & 45,178 \\
    Qwen 2.5 72B $\rightarrow$ Qwen3.5-27B & 86.46 & 89.39 & 167,359 & 93.90 & 29,600 \\
    Gemini 2.0 Flash $\rightarrow$ Gemini 3.6 Flash & 86.10 & 95.85 & 194,112 & 97.68 & 27,397 \\
    Claude 3.7 Sonnet $\rightarrow$ Claude Sonnet 5 & 87.07 & 95.37 & 186,689 & 97.20 & 33,895 \\
    Grok 3 $\rightarrow$ Grok 4.1 & 88.41 & 95.98 & 190,944 & 97.68 & 21,726 \\
    \bottomrule
    \end{tabular}
    }
    \label{tab:incremental_model_update_pool}
\end{table}

\subsection{Evaluations on the Updated Models}
\label{sec:model_update_correctness}

We replace the five models used in Section~\ref{sec:evaluation_overall} with Claude Sonnet 5, GPT-5.6-Luna, Gemini 3.6 Flash, Grok 4.1, and Qwen3.5-27B, following the transitions in Table~\ref{tab:incremental_model_update_pool}.
We update Memory for the upgraded pool.
The orchestration logic is unchanged.
We evaluate correctness and performance refinement on HumanEval-X, EffiBench-X, BigCodeBench, and APPS using the settings in Section~\ref{sec:experiment_settings}.
As in Section~\ref{sec:evaluation_overall}, we compare with PCG using each candidate model separately and with MapCoder and CodeSIM, now using GPT-5.6-Luna.

\begin{table}[!ht]
    \caption{Pass@1 (\%) on the four benchmarks with the upgraded pool.
Shading marks \cellfirst{first}, \cellsecond{second}, and \cellthird{third} per column.
Tied values share a rank.}
    \centering
    \adjustbox{max width=\textwidth}{
    \begin{tabular}{l|ccccc|ccccc|c|c}
    \toprule
    \multirow{2}{*}{\textbf{Model}} & \multicolumn{5}{c|}{\textbf{HumanEval-X}} & \multicolumn{5}{c|}{\textbf{EffiBench-X}} & \multirow{2}{*}{\textbf{BigCodeBench}} & \multirow{2}{*}{\textbf{APPS}} \\
     & \textbf{Python} & \textbf{Java} & \textbf{C++} & \textbf{Go} & \textbf{Rust} & \textbf{Python} & \textbf{Java} & \textbf{C++} & \textbf{Go} & \textbf{Rust} & & \\
    \midrule
    PCG + Claude Sonnet 5 & \cellfirst{100.00} & 97.56 & 96.34 & \cellthird{98.17} & 85.37 & 88.69 & 95.24 & 87.20 & 86.61 & 85.42 & 67.11 & 70.10 \\
    PCG + GPT-5.6-Luna & \cellthird{98.17} & 96.95 & 97.56 & \cellthird{98.17} & 80.49 & 94.05 & 99.11 & 93.75 & 93.75 & 80.65 & 65.35 & 55.10 \\
    PCG + Gemini 3.6 Flash & \cellfirst{100.00} & \cellsecond{99.39} & \cellsecond{99.39} & 97.56 & 85.37 & \cellfirst{99.70} & \cellfirst{100.00} & \cellfirst{99.70} & \cellthird{98.51} & \cellthird{91.07} & 66.75 & 88.10 \\
    PCG + Grok 4.1 & \cellfirst{100.00} & \cellthird{98.17} & \cellfirst{100.00} & 95.12 & \cellsecond{88.41} & \cellthird{97.32} & 98.21 & 97.62 & 84.82 & 66.67 & 64.47 & 85.70 \\
    PCG + Qwen3.5-27B & 97.56 & 92.68 & 94.51 & 90.24 & 79.27 & 85.71 & 88.39 & 85.71 & 74.11 & 75.60 & 58.42 & 77.90 \\
    \midrule
    MapCoder + GPT-5.6-Luna & \cellsecond{98.78} & \cellthird{98.17} & \cellthird{98.78} & \cellsecond{98.78} & \cellthird{87.80} & \cellsecond{98.51} & \cellsecond{99.70} & \cellthird{98.51} & 97.62 & \cellsecond{92.56} & \cellsecond{71.14} & \cellsecond{88.70} \\
    CodeSIM + GPT-5.6-Luna & \cellsecond{98.78} & 97.56 & \cellthird{98.78} & \cellfirst{99.39} & 86.59 & 97.02 & \cellthird{99.40} & \cellsecond{98.81} & \cellsecond{98.81} & 89.29 & \cellthird{69.56} & \cellthird{88.40} \\
    \midrule
    PerfOrch & \cellfirst{100.00} & \cellfirst{100.00} & \cellfirst{100.00} & \cellfirst{99.39} & \cellfirst{90.85} & \cellfirst{99.70} & \cellfirst{100.00} & \cellfirst{99.70} & \cellfirst{99.40} & \cellfirst{99.70} & \cellfirst{82.11} & \cellfirst{97.40} \\
    \bottomrule
    \end{tabular}
    }
    \label{tab:model_update_correctness}
\end{table}

\textit{Correctness.} PerfOrch matches or exceeds the strongest baseline in every column of Table~\ref{tab:model_update_correctness}.
On eight of the ten multilingual columns, its margin is at most 0.61~pp, and six are ties.
The remaining two columns are Rust, where PerfOrch leads by 2.44~pp on HumanEval-X (90.85\% vs.\ PCG~+~Grok~4.1's 88.41\%) and 7.14~pp on EffiBench-X (99.70\% vs.\ MapCoder's 92.56\%).
The strongest baselines are near the correctness ceiling on the other languages, limiting the possible absolute gains.
On BigCodeBench and APPS, PerfOrch leads by 10.97~pp (82.11\% vs.\ MapCoder's 71.14\%) and 8.70~pp (97.40\% vs.\ MapCoder's 88.70\%).
The strongest baseline still varies across workloads.
Gemini~3.6~Flash is the strongest PCG model when averaged across languages on HumanEval-X and EffiBench-X, and on APPS.
On BigCodeBench, it trails PCG~+~Claude~Sonnet~5.
MapCoder is the strongest baseline overall on both BigCodeBench and APPS.

\textit{Performance refinement.} Tables~\ref{tab:model_update_opt_sp} and~\ref{tab:model_update_opt_imp} repeat the refinement analysis from Section~\ref{sec:quantitative_results}.
PerfOrch ranks first in OPT, SP, and average-memory $\overline{IMP}$ in all twelve columns.
Its execution-time OPT ranges from 31.85\% to 89.70\%, compared with 23.60--61.59\% for the strongest single-model configuration in each column.
The corresponding average-memory OPT ranges are 47.02--78.05\% and 3.66--61.59\%.
C++ and Rust have the largest speedups on each multilingual benchmark.
Their SP values are 1.444$\times$ and 1.572$\times$ on HumanEval-X, and 1.656$\times$ and 1.296$\times$ on EffiBench-X.
Unlike correctness, refinement does not saturate with the stronger pool.

\begin{table}[!ht]
    \caption{Execution-time refinement with the upgraded pool, on the four benchmarks.
OPT and SP follow Equations~\ref{eq:opt} and~\ref{eq:sp}.
Shading marks \cellfirst{first}, \cellsecond{second}, and \cellthird{third} per column.}
    \centering
    \adjustbox{max width=\textwidth}{
    \begin{tabular}{l|ccccc|ccccc|c|c}
    \toprule
    \multirow{2}{*}{\textbf{Model}} & \multicolumn{5}{c|}{\textbf{HumanEval-X}} & \multicolumn{5}{c|}{\textbf{EffiBench-X}} & \multirow{2}{*}{\textbf{BigCodeBench}} & \multirow{2}{*}{\textbf{APPS}} \\
     & \textbf{Python} & \textbf{Java} & \textbf{C++} & \textbf{Go} & \textbf{Rust} & \textbf{Python} & \textbf{Java} & \textbf{C++} & \textbf{Go} & \textbf{Rust} & & \\
    \midrule
    \multicolumn{13}{c}{\textbf{OPT (Execution Time)}} \\
    \midrule
    PCG + Claude Sonnet 5 & \cellsecond{53.05} & 37.80 & \cellsecond{61.59} & \cellsecond{42.07} & 47.56 & 38.99 & \cellthird{24.70} & 46.43 & \cellthird{28.57} & \cellthird{47.62} & \cellthird{22.19} & 44.70 \\
    PCG + GPT-5.6-Luna & 43.29 & 35.98 & 34.76 & 28.66 & 35.37 & 40.77 & 23.51 & 37.50 & 26.49 & 36.31 & \cellsecond{23.60} & \cellsecond{61.40} \\
    PCG + Gemini 3.6 Flash & \cellthird{50.61} & \cellsecond{42.68} & \cellthird{42.68} & \cellthird{41.46} & \cellthird{48.17} & \cellthird{44.35} & 20.83 & \cellsecond{50.89} & \cellsecond{32.14} & \cellthird{47.62} & 17.63 & 56.90 \\
    PCG + Grok 4.1 & 49.39 & \cellthird{39.02} & 40.85 & 36.59 & \cellsecond{52.44} & \cellsecond{47.92} & \cellsecond{25.60} & \cellthird{49.11} & 24.70 & \cellsecond{55.65} & 20.53 & 53.20 \\
    PCG + Qwen3.5-27B & 46.34 & 32.32 & 31.71 & 39.63 & 42.68 & 38.39 & 16.67 & 40.18 & 16.96 & 42.26 & 16.05 & \cellthird{57.50} \\
    \midrule
    PerfOrch & \cellfirst{58.54} & \cellfirst{56.71} & \cellfirst{80.49} & \cellfirst{59.15} & \cellfirst{81.71} & \cellfirst{63.69} & \cellfirst{31.85} & \cellfirst{73.51} & \cellfirst{56.85} & \cellfirst{87.80} & \cellfirst{39.65} & \cellfirst{89.70} \\
    \midrule
    \midrule
    \multicolumn{13}{c}{\textbf{SP (Execution Time)}} \\
    \midrule
    PCG + Claude Sonnet 5 & \cellsecond{1.082} & 1.096 & \cellthird{1.029} & 1.013 & \cellsecond{1.066} & 1.065 & \cellthird{1.077} & 1.041 & \cellsecond{1.022} & \cellthird{1.153} & \cellsecond{1.022} & 1.071 \\
    PCG + GPT-5.6-Luna & 1.052 & 1.090 & 1.028 & 1.011 & 1.039 & \cellthird{1.078} & 1.072 & \cellthird{1.045} & \cellthird{1.017} & 1.061 & \cellsecond{1.022} & 1.045 \\
    PCG + Gemini 3.6 Flash & \cellthird{1.072} & \cellthird{1.106} & \cellsecond{1.037} & 1.012 & 1.038 & 1.069 & 1.071 & \cellsecond{1.068} & 1.010 & \cellsecond{1.183} & \cellthird{1.019} & 1.066 \\
    PCG + Grok 4.1 & 1.066 & \cellsecond{1.108} & \cellsecond{1.037} & \cellthird{1.014} & \cellthird{1.055} & \cellsecond{1.098} & \cellsecond{1.093} & 1.043 & 1.015 & 1.140 & 1.018 & \cellthird{1.096} \\
    PCG + Qwen3.5-27B & 1.063 & 1.069 & 1.028 & \cellsecond{1.015} & 1.011 & 1.064 & 1.046 & 1.032 & 1.004 & 1.018 & 1.013 & \cellsecond{1.113} \\
    \midrule
    PerfOrch & \cellfirst{1.121} & \cellfirst{1.136} & \cellfirst{1.444} & \cellfirst{1.019} & \cellfirst{1.572} & \cellfirst{1.104} & \cellfirst{1.099} & \cellfirst{1.656} & \cellfirst{1.026} & \cellfirst{1.296} & \cellfirst{1.040} & \cellfirst{1.222} \\
    \bottomrule
    \end{tabular}
    }
    \label{tab:model_update_opt_sp}
\end{table}

\begin{table}[!ht]
    \caption{Average-memory refinement with the upgraded pool, on the four benchmarks.
OPT and $\overline{IMP}$ follow Equations~\ref{eq:opt} and~\ref{eq:imp}.
Shading marks \cellfirst{first}, \cellsecond{second}, and \cellthird{third} per column.}
    \centering
    \adjustbox{max width=\textwidth}{
    \begin{tabular}{l|ccccc|ccccc|c|c}
    \toprule
    \multirow{2}{*}{\textbf{Model}} & \multicolumn{5}{c|}{\textbf{HumanEval-X}} & \multicolumn{5}{c|}{\textbf{EffiBench-X}} & \multirow{2}{*}{\textbf{BigCodeBench}} & \multirow{2}{*}{\textbf{APPS}} \\
     & \textbf{Python} & \textbf{Java} & \textbf{C++} & \textbf{Go} & \textbf{Rust} & \textbf{Python} & \textbf{Java} & \textbf{C++} & \textbf{Go} & \textbf{Rust} & & \\
    \midrule
    \multicolumn{13}{c}{\textbf{OPT (Average Memory)}} \\
    \midrule
    PCG + Claude Sonnet 5 & \cellthird{49.39} & 56.10 & \cellsecond{3.66} & 48.78 & \cellthird{4.27} & 46.13 & 54.76 & \cellsecond{25.30} & 38.99 & \cellthird{22.32} & \cellsecond{31.32} & 37.10 \\
    PCG + GPT-5.6-Luna & 47.56 & 56.10 & \cellthird{3.05} & 46.34 & 1.83 & 49.11 & \cellthird{58.33} & 10.12 & \cellthird{48.51} & 8.63 & 29.82 & \cellthird{44.00} \\
    PCG + Gemini 3.6 Flash & 48.17 & \cellthird{57.93} & \cellsecond{3.66} & \cellthird{50.00} & 1.83 & \cellsecond{53.87} & 54.76 & \cellthird{22.32} & \cellsecond{52.98} & 14.58 & \cellthird{30.61} & 35.80 \\
    PCG + Grok 4.1 & 44.51 & \cellsecond{61.59} & \cellsecond{3.66} & \cellsecond{56.10} & \cellsecond{4.88} & \cellthird{51.19} & \cellsecond{61.31} & 18.75 & 32.44 & \cellsecond{22.92} & 30.35 & \cellsecond{46.80} \\
    PCG + Qwen3.5-27B & \cellsecond{50.00} & 57.32 & \cellsecond{3.66} & 44.51 & 3.05 & 42.26 & 44.64 & 14.88 & 29.17 & 14.58 & 30.44 & 43.80 \\
    \midrule
    PerfOrch & \cellfirst{64.02} & \cellfirst{78.05} & \cellfirst{56.10} & \cellfirst{64.02} & \cellfirst{57.93} & \cellfirst{64.29} & \cellfirst{74.70} & \cellfirst{47.02} & \cellfirst{60.12} & \cellfirst{60.71} & \cellfirst{62.98} & \cellfirst{71.70} \\
    \midrule
    \multicolumn{13}{c}{\textbf{$\overline{IMP}$ (Average Memory)}} \\
    \midrule
    PCG + Claude Sonnet 5 & 2.43 & 3.46 & \cellsecond{1.92} & \cellthird{1.27} & \cellsecond{1.25} & \cellthird{3.33} & 4.68 & \cellthird{4.99} & 1.50 & \cellsecond{3.34} & \cellthird{1.46} & \cellsecond{3.17} \\
    PCG + GPT-5.6-Luna & 2.16 & 3.74 & \cellthird{0.77} & 1.23 & 0.05 & 2.82 & 4.94 & 1.34 & \cellsecond{1.67} & 2.22 & \cellsecond{1.54} & 1.71 \\
    PCG + Gemini 3.6 Flash & 2.21 & \cellthird{4.03} & 0.71 & \cellthird{1.27} & 0.14 & 3.02 & \cellthird{5.19} & 3.41 & \cellthird{1.57} & 2.73 & 1.44 & 1.88 \\
    PCG + Grok 4.1 & \cellthird{2.60} & \cellsecond{4.35} & 0.37 & \cellsecond{1.68} & \cellthird{0.46} & \cellsecond{3.39} & \cellsecond{6.60} & \cellsecond{5.44} & 1.20 & \cellthird{2.83} & 1.38 & \cellthird{2.46} \\
    PCG + Qwen3.5-27B & \cellsecond{2.67} & 3.47 & 0.25 & 1.10 & 0.43 & 2.40 & 3.31 & 3.51 & 0.86 & 2.40 & 1.16 & 2.21 \\
    \midrule
    PerfOrch & \cellfirst{2.86} & \cellfirst{5.06} & \cellfirst{5.19} & \cellfirst{1.89} & \cellfirst{11.39} & \cellfirst{3.73} & \cellfirst{6.61} & \cellfirst{7.01} & \cellfirst{1.83} & \cellfirst{8.76} & \cellfirst{2.80} & \cellfirst{3.24} \\
    \bottomrule
    \end{tabular}
    }
    \label{tab:model_update_opt_imp}
\end{table}

\textit{Token cost.} Table~\ref{tab:token_comparison_model_update} repeats the cost analysis of Section~\ref{sec:token_analysis}, including MapCoder and CodeSIM.
For execution-time optimization, PerfOrch uses 178.9--1{,}808.5 output tokens per task on the multilingual benchmarks, compared with 443.0--2{,}306.2 for the original pool.
The lowest cost is 178.9 tokens on HumanEval-X C++, within 1.5$\times$ of the cheapest PCG configuration.
PCG~+~Qwen3.5-27B remains the most expensive PCG configuration on EffiBench-X (1{,}724.6--2{,}145.6 tokens, 2.7--5.0$\times$ the next most expensive PCG configuration per language).
BigCodeBench remains PerfOrch's most expensive benchmark, at 3{,}719.8 tokens per task compared with 3{,}524.0 for the original pool.
The 204 unsolved tasks exhaust the generate--debug retry budget.
These tasks average 12{,}054 tokens each, compared with 1{,}903 for the 936 solved tasks.
On APPS, token usage falls from 3{,}368.9 to 1{,}761.2 as the number of unsolved tasks falls from 66 (6.6\%) to 26 (2.6\%).
The upgraded-pool cost is 1.2$\times$ PCG~+~Qwen3.5-27B (1{,}423.8 tokens) and 4.0--4.8$\times$ the other PCG configurations (367.4--442.6).
Across all twelve columns, PerfOrch's execution-time configuration uses 5--45\% of the brute-force estimate.

\textit{Effect of the model-pool update.} On EffiBench-X Rust, PerfOrch improves from 93.15\% to 99.70\%.
The APPS gain is 4.00~pp (93.40\%~$\to$~97.40\%).
PerfOrch gains 1.67~pp on BigCodeBench (80.44\% $\to$~82.11\%), alongside improvements in four of the five PCG configurations (PCG~+~Qwen3.5-27B declines from 60.26\% to 58.42\%).
After its underlying model is updated, MapCoder gains 5.00~pp on this benchmark (66.14\%~$\to$~71.14\%).
On HumanEval-X Rust, pass@1 changes from 91.46\% to 90.85\%.
The highest execution-time speedup rises from 1.427$\times$ to 1.656$\times$, and on BigCodeBench execution-time OPT rises from 21.32\% to 39.65\%, while average-memory OPT rises from 31.67\% to 62.98\%.
Token usage decreases in eight of the twelve execution-time settings, including a 48\% reduction on APPS, while it increases on HumanEval-X Python and Rust, EffiBench-X Java, and BigCodeBench.
These results show that complementary model strengths remain useful after all five candidate models are replaced: PerfOrch matches or exceeds the strongest correctness baseline and leads the reported refinement metrics with the updated pool.
The same orchestration framework can therefore benefit from stronger models through updated Memory rankings, supporting its applicability across the two evaluated model generations.

\begin{table}[!ht]
    \caption{Mean output code-token count per task with the upgraded pool, on the four benchmarks.
The brute-force row is an estimate that queries all five models at every stage.}
    \centering
    \adjustbox{max width=\textwidth}{
    \begin{tabular}{l|ccccc|ccccc|c|c}
    \toprule
    \multirow{2}{*}{\textbf{Configuration}} & \multicolumn{5}{c|}{\textbf{HumanEval-X}} & \multicolumn{5}{c|}{\textbf{EffiBench-X}} & \multirow{2}{*}{\textbf{BigCodeBench}} & \multirow{2}{*}{\textbf{APPS}} \\
     & \textbf{Python} & \textbf{Java} & \textbf{C++} & \textbf{Go} & \textbf{Rust} & \textbf{Python} & \textbf{Java} & \textbf{C++} & \textbf{Go} & \textbf{Rust} & & \\
    \midrule
    PCG + Claude Sonnet 5 & 340.6 & 256.0 & 158.6 & 208.8 & 176.5 & 406.7 & 619.4 & 366.8 & 646.2 & 395.7 & 728.0 & 378.6 \\
    PCG + GPT-5.6-Luna & 157.7 & 254.7 & 133.5 & 224.6 & 147.9 & 331.2 & 506.4 & 312.7 & 494.9 & 338.7 & 370.6 & 391.8 \\
    PCG + Gemini 3.6 Flash & 338.0 & 251.5 & 121.8 & 235.2 & 148.9 & 334.1 & 545.5 & 263.4 & 516.4 & 284.1 & 725.3 & 367.4 \\
    PCG + Grok 4.1 & 309.2 & 225.6 & 128.2 & 204.5 & 134.8 & 325.4 & 483.8 & 273.1 & 405.1 & 247.3 & 698.0 & 442.6 \\
    PCG + Qwen3.5-27B & 414.1 & 347.5 & 204.0 & 325.4 & 259.0 & 1855.3 & 2145.6 & 1830.7 & 1724.6 & 1838.3 & 858.3 & 1423.8 \\
    \midrule
    MapCoder + GPT-5.6-Luna & 2091.6 & 3589.2 & 3920.3 & 3065.4 & 1470.4 & 3672.5 & 1150.7 & 3359.5 & 2588.7 & 2245.1 & 3231.2 & 3184.8 \\
    CodeSIM + GPT-5.6-Luna & 1024.4 & 1837.9 & 1864.7 & 2568.7 & 1535.9 & 5899.0 & 3769.8 & 6291.4 & 6757.3 & 7014.7 & 3212.2 & 4356.9 \\
    \midrule
    PerfOrch (exec time) & 578.9 & 526.5 & 178.9 & 499.6 & 860.1 & 1011.9 & 1808.5 & 584.5 & 1364.8 & 900.9 & 3719.8 & 1761.2 \\
    PerfOrch (avg memory) & 568.2 & 541.3 & 181.2 & 521.0 & 872.6 & 1261.7 & 1874.9 & 692.6 & 1231.1 & 1134.4 & 3637.3 & 1803.4 \\
    \midrule
    Brute-force search & 1278.6 & 1270.3 & 1521.3 & 3284.0 & 8167.6 & 3061.7 & 4646.0 & 4175.2 & 16410.2 & 17330.9 & 13468.1 & 12726.3 \\
    \bottomrule
    \end{tabular}
    }
    \label{tab:token_comparison_model_update}
\end{table}

%% file: section/07.discussion.tex
\section{Threats to Validity and Discussion}
\label{sec:discussion_overall}
We first discuss threats to validity (Section~\ref{sec:threats}).
Sections~\ref{sec:k_sensitivity}--\ref{sec:sampling_strategies} examine candidate set size, refinement strategy, and decoding strategy.
Sections~\ref{sec:model_specialization}--\ref{sec:real_world_usage} discuss model capability differences, repository-level integration, and practical applicability.

\subsection{Threats to Validity}
\label{sec:threats}

\textbf{Internal validity.}
The offline profiling that builds Memory rankings must correctly identify which models perform best for each language--category--stage combination.
Noisy measurements here could lead to suboptimal model selection, causing more retry iterations and higher token cost.
Category labels also guide model selection, so incorrect labels can affect the order in which models are tried.
We assess automated categorization against human annotation: the predicted labels contain no incorrect categories on 96.95\% of problems, and micro-recall reaches 94.81\% (Section~\ref{sec:category}).

\textbf{Construct validity.}
The runtime performance profiling that guides accept/reject decisions in the Refinement Agent (Section~\ref{sec:refinement}) must reliably compare execution time or memory usage between the original and refined code.
Measurement noise here could cause the agent to accept a slower solution or reject a faster one, resulting in non-optimized or even degraded performance.
We reduce measurement noise through repeated execution and controlled hardware settings.
For HumanEval-X, we execute each task's test cases sequentially in a loop of 100{,}000 repetitions and time the entire loop with \texttt{perf stat} because individual calls can be too short for reliable process-level timing.
EffiBench-X, BigCodeBench, and APPS use their existing test suites without an additional repetition loop.
Average memory utilization is the mean of three independent CMDBench process invocations per solution.
During profiling, we disable Turbo Boost, fix CPU frequency, and isolate execution (Section~\ref{sec:experiment_settings}).
We assessed the profiler on synthetic single-process workloads running 10\,ms--5\,s, using hyperfine~\cite{hyperfine} as the reference for wall-clock timing.
The relative measurement errors were 0.17--0.36\% for average memory utilization and 0.77--1.12\% for execution time.
Based on these measurements, we use refinement thresholds of 0.4\% for average memory utilization and 1.2\% for execution time, each above the largest observed error for the corresponding metric.
Under this rule, a candidate is accepted only if it preserves correctness and its relative reduction exceeds the corresponding threshold (Section~\ref{sec:refinement}).
Greedy decoding (Section~\ref{sec:experiment_settings}) limits sampling variation, but does not explore alternative outputs from the same model.
Section~\ref{sec:sampling_strategies} measures the effect of provider-default sampling on BigCodeBench.

\textbf{External validity.}
Our four benchmarks may not cover all real-world coding scenarios.
However, they span 10 algorithmic categories across 5 languages and include compositional library use and competitive-programming tasks.
Memory rankings are built entirely from HumanEval-X, while EffiBench-X, BigCodeBench, and APPS are never seen during profiling.
PerfOrch's consistent gains on all three (Section~\ref{sec:quantitative_results}) show that the profiled model rankings remain effective beyond the profiling set.
As LLMs continue to improve, HumanEval-X may eventually become too easy to tell models apart.
PerfOrch does not depend on any specific benchmark, and switching to a harder one requires only re-running the offline profiling.
LLM capabilities evolve rapidly, so these rankings are a snapshot.
Section~\ref{sec:model_update_case_study} evaluates one complete model-pool update and quantifies the cost and estimation error of incremental profiling.
PerfOrch works at the function level, matching the scope of all four benchmarks.
Section~\ref{sec:repo_level} sets out what integration with a repository-level pipeline would require.

\subsection{Candidate Set Size}
\label{sec:k_sensitivity}

\begin{figure}[t]
    \centering
    \includegraphics[width=\linewidth]{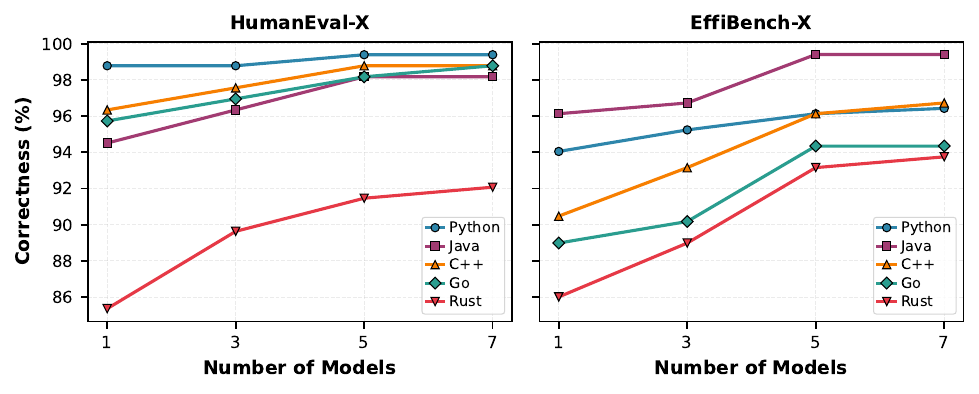}
    \caption{Sensitivity of PerfOrch pass@1 (\%) to the candidate set size $k$ on HumanEval-X (left) and EffiBench-X (right), one curve per language.}
    \Description{}
    \label{fig:k_models}
\end{figure}

PerfOrch tries up to $k$ ranked LLMs at each stage before accepting a solution.
Figure~\ref{fig:k_models} shows how correctness changes with $k$ on both benchmarks.
Each candidate is a distinct LLM; the $k=7$ sensitivity experiment adds Llama 3.1 8B and Codestral Mamba to the five-model pool used in the main evaluation.

On HumanEval-X, correctness rises with $k$ and largely flattens at $k=5$.
Rust sees the largest gain, climbing from about 85\% ($k=1$) to over 91\% ($k=5$); Python, already near 99\%, improves by less than 1~pp.
Going from $k=5$ to $k=7$ adds negligible benefit.
On EffiBench-X, the pattern is the same but stronger: Java rises from roughly 96\% to over 99\%, and C++ gains about 4~pp, again with diminishing returns past $k=5$.
This sensitivity analysis shows that adding more models beyond five incurs extra cost with minimal return on correctness, motivating us to set $k=5$.

\subsection{Refinement Strategy}
\label{sec:refinement_strategy}
Section~\ref{sec:token_analysis} compares PerfOrch's overall token cost against single-model PCG and brute-force search across the full pipeline.
This section zooms into the \emph{refinement stage} specifically, evaluating whether PerfOrch's sequential acceptance design achieves good performance refinement at lower cost than trying all five models exhaustively.
We compare three configurations: (1)~\textbf{Top-1}, which uses only the highest-ranked refinement model; (2)~\textbf{PerfOrch} (sequential acceptance with $k{=}5$): tries models in rank order and accepts the first one that passes all tests and whose relative performance improvement exceeds the target threshold (Section~\ref{sec:refinement}); and (3)~\textbf{Exhaustive}, which tries all five refinement models and keeps the best result.

\begin{table}[!ht]
    \caption{Execution-time refinement under the three strategies on HumanEval-X and EffiBench-X.}
    \centering
    \adjustbox{max width=\textwidth}{
    \begin{tabular}{l|ccccc|ccccc}
    \toprule
    \textbf{Strategy} & \multicolumn{5}{c|}{\textbf{HumanEval-X}} & \multicolumn{5}{c}{\textbf{EffiBench-X}} \\
     & \textbf{Python} & \textbf{Java} & \textbf{C++} & \textbf{Go} & \textbf{Rust} & \textbf{Python} & \textbf{Java} & \textbf{C++} & \textbf{Go} & \textbf{Rust} \\
    \midrule
    \multicolumn{11}{c}{\textbf{OPT}} \\
    \midrule
    \textbf{Top-1} & 48.78 & 51.83 & 70.73 & 50.00 & 59.76 & 51.79 & 57.44 & 54.76 & 26.19 & 45.83 \\
    \textbf{PerfOrch} & 78.05 & 78.05 & 86.59 & 85.37 & 81.10 & 77.08 & 90.18 & 80.95 & 60.71 & 76.79 \\
    \textbf{Exhaustive} & 78.05 & 78.05 & 86.59 & 85.37 & 81.10 & 77.08 & 90.18 & 80.95 & 60.71 & 76.79 \\
    \midrule
    \midrule
    \multicolumn{11}{c}{\textbf{SP}} \\
    \midrule
    \textbf{Top-1} & 1.102x & 1.100x & 1.302x & 1.037x & 1.352x & 1.059x & 1.082x & 1.164x & 1.026x & 1.155x \\
    \textbf{PerfOrch} & 1.141x & 1.146x & 1.335x & 1.050x & 1.427x & 1.075x & 1.139x & 1.240x & 1.046x & 1.250x \\
    \textbf{Exhaustive} & 1.196x & 1.165x & 1.389x & 1.072x & 1.542x & 1.093x & 1.172x & 1.279x & 1.051x & 1.279x \\
    \bottomrule
    \end{tabular}
    }
    \label{tab:paper_refine_strategy_exectime}
\end{table}

\textbf{Refinement coverage (OPT).}
Table~\ref{tab:paper_refine_strategy_exectime} (OPT rows) shows the percentage of problems with an accepted execution-time improvement.
PerfOrch and Exhaustive produce \emph{identical} OPT in every cell across both benchmarks and all five languages.
Both substantially outperform Top-1, for example, on EffiBench-X Java, Top-1 improves 57.44\% of problems while PerfOrch improves 90.18\% (+32.74~pp).

\textbf{Speedup magnitude (SP).}
Table~\ref{tab:paper_refine_strategy_exectime} (SP rows) reports the speedup across all problems.
Exhaustive yields moderately higher SP than PerfOrch because it picks the \emph{best} result among all five models, while PerfOrch accepts the \emph{first} successful one.
The largest gap is 0.115x (HumanEval-X Rust), and most language--benchmark pairs fall below 0.05x.

\begin{table}[!ht]
    \centering
    \caption{Refinement-stage output tokens per task for PerfOrch and Exhaustive on HumanEval-X and EffiBench-X.}
    \adjustbox{max width=\textwidth}{%
    \begin{tabular}{l|l|ccccc|ccccc}
    \toprule
    \multirow{2}{*}{\textbf{Metric}} & \multirow{2}{*}{\textbf{Models}}
    & \multicolumn{5}{c|}{\textbf{HumanEval-X}}
    & \multicolumn{5}{c}{\textbf{EffiBench-X}} \\
    \cmidrule(lr){3-7}\cmidrule(lr){8-12}
    & & \textbf{Python} & \textbf{Java} & \textbf{C++} & \textbf{Go} & \textbf{Rust}
    & \textbf{Python} & \textbf{Java} & \textbf{C++} & \textbf{Go} & \textbf{Rust} \\
    \midrule
    \multirow{2}{*}{\makecell{\textbf{Exec}\\\textbf{Time}}} & PerfOrch  & 214.01 & 282.26 & 253.10 & 257.76 & 233.88 & 355.60 & 543.96 & 502.15 & 722.82 & 527.56 \\
    & Exhaustive & 794 & 769 & 788 & 722 & 745 & 1452 & 1607 & 1510 & 1394 & 1482 \\
    \midrule
    \multirow{2}{*}{\makecell{\textbf{Avg}\\\textbf{Mem}}} & PerfOrch  & 191.65 & 295.09 & 389.10 & 324.30 & 306.79 & 434.30 & 662.51 & 722.49 & 683.35 & 778.16 \\
    & Exhaustive & 716 & 744 & 788 & 703 & 691 & 1386 & 1587 & 1505 & 1336 & 1441 \\
    \bottomrule
    \end{tabular}
    }
    \label{tab:agent_cost}
\end{table}

\textbf{Refinement-stage cost savings.}
Table~\ref{tab:agent_cost} shows the output token counts in the refinement stage.
Because PerfOrch stops at the first successful candidate, it uses fewer tokens than Exhaustive.
For example, on HumanEval-X Go (execution time), PerfOrch uses 257.76 versus 722 tokens (64\% fewer); on EffiBench-X Java, it uses 543.96 versus 1{,}607 tokens (66\% fewer).
This pattern holds across all languages and both metrics.
PerfOrch thus matches Exhaustive's execution-time refinement coverage while reducing refinement-stage token use by 66\% on average across the ten benchmark--language settings.
For average-memory optimization, the mean refinement-stage token reduction is 57\%.

\subsection{Decoding Strategy}
\label{sec:sampling_strategies}

As described in Section~\ref{sec:experiment_settings}, both the main evaluation and the model-update case study use greedy decoding ($temperature=0$).
This setting limits sampling variation during offline profiling and keeps PerfOrch and its baselines under the same decoding conditions.
Table~\ref{tab:temperature_sensitivity} compares greedy decoding with each model's default sampling configuration on BigCodeBench, keeping one output per candidate.
We choose BigCodeBench because PerfOrch has its lowest correctness on this benchmark, leaving more room to observe the effect of decoding strategy.
Every candidate model's default-sampling pass@1 is within 0.18 percentage points of its greedy counterpart, corresponding to at most two of the 1{,}140 tasks.
PerfOrch achieves 80.44\% pass@1 under greedy decoding and 79.30\% under default sampling, remaining above all five individual models under both settings.
Greedy decoding therefore retains PerfOrch's higher correctness in this comparison, consistent with Song et al.~\cite{song2024good}, who find that greedy decoding often outperforms single-output sampling on their evaluated tasks.
The comparison includes both hosted commercial models and self-hosted Qwen, and evaluates the decoding choice with one output per candidate.

\begin{table}[!ht]
    \caption{Pass@1 (\%) on BigCodeBench under greedy and provider-default sampling ($temperature=1.0$, or $0.7$ for Qwen 2.5 72B). $\Delta$ is default minus greedy, in percentage points.}
    \centering
    \begin{tabular}{l|ccc}
    \toprule
    \textbf{Model} & \textbf{Greedy} & \textbf{Default} & $\Delta$ \\
    \midrule
    Claude 3.7 Sonnet & 55.88 & 55.70 & -0.18 \\
    GPT-4o & 56.58 & 56.75 & +0.17 \\
    Gemini 2.0 Flash & 63.33 & 63.51 & +0.18 \\
    Grok 3 & 57.63 & 57.63 & 0.00 \\
    Qwen 2.5 72B & 56.58 & 56.67 & +0.09 \\
    \midrule
    PerfOrch & 80.44 & 79.30 & -1.14 \\
    \bottomrule
    \end{tabular}
    \label{tab:temperature_sensitivity}
\end{table}

For self-hosted open-source models, PerfOrch could also incorporate test-time scaling by generating multiple candidates per model and using execution results to select among them, as explored in code-generation methods such as S*~\cite{li2025sstar}.
At each stage, the agent could try several samples from a ranked model and accept the first that satisfies its stage-specific criteria before moving to the next model.
Generation and debugging require a candidate to pass the tests, while refinement additionally requires an improvement in the target metric.
This extension could combine broader exploration within a model with complementary strengths across models.
Generating additional samples increases decoding compute, and parallel sampling also requires GPU memory for the continuations' key-value caches.
Serving systems such as vLLM can share the prompt's key-value cache across samples to reduce duplicated storage~\cite{kwon2023efficient}.

Multiple samples per model would also require reassessing the candidate pool size studied in Section~\ref{sec:k_sensitivity}.
Under a fixed inference budget, the number of samples per model and the number of distinct models are joint allocation choices.
Their balance depends on how much additional sampling improves each model's coverage and how much the models' failures overlap.
Memory would need to be re-profiled under the chosen sampling configuration, and evaluating this allocation remains future work.

\subsection{Differences in Model Capability}
\label{sec:model_specialization}

Differences in training data and model design offer possible explanations for the complementary strengths observed in our profiles (Section~\ref{sec:experiment_memory}).
\textit{Training data} determines which programming languages and coding patterns a model encounters, so differences in this exposure can lead to different strengths across coding tasks.
Baltaji et al.~\cite{baltaji2025crosslingual} identify Rust as a low-resource language whose limited publicly available code constrains LLM training and performance.
This is consistent with the lower generation and debugging rates we observe for Rust than for Python.

\textit{Model architecture and training objectives} can also shape how well a model understands and repairs code.
Hellendoorn et al.~\cite{hellendoorn2020global} show that incorporating program relations into attention improves variable-misuse localization and repair.
Ding et al.~\cite{ding2024semcoder} train SemCoder to reason about execution behavior, improving code generation, execution reasoning, and debugging.
PerfOrch measures these capabilities by language, category, and stage to select models with complementary strengths.

These complementary strengths appear in our profiles and cases as partly non-overlapping errors and missed optimization opportunities.
The debugging examples show how switching models can escape local solution patterns that trap a single model, and the refinement examples illustrate algorithmic rather than superficial optimization (Section~\ref{sec:qualitative_analysis}).
PerfOrch exploits these differences by selecting models independently for generation, debugging, and refinement, allowing one model to address errors or inefficiencies left by another.

\subsection{Integration with Repository-Level Pipelines}
\label{sec:repo_level}

The complementary strengths measured in our generation, debugging, and refinement profiles (Section~\ref{sec:experiment_memory}) motivate investigating multi-model orchestration for repository-level coding.
Repository level systems combine localization or retrieval with code generation.
Agentless~\cite{xia2024agentless} runs three fixed phases: it localizes the fault, generates a patch, and validates the patch against tests.
AutoCodeRover~\cite{zhang2024autocoderover} retrieves the relevant methods through project-structure-aware search and then generates a patch over them.
RepoCoder~\cite{zhang2023repocoder} alternates retrieval and generation, using each draft to retrieve better context for the next.
These pipelines provide potential integration points when the requested change can be expressed as a function-level generation task.
At such a point, PerfOrch could replace the generation step with its generate--debug--refine pipeline.
The host system would retain localization, retrieval, and integration, and supply the function specification and retrieved context.
PerfOrch would also need executable tests during candidate selection, so integration must make the host's tests available at that point.

We evaluate repository-level task categorization in a pilot on database, API, and UI tasks from SWE-Bench Pro~\cite{deng2025swebenchpro} (Table~\ref{tab:repo-pilot}).
The vocabulary includes domain-specific concerns alongside the algorithmic categories of Section~\ref{sec:category}.
We seed the domain vocabularies from studies of database schema changes and testing~\cite{hernandez2022nosqlschema,gobert2023dbtesting}, API usage and developer needs~\cite{venkatesh2016webapi,liu2022apineeds}, and GUI programming concerns~\cite{mijailovic2014gui}.
GPT-5.6-Sol assigns multiple labels to 60 tasks per domain using their reference patches, and we retain categories assigned to at least five tasks.
On 30 sampled tasks per domain, the ten algorithmic categories leave 60.0--66.7\% without an applicable label.
The retained literature-derived categories cover 83.3--93.3\% of tasks, and the pilot identifies three additional categories: cache, error status, and endpoint.
An author checks the assigned labels on 30 tasks and finds all 52 assignments correct.
This patch-based labeling provides concrete domain vocabularies for extending categorization beyond algorithmic problems.
Assigning labels at inference time without reference patches, profiling candidate models on these categories, and evaluating category-aware routing remain future work.

\begin{table}[!htbp]
\centering
\small
\caption{Repository-level pilot on SWE-Bench Pro. \emph{LC-10 none}: share of 30 sampled tasks per domain with no applicable Table~\ref{Tab:category} category. \emph{Retained}: categories GPT-5.6-Sol assigned to $\geq$5 of 60 tasks given the reference patch (task counts); $\dagger$: added beyond the source taxonomies. \emph{Lit.\ cov.}: tasks with $\geq$1 retained literature category.}
\label{tab:repo-pilot}
\setlength{\tabcolsep}{4pt}
\begin{tabularx}{\linewidth}{@{}l r r >{\raggedright\arraybackslash}X r@{}}
\toprule
Domain & Domain tasks & LC-10 none & Retained categories & Lit.\ cov. \\
\midrule
Database & 78 & 60.0\% & Data access (40), Query (20), Migration (9), Schema field (8), Cache$^\dagger$ (7), Performance (7), Connection config (6) & 93.3\% \\
\midrule
API & 205 & 66.7\% & Payload (25), Resource ops (18), Input validation (10), Error status$^\dagger$ (9), Endpoint$^\dagger$ (8), Auth (6), Config/build (6), Transport (6) & 83.3\% \\
\midrule
UI & 117 & 66.7\% & Data binding (25), Structure (23), Event (16), Status feedback (13), Style (11), Formatting (8), Layout (8), Validation (8), Coupling (7) & 90.0\% \\
\bottomrule
\end{tabularx}
\end{table}

\subsection{Practical Applicability}
\label{sec:real_world_usage}

PerfOrch requires test cases to validate generated code at each pipeline stage.
This fits naturally into several common development workflows.
In test-driven development (TDD), tests are written before code, directly providing the pass/fail feedback PerfOrch needs.
In legacy system refactoring, existing test suites validate that the new code preserves the original behavior.
CI/CD pipelines rely on automated tests as quality gates, and PerfOrch can serve as a code generation step within such pipelines.
Where no tests exist yet, automated test generation can supply them from a specification or a function signature~\cite{Chen2022CodeTCG,mathews2024test}.
PerfOrch uses the resulting pass/fail outcomes for routing, and the correctness assessment depends on the faults those tests detect.

\FloatBarrier

%% file: section/08.relatedwork.tex
\section{Related Work}
\label{sec:related_overall}

This section reviews direct prompting of a single model for code generation (Section~\ref{sec:related_prompting}), agentic systems for code generation (Section~\ref{sec:related_agentic}), LLM-based performance optimization (Section~\ref{sec:related_performance}), and benchmarks beyond the four used in this paper (Section~\ref{sec:related_benchmarks}).

\subsection{Direct Prompting for Code Generation}
\label{sec:related_prompting}

Functional correctness remains a primary challenge in LLM-based code generation. While LLMs are increasingly embedded in software development workflows~\cite{bairi2024codeplan, sheng2025solsearch, xia2024agentless}, they often produce outputs affected by nondeterminism~\cite{liu2024no, ouyang2025empirical, zhu2024hot}, hallucinations~\cite{liu2024exploring}, security vulnerabilities~\cite{guo2024risky, kim2024codexity, liu2024no}, and overestimated capability~\cite{herrera2023large}. Wang et al.~\cite{wang2025towards} build a taxonomy of these errors on HumanEval and find that they are often non-trivial, multi-line mistakes with varied root causes.
One line of work improves the prompt without changing the model: few-shot prompting~\cite{li2024assessing, kouemo2024chain}, in-context learning~\cite{acharya2025optimizing, gao2023makes}, and Chain-of-Thought prompting~\cite{wei2022chain, niu2024evaluating} raise initial generation quality.

A second line samples several candidates from one model and selects among them by execution. Shi et al.~\cite{shi2022natural} rank candidates by agreement on executed outputs, and Chen et al.~\cite{Chen2022CodeTCG} generate test cases alongside the code and select the candidate whose outputs agree with both the generated tests and the other candidates.

A third line repairs a failing solution from execution feedback. Xia et al.~\cite{xia2023automated} apply pre-trained models to automated program repair. Olausson et al.~\cite{olausson2023self} measure the conditions under which a model can repair its own output, and Hidv\'egi et al.~\cite{hidvegi2024cigar} reduce the token cost of that loop. Zhong et al.~\cite{zhong2024debug} guide it with runtime execution traces, Shi et al.~\cite{shi2024code} debug hierarchically after decomposing the code into subfunctions, and Mathews and Nagappan~\cite{mathews2024test} drive it from developer-written tests. In each of these loops the model that diagnoses the fault is the model that produced it.

\subsection{Agentic Systems for Code Generation}
\label{sec:related_agentic}

Qian et al.~\cite{qian2024chatdev} organize agents into the roles of a software company, covering design, coding, and testing. Hong et al.~\cite{hong2024metagpt} encode standard operating procedures into prompt sequences and have agents communicate through structured documents and diagrams rather than dialogue. He et al.~\cite{he2025review} survey this literature across lifecycle phases and report that cross-examination among agents reduces hallucination. More recent frameworks specialize their agents for code generation itself. Zhou et al.~\cite{zhou2024lats} search over reasoning and acting steps with Monte Carlo tree search, Zhang et al.~\cite{zhang2024paircoder} pair a navigator and a driver to explore several plans, Zhang et al.~\cite{zhang2025thinkcoder} add an exploration phase before committing to a refinement, and Hosain et al.~\cite{hosain2025xolver} let a team of agents accumulate shared experience across problems. Agent-based repair takes the same shape. Bouzenia et al.~\cite{bouzenia2025repairagent} let RepairAgent choose repair tools autonomously, and Xia et al.~\cite{xia2024agentless} report with Agentless that a fixed three-phase procedure is competitive with such agents. In all of these frameworks the agents differ in role and draw on one underlying model.

A separate line of work varies the model instead.
Shazeer et al.~\cite{shazeer2017outrageously} route tokens to sparsely activated experts inside a single network, and later systems route whole requests across separate models.
Jiang et al.~\cite{jiang2023llm} rank and fuse the outputs of several LLMs after generation.
Chen et al.~\cite{chen2023frugalgpt} cascade from cheaper to more expensive models until a learned scorer accepts an answer.
The cascade responds to generated outputs, ordering models by cost and evaluating each complete answer with a learned score.
Mei et al.~\cite{mei2025omnirouter} predict per-model accuracy and cost and allocate queries under capacity constraints, on knowledge and mathematical reasoning benchmarks with no code-generation task.
Vangala and Malik~\cite{vangala2025efficient} combine keyword and classifier routing with a deployment layer that manages latency and GPU cost.
OmniRouter and Pick-and-Spin decide once per request from a predicted property of the query, and the selected model then produces the entire answer.
Su et al.~\cite{su2025difficulty} propose DAAO, which allocates workflow operators and models using estimated query difficulty and operator context to balance task accuracy against inference cost.
Its difficulty estimator learns from workflow-level success.
PerfOrch measures each model's competence separately by stage, language, and category, uses execution-based acceptance to determine whether to try another candidate, and ranks refiners by improvements in the generated program's runtime or memory utilization.

\subsection{LLM-Based Performance Optimization}
\label{sec:related_performance}

Fewer studies measure the runtime performance of generated code.
Li et al.~\cite{li2024assessing, li2026performance} find that LLM-generated code often exhibits suboptimal performance compared to human-written solutions.
Huang et al.~\cite{huang2024effibench} find in EffiBench that GPT-4's code takes 3.12 times the execution time of the human-written canonical solutions on average, and that correctness does not predict efficiency, since open-source models with lower pass@1 often score better on efficiency by solving only the simpler problems. Niu et al.~\cite{niu2024evaluating} report that model size does not predict the efficiency of generated code, whereas training strategy and prompting do.
Two directions improve efficiency directly. Shypula et al.~\cite{shypula2024learning} train on a corpus of human performance-improving edits and show that retrieval and self-play raise the speedup a model produces, and Acharya et al.~\cite{acharya2025optimizing} retrieve optimization examples at inference time to guide runtime improvement.
Huang et al.~\cite{huang2024effilearner} close the loop with measurement, feeding profiled runtime and memory readings back to the model for self-optimization.
The frameworks of Section~\ref{sec:related_agentic} accept a solution once it passes its tests, so the execution time and memory utilization of what they accept stay unmeasured.

\subsection{Code Generation Benchmarks}
\label{sec:related_benchmarks}

\noindent\textit{Function-level and program-level benchmarks.} Beyond the four benchmarks of Section~\ref{sec:background_bench}, function-level and standard-I/O benchmarks differ mainly in where their problems come from. Austin et al.~\cite{austin2021program} release MBPP, 974 crowd-sourced entry-level Python problems, and Liu et al.~\cite{liu2023your} show that the original HumanEval test suites accept many incorrect solutions and release EvalPlus with far larger suites. Contest problems are a harder source. Li et al.~\cite{li2022competition} collect CodeContests from Codeforces and other contest platforms to train and evaluate AlphaCode, and Khan et al.~\cite{khan2024xcodeeval} build xCodeEval from about 7.5K unique problems in 11 languages with seven tasks, including program synthesis, repair, and translation. Trentini et al.~\cite{trentini2025advancing} use USACO contest problems, and Dingle and Krulis~\cite{dingle2024tackling} use university programming assignments. Other benchmarks target contamination. Jain et al.~\cite{jain2025livecodebench} keep adding newly published contest problems to LiveCodeBench, and Xia et al.~\cite{xia2025leetcodedataset} date each LeetCode problem so that a model is evaluated only on problems released after its training cutoff. Zhang et al.~\cite{zhang2025unseen} obfuscate the descriptions, code, and dependencies of functions from real projects in ObfusEval and observe pass-rate decreases of up to 62.5\% after obfuscation. A further group measures properties other than the pass rate. Peng et al.~\cite{peng2025coffe} add worst-case inputs in COFFE so that efficiency is measured where algorithmic complexity becomes visible, Guo et al.~\cite{guo2024risky} measure risky execution behavior, and Xu et al.~\cite{xu2025licoeval} measure license compliance in generated code.

\noindent\textit{Repository-level benchmarks.} Repository-level benchmarks ask for code that fits into an existing project. Du et al.~\cite{du2024classeval} build ClassEval, 100 Python tasks that each ask for a whole class whose methods depend on one another, and find all eleven evaluated LLMs weaker at class level than at method level. Yu et al.~\cite{yu2024codereval} draw the 230 Python and 230 Java functions of CoderEval from open-source projects, and many of them call other code in the same project. CrossCodeEval~\cite{ding2023crosscodeeval} and RepoBench~\cite{liu2024repobench} evaluate code completion that needs context from other files of the repository. Jimenez et al.~\cite{jimenez2024swebench} pair real GitHub issues in SWE-bench with the pull requests that resolved them and check a generated patch against the repository's tests. Li et al.~\cite{li2024devbench} cover the software development lifecycle in DevEval (originally DevBench), with 22 repositories in four languages and tasks from software design and environment setup to implementation and testing. Two recent benchmarks measure performance at this level. He et al.~\cite{he2025sweperf} build SWE-Perf from 140 performance-improving pull requests, and Shetty et al.~\cite{shetty2025gso} pose 102 optimization tasks across 10 codebases in GSO, on which leading software-engineering agents succeed on fewer than 5\% of the tasks. PerfOrch takes a function-level specification and its tests as input, so applying it to these benchmarks requires the localization or retrieval front end discussed in Section~\ref{sec:repo_level}.

%% file: section/09.conclusion.tex
\section{Conclusion and Future Work}
\label{sec:conclusion_overall}

No single LLM dominates code generation across programming languages, algorithmic categories, and generation stages, and competence inverts across those stages.
By combining models with complementary strengths, PerfOrch can repair errors and improve execution performance that persist when a single model handles the entire pipeline.
PerfOrch turns that observation into a system that decomposes code generation into categorization, generation, debugging, and refinement agents, each selecting from a pool of candidate LLMs through Memory rankings profiled offline per language and category.

Our evaluation on four benchmarks across five languages produced four findings.
First, per-stage cross-model composition outperforms every single-model pipeline on all four benchmarks.
Second, PerfOrch reduces execution time for an average of 75.03\% of solved problems across the twelve benchmark--language settings, with an average speedup of 1.18$\times$.
Third, Memory rankings built entirely from HumanEval-X profiling remain effective on EffiBench-X, BigCodeBench, and APPS without re-profiling, showing that the profiled model capabilities support collaboration beyond the profiling benchmark.
Fourth, sequential acceptance matches the execution-time refinement coverage of exhaustive multi-model evaluation with 66\% fewer refinement-stage tokens; the refinement-stage token saving for average-memory optimization is 57\%.

PerfOrch separates which models to use from how the pipeline proceeds. Adding or replacing a model updates the Memory rankings and leaves the four agents and their routing logic unchanged. Section~\ref{sec:model_update_case_study} exercises that separation. After all five models are replaced with newer versions, PerfOrch matches or exceeds the strongest correctness baseline and leads the reported refinement metrics, and incremental generation-stage profiling reaches those rankings at roughly one sixth of the cost of a full pass, within 1.70--4.51~pp of the measured values.

Several directions remain open. PerfOrch operates at the function level, matching the scope of all four benchmarks. Section~\ref{sec:repo_level} identifies the point inside a repository-level pipeline where it would attach and the conditions that integration would have to meet. The failure analysis in Section~\ref{sec:qualitative_analysis} marks two boundaries of the current approach: specification misunderstandings shared by every candidate model, and problems whose correct algorithm lies beyond the reach of iterative debugging. The first points toward specification enrichment such as test amplification or formal preconditions, the second toward search-based program synthesis. 

\textbf{Data Availability:} Replication package (all code and data) can be accessed at our repository: \url{https://figshare.com/s/2e6ba88f5be604c28b17}. Upon acceptance, we will make this repository publicly available. 

%% file: section/Appendix.tex
\appendix
\clearpage

\section{Appendix}
\label{sec:appendix}

\subsection{LLM Profiling Details}
\label{sec:appendix_profiling}

Section~\ref{sec:experiment_memory} reports category-level execution-time OPT and SP that populate the refinement-stage memory rankings. Tables~\ref{tab:memory_refine_opt_average_memory_by_language_category} and~\ref{tab:memory_refine_imp_average_memory_by_language_category} present the corresponding results for average memory utilization, measured by OPT and $\overline{IMP}$, respectively.

Three patterns appear in the average-memory profiling data.
\textit{First}, model rankings shift between the two metrics. In Java, Gemini~2.0~Flash leads execution-time OPT in most categories (Table~\ref{tab:memory_refine_opt_execution_time_by_language_category}), yet Claude~3.7~Sonnet dominates average-memory OPT across the board (e.g., 77.92\% Array, 82.76\% Sorting in Table~\ref{tab:memory_refine_opt_average_memory_by_language_category}). Speed and memory therefore rank models differently, and the Refinement Agent keeps a separate ranking for each.
\textit{Second}, Go yields the least memory refinement of the five languages: its OPT values cluster in the 9--40\% range and its $\overline{IMP}$ rarely exceeds 0.50. Rust Number~Theory sits at the other end, with the largest absolute memory reductions in the table ($\overline{IMP}$ up to 8.68 for GPT-4o).
\textit{Third}, isolated category--model spikes persist: Grok~3 reaches 72.73\% OPT on C++--Hash~Table yet falls to 30.26\% on C++--Math, and Gemini~2.0~Flash achieves 90.91\% OPT on Python--Hash~Table while scoring 62.50\% on Python--Simulation. Aggregating to the language or the model would average these cells away, which is why Memory is indexed per category.

\begin{table}[!ht]
    \caption{Category-level average-memory OPT (\%) on HumanEval-X (Equation~\ref{eq:opt}). Shading marks \cellfirst{first}, \cellsecond{second}, and \cellthird{third} per column. With Table~\ref{tab:memory_refine_imp_average_memory_by_language_category}, these values populate the Refinement Agent's average-memory rankings.}
    \vspace{-0.3cm}
    \centering
    \renewcommand{\arraystretch}{0.85}
    \setlength{\aboverulesep}{0.2ex}
    \setlength{\belowrulesep}{0.2ex}
    \adjustbox{max width=0.98\textwidth}{
    \begin{tabular}{l|l|cccccccccc}
    \toprule
    \textbf{Lang.} & \textbf{Model} & \textbf{Array} & \textbf{Math} & \textbf{String} & \textbf{Count.} & \makecell{\textbf{Num.} \\ \textbf{Theory}} & \textbf{Sim.} & \textbf{Sort.} & \textbf{Enum.} & \makecell{\textbf{Greedy} \\ \textbf{Algo.}} & \makecell{\textbf{Hash} \\ \textbf{Table}} \\
    \midrule
    \multirow{5}{*}{\rotatebox{90}{Python}} & GPT-4o & \cellsecond{58.44} & \cellsecond{61.84} & \cellsecond{51.35} & \cellfirst{68.00} & \cellfirst{66.67} & \cellthird{50.00} & \cellthird{62.07} & \cellfirst{83.33} & \cellthird{63.64} & 36.36 \\
     & Claude 3.7 Sonnet & \cellsecond{58.44} & 56.58 & \cellfirst{58.11} & \cellsecond{62.00} & \cellsecond{60.61} & \cellfirst{62.50} & \cellsecond{65.52} & \cellthird{58.33} & \cellsecond{72.73} & \cellsecond{72.73} \\
     & Gemini 2.0 Flash & \cellfirst{67.53} & \cellfirst{64.47} & \cellfirst{58.11} & \cellthird{58.00} & \cellsecond{60.61} & \cellfirst{62.50} & \cellfirst{68.97} & \cellsecond{75.00} & \cellthird{63.64} & \cellfirst{90.91} \\
     & Grok 3 & \cellthird{57.14} & 52.63 & \cellthird{50.00} & 50.00 & \cellthird{54.55} & \cellsecond{53.12} & 58.62 & \cellthird{58.33} & \cellfirst{81.82} & \cellthird{45.45} \\
     & Qwen 2.5 72B & 54.55 & \cellthird{60.53} & \cellsecond{51.35} & 52.00 & \cellsecond{60.61} & \cellthird{50.00} & 51.72 & \cellthird{58.33} & \cellthird{63.64} & 36.36 \\
    \midrule
    \multirow{5}{*}{\rotatebox{90}{Java}} & GPT-4o & 58.44 & 59.21 & 51.35 & 54.00 & \cellthird{63.64} & \cellthird{65.62} & 48.28 & \cellsecond{58.33} & 63.64 & 36.36 \\
     & Claude 3.7 Sonnet & \cellfirst{77.92} & \cellfirst{76.32} & \cellfirst{67.57} & \cellsecond{76.00} & \cellsecond{69.70} & \cellfirst{81.25} & \cellfirst{82.76} & \cellfirst{75.00} & \cellfirst{90.91} & \cellfirst{81.82} \\
     & Gemini 2.0 Flash & \cellthird{66.23} & 57.89 & \cellthird{58.11} & 58.00 & \cellthird{63.64} & 62.50 & \cellthird{55.17} & \cellthird{50.00} & \cellthird{72.73} & \cellsecond{63.64} \\
     & Grok 3 & \cellsecond{72.73} & \cellsecond{72.37} & \cellsecond{63.51} & \cellfirst{78.00} & \cellsecond{69.70} & \cellsecond{78.12} & \cellsecond{65.52} & \cellthird{50.00} & \cellsecond{81.82} & \cellthird{54.55} \\
     & Qwen 2.5 72B & 61.04 & \cellthird{64.47} & 48.65 & \cellthird{60.00} & \cellfirst{72.73} & 59.38 & \cellthird{55.17} & \cellsecond{58.33} & \cellsecond{81.82} & 27.27 \\
    \midrule
    \multirow{5}{*}{\rotatebox{90}{C++}} & GPT-4o & \cellsecond{42.86} & \cellsecond{38.16} & \cellthird{58.11} & \cellthird{50.00} & \cellsecond{45.45} & 37.50 & \cellsecond{48.28} & \cellthird{16.67} & 36.36 & \cellthird{45.45} \\
     & Claude 3.7 Sonnet & \cellfirst{58.44} & \cellfirst{50.00} & \cellfirst{74.32} & \cellfirst{68.00} & \cellfirst{54.55} & \cellfirst{59.38} & \cellfirst{62.07} & \cellfirst{58.33} & \cellfirst{63.64} & \cellsecond{54.55} \\
     & Gemini 2.0 Flash & \cellthird{41.56} & \cellthird{34.21} & \cellthird{58.11} & 38.00 & 36.36 & \cellsecond{46.88} & \cellthird{31.03} & 8.33 & \cellsecond{54.55} & \cellthird{45.45} \\
     & Grok 3 & \cellthird{41.56} & 30.26 & \cellsecond{62.16} & \cellsecond{52.00} & \cellthird{39.39} & \cellthird{43.75} & \cellsecond{48.28} & \cellsecond{25.00} & \cellthird{45.45} & \cellfirst{72.73} \\
     & Qwen 2.5 72B & 37.66 & 27.63 & \cellthird{58.11} & 40.00 & 33.33 & 37.50 & \cellthird{31.03} & \cellthird{16.67} & 27.27 & \cellthird{45.45} \\
    \midrule
    \multirow{5}{*}{\rotatebox{90}{Go}} & GPT-4o & \cellfirst{31.17} & \cellfirst{32.89} & \cellthird{25.68} & \cellsecond{28.00} & \cellsecond{36.36} & \cellthird{31.25} & \cellsecond{24.14} & \cellfirst{41.67} & \cellsecond{9.09} & \cellfirst{36.36} \\
     & Claude 3.7 Sonnet & 22.08 & 23.68 & \cellfirst{33.78} & \cellfirst{30.00} & \cellsecond{36.36} & 25.00 & \cellfirst{34.48} & \cellsecond{25.00} & \cellfirst{27.27} & \cellfirst{36.36} \\
     & Gemini 2.0 Flash & 22.08 & \cellsecond{28.95} & 22.97 & \cellthird{18.00} & \cellfirst{39.39} & \cellsecond{34.38} & \cellthird{17.24} & \cellsecond{25.00} & \cellsecond{9.09} & \cellthird{9.09} \\
     & Grok 3 & \cellsecond{25.97} & \cellthird{26.32} & \cellsecond{29.73} & \cellthird{18.00} & 27.27 & \cellfirst{43.75} & \cellsecond{24.14} & \cellsecond{25.00} & \cellfirst{27.27} & \cellsecond{18.18} \\
     & Qwen 2.5 72B & \cellthird{23.38} & \cellthird{26.32} & 20.27 & 12.00 & \cellthird{33.33} & 21.88 & \cellthird{17.24} & \cellthird{16.67} & \cellfirst{27.27} & 0.00 \\
    \midrule
    \multirow{5}{*}{\rotatebox{90}{Rust}} & GPT-4o & \cellsecond{46.75} & \cellsecond{42.11} & \cellthird{58.11} & 50.00 & \cellfirst{54.55} & \cellsecond{46.88} & \cellsecond{41.38} & \cellthird{33.33} & \cellsecond{45.45} & \cellthird{54.55} \\
     & Claude 3.7 Sonnet & \cellfirst{48.05} & \cellfirst{44.74} & \cellsecond{67.57} & \cellsecond{58.00} & \cellsecond{51.52} & \cellfirst{50.00} & \cellfirst{44.83} & \cellfirst{50.00} & \cellsecond{45.45} & \cellfirst{72.73} \\
     & Gemini 2.0 Flash & \cellthird{44.16} & \cellsecond{42.11} & \cellfirst{70.27} & \cellfirst{60.00} & \cellsecond{51.52} & \cellthird{37.50} & \cellthird{37.93} & \cellsecond{41.67} & \cellsecond{45.45} & \cellsecond{63.64} \\
     & Grok 3 & 38.96 & \cellthird{36.84} & \cellthird{58.11} & \cellthird{52.00} & \cellthird{42.42} & \cellsecond{46.88} & 34.48 & 25.00 & \cellfirst{54.55} & 45.45 \\
     & Qwen 2.5 72B & 28.57 & 28.95 & 41.89 & 40.00 & 36.36 & 31.25 & 24.14 & \cellthird{33.33} & \cellthird{36.36} & 36.36 \\
    \bottomrule
    \end{tabular}
    }
    \vspace{-0.3cm}
    \label{tab:memory_refine_opt_average_memory_by_language_category}
\end{table}

\begin{table}[!ht]
    \caption{Category-level average-memory $\overline{IMP}$ (\%) on HumanEval-X (Equation~\ref{eq:imp}). Shading marks \cellfirst{first}, \cellsecond{second}, and \cellthird{third} per column.}
    \vspace{-0.3cm}
    \centering
    \renewcommand{\arraystretch}{0.85}
    \setlength{\aboverulesep}{0.2ex}
    \setlength{\belowrulesep}{0.2ex}
    \adjustbox{max width=0.98\textwidth}{
    \begin{tabular}{l|l|cccccccccc}
    \toprule
    \textbf{Lang.} & \textbf{Model} & \textbf{Array} & \textbf{Math} & \textbf{String} & \textbf{Count.} & \makecell{\textbf{Num.} \\ \textbf{Theory}} & \textbf{Sim.} & \textbf{Sort.} & \textbf{Enum.} & \makecell{\textbf{Greedy} \\ \textbf{Algo.}} & \makecell{\textbf{Hash} \\ \textbf{Table}} \\
    \midrule
    \multirow{5}{*}{\rotatebox{90}{Python}} & GPT-4o & \cellfirst{1.67} & \cellsecond{3.37} & \cellthird{0.71} & \cellsecond{1.65} & \cellthird{4.16} & 1.56 & \cellsecond{1.67} & \cellfirst{2.38} & 1.81 & 0.60 \\
     & Claude 3.7 Sonnet & \cellfirst{1.67} & \cellfirst{3.50} & \cellsecond{0.91} & \cellfirst{1.74} & \cellsecond{4.24} & \cellsecond{1.70} & \cellfirst{1.70} & 1.56 & \cellfirst{2.47} & \cellsecond{1.21} \\
     & Gemini 2.0 Flash & \cellsecond{1.41} & \cellthird{3.22} & \cellfirst{0.94} & 1.33 & \cellfirst{4.32} & \cellfirst{1.77} & 1.49 & \cellsecond{1.95} & \cellthird{1.83} & \cellfirst{1.27} \\
     & Grok 3 & \cellsecond{1.41} & 2.75 & 0.69 & 1.21 & 3.21 & 1.49 & \cellthird{1.57} & \cellthird{1.87} & 1.66 & \cellthird{0.61} \\
     & Qwen 2.5 72B & \cellsecond{1.41} & 3.06 & 0.65 & \cellthird{1.47} & 4.02 & \cellthird{1.61} & 1.41 & 1.46 & \cellsecond{2.03} & 0.58 \\
    \midrule
    \multirow{5}{*}{\rotatebox{90}{Java}} & GPT-4o & \cellthird{1.68} & \cellthird{1.97} & 1.20 & \cellthird{1.74} & \cellthird{2.25} & \cellthird{1.90} & 1.03 & \cellfirst{1.82} & 1.05 & \cellsecond{1.03} \\
     & Claude 3.7 Sonnet & \cellfirst{2.18} & \cellfirst{2.50} & \cellfirst{2.15} & \cellfirst{2.51} & \cellsecond{2.45} & \cellfirst{2.78} & \cellfirst{1.75} & \cellsecond{1.51} & \cellthird{1.63} & \cellfirst{2.04} \\
     & Gemini 2.0 Flash & 1.64 & 1.61 & \cellthird{1.38} & 1.45 & 1.91 & 1.85 & 1.11 & 0.83 & 1.02 & \cellthird{0.99} \\
     & Grok 3 & \cellsecond{1.88} & \cellsecond{2.19} & \cellsecond{1.74} & \cellsecond{1.81} & \cellfirst{2.56} & \cellsecond{2.47} & \cellsecond{1.57} & 1.16 & \cellsecond{1.65} & 0.87 \\
     & Qwen 2.5 72B & 1.56 & 1.67 & 1.10 & 1.39 & 2.09 & 1.47 & \cellthird{1.20} & \cellthird{1.47} & \cellfirst{2.01} & 0.95 \\
    \midrule
    \multirow{5}{*}{\rotatebox{90}{C++}} & GPT-4o & \cellthird{0.97} & \cellthird{2.43} & 2.12 & \cellthird{2.20} & \cellsecond{4.29} & \cellthird{2.10} & \cellthird{0.58} & \cellthird{0.64} & 0.48 & \cellthird{1.00} \\
     & Claude 3.7 Sonnet & \cellfirst{1.53} & \cellfirst{3.67} & \cellfirst{3.48} & \cellfirst{3.34} & \cellfirst{5.48} & \cellfirst{2.73} & \cellfirst{0.92} & \cellfirst{2.10} & \cellfirst{2.89} & \cellsecond{1.19} \\
     & Gemini 2.0 Flash & \cellsecond{1.23} & \cellsecond{2.57} & \cellthird{2.39} & \cellsecond{2.52} & 3.53 & \cellsecond{2.70} & 0.39 & 0.02 & \cellthird{1.60} & 0.78 \\
     & Grok 3 & 0.92 & 2.17 & \cellsecond{2.51} & 1.95 & 3.85 & 1.90 & \cellsecond{0.69} & \cellsecond{0.72} & \cellsecond{1.77} & \cellfirst{1.44} \\
     & Qwen 2.5 72B & 0.69 & 2.05 & 2.20 & 1.94 & \cellthird{3.96} & 1.52 & 0.36 & 0.45 & 0.24 & 0.52 \\
    \midrule
    \multirow{5}{*}{\rotatebox{90}{Go}} & GPT-4o & \cellfirst{0.31} & \cellfirst{0.22} & \cellthird{0.12} & \cellsecond{0.15} & \cellfirst{0.25} & \cellthird{0.23} & \cellsecond{0.39} & \cellfirst{0.65} & \cellthird{0.11} & 0.04 \\
     & Claude 3.7 Sonnet & \cellthird{0.23} & \cellsecond{0.17} & \cellfirst{0.18} & \cellfirst{0.20} & \cellthird{0.21} & 0.10 & \cellfirst{0.49} & \cellsecond{0.51} & \cellfirst{0.17} & \cellfirst{0.46} \\
     & Gemini 2.0 Flash & 0.20 & 0.11 & \cellsecond{0.13} & \cellthird{0.13} & 0.17 & \cellsecond{0.26} & 0.24 & 0.20 & 0.09 & \cellsecond{0.37} \\
     & Grok 3 & \cellthird{0.23} & \cellthird{0.15} & \cellthird{0.12} & 0.09 & \cellsecond{0.23} & \cellfirst{0.32} & 0.23 & \cellthird{0.30} & \cellsecond{0.12} & \cellthird{0.05} \\
     & Qwen 2.5 72B & \cellsecond{0.24} & \cellthird{0.15} & 0.09 & 0.05 & 0.20 & \cellsecond{0.26} & \cellthird{0.25} & 0.03 & \cellthird{0.11} & 0.00 \\
    \midrule
    \multirow{5}{*}{\rotatebox{90}{Rust}} & GPT-4o & \cellfirst{1.41} & \cellfirst{4.63} & \cellsecond{3.39} & 2.23 & \cellfirst{8.68} & \cellfirst{3.61} & \cellsecond{1.17} & 0.32 & \cellfirst{0.90} & \cellsecond{0.64} \\
     & Claude 3.7 Sonnet & \cellsecond{1.40} & \cellsecond{4.53} & \cellfirst{3.57} & \cellfirst{2.99} & \cellsecond{7.86} & \cellsecond{3.57} & \cellfirst{1.44} & \cellfirst{3.94} & \cellfirst{0.90} & \cellfirst{1.42} \\
     & Gemini 2.0 Flash & \cellthird{1.06} & \cellthird{4.20} & \cellthird{3.16} & \cellsecond{2.62} & \cellthird{7.22} & 2.22 & 0.58 & \cellsecond{0.83} & \cellthird{0.69} & \cellthird{0.51} \\
     & Grok 3 & 0.95 & 3.47 & 2.85 & \cellthird{2.25} & 5.70 & \cellthird{3.08} & \cellthird{0.73} & \cellthird{0.60} & \cellsecond{0.83} & 0.47 \\
     & Qwen 2.5 72B & 0.56 & 2.65 & 1.59 & 1.63 & 4.94 & 2.01 & 0.60 & 0.32 & 0.45 & 0.06 \\
    \bottomrule
    \end{tabular}
    }
    \vspace{-0.3cm}
    \label{tab:memory_refine_imp_average_memory_by_language_category}
\end{table}

\subsection{Prompt Templates}
\label{sec:appendix_prompts}

This subsection lists the prompt templates used by the four PerfOrch agents (Sections~\ref{sec:category}--\ref{sec:refinement}), in pipeline order.

\textbf{Categorizing Agent Prompt.} The Categorizing Agent receives a problem description and assigns one or more of the ten predefined algorithmic categories (Table~\ref{Tab:category}).

\begin{lstlisting}[style=prompt,caption={Categorizing Agent Prompt Template}]
Please classify the following coding problem by assigning the appropriate categories based on the problem description:

Problem Description: [problem_description]

Available categories: [available_cats]

Please select the appropriate categories from the available categories based on the nature of the problem.
If the problem is not related to any of the available categories, please select the closest categories.
\end{lstlisting}

\textbf{Generation Agent Prompt.} The Generation Agent receives a task description, function signature, and example test case, and produces a complete implementation.

\begin{lstlisting}[style=prompt,caption={Generation Agent Prompt Template}]
Please complete [language] code [signature] based on the task description and test cases.

Task Description: [task description]

Test Case: [example test]

Rules:
- Encapsulate the code within a [language] code block
- Do not include the test case within the code block
- Ensure the provided test case passes with your solution
- Implement all logic strictly within a single method
- Do not split code into multiple methods, helper functions, or classes
- Do not change the function signature

Solution Code:
[Your Code Here]
\end{lstlisting}

\textbf{Debugging Agent Prompt.} The Debugging Agent receives faulty code that failed correctness verification and attempts to repair it.

\begin{lstlisting}[style=prompt,caption={Debugging Agent Prompt Template}]
Please fix the following code:
[buggy solution]
\end{lstlisting}

\textbf{Refinement Agent Prompt.} The Refinement Agent receives a correct solution together with an overhead analysis report and optimizes for the user-specified metric (execution time or average memory).

\begin{lstlisting}[style=prompt,caption={Refinement Agent Prompt Template}]
Optimize the [metric] of the following [language] code based on the task, test case, and overhead analysis provided.
Ensure the optimized code can pass the given test case.

Task Description: [task description]

Test Case: [small test cases]

Original Code: [Original Code]

Overhead Analysis: [overhead analysis report]

Optimization Rules:
- Focus solely on code optimization
- Encapsulate the code within a [language] code block
- Do not include the test case within the code block
- Ensure the provided test case passes with your solution
- Implement all logic strictly within a single method
- Do not split code into multiple methods, helper functions, or classes
- Do not change the function signature

Optimized Code:
[Your Code Here]
\end{lstlisting}